%% file: azcyg.tex
\documentclass[preprint,times,tighten]{aastex63}
\pdfoutput=1 
\usepackage{amsmath,amstext}
\usepackage{newtxmath} 
\usepackage{bbold}
\newcommand{\V}[1]{\boldsymbol{#1}}
\newcommand\norm[1]{\left\lVert#1\right\rVert^2}
\shorttitle{Interferometry of AZ~Cyg}
\shortauthors{Norris et al.}

\begin{document}

\title{Long Term Evolution of Surface Features on the Red Supergiant AZ~Cyg}
\author{Ryan P. Norris}
\correspondingauthor{Ryan Norris}
\email{ryan.norris@nmt.edu}
\affil{Department of Physics, New Mexico Institute of Mining and Technology, 801 Leroy Place, Socorro, NM 87801, USA}
\author{Fabien R. Baron}
\affil{Center for High Angular Resolution Astronomy, Department of Physics and Astronomy, Georgia State University, P.O. Box 5060, Atlanta, GA 30302-5060, USA}
\author{John D. Monnier}
\affil{Department of Astronomy, University of Michigan, 1085 S. University Ave., Ann Arbor, MI 48109-1090, USA}
\author{Claudia Paladini}
\affil{European Southern Observatory, Santiago, Chile}
\author{Matthew D. Anderson}
\affil{Center for High Angular Resolution Astronomy, Department of Physics and Astronomy, Georgia State University, P.O. Box 5060, Atlanta, GA 30302-5060, USA}
\affil{The CHARA Array of Georgia State University, Mount Wilson Observatory, Mount Wilson, CA 91023, USA}
\author{Arturo O. Martinez}
\affil{Center for High Angular Resolution Astronomy, Department of Physics and Astronomy, Georgia State University, P.O. Box 5060, Atlanta, GA 30302-5060, USA}
\author{Gail H. Schaefer}
\affil{The CHARA Array of Georgia State University, Mount Wilson Observatory, Mount Wilson, CA 91023, USA}

\author{Xiao Che}
\affil{Department of Astronomy, University of Michigan, 918 Dennison Building, Ann Arbor, MI 48109-1090, USA}

\author{Andrea Chiavassa}
\affil{Universit\'{e} C\^{o}te d'Azur, Observatoire de la C\^{o}te d'Azur, CNRS, Lagrange, CS F-34229, Nice, France}
\author{Michael S. Connelley}
\affil{University of Hawaii, Institute for Astronomy, 640 N. Aohoku Place, Hilo, HI 96720, USA}
\author{Christopher D. Farrington}
\affil{The CHARA Array of Georgia State University, Mount Wilson Observatory, Mount Wilson, CA 91023, USA}
\author{Douglas R. Gies}
\affil{Center for High Angular Resolution Astronomy, Department of Physics and Astronomy, Georgia State University, P.O. Box 5060, Atlanta, GA 30302-5060, USA}
\author{L\'aszl\'o L. Kiss}
\affil{Konkoly  Observatory,  Research  Centre  for  Astronomy  and  Earth Sciences, 1121 Budapest, Konkoly Thege Mikl\'{o}s \'{u}t  15-17, Hungary}
\affil{Institute of Astronomy, School of Physics, University of Sydney, NSW 2006, Australia}
\affil{ELTE E\"{o}tv\"{o}s Lor\'{a}nd University, Institute of Physics, P\'{a}zm\'{a}ny P\'{e}ter s\'{e}t\'{a}ny 1/A, 1117 Budapest, Hungary}

\author{John B. Lester}
\affil{ David A. Dunlap Department of Astronomy \& Astrophysics, University of
Toronto, 50 St. George Street, Toronto, ON, M5S 3H4, Canada}
\affil{Department of Chemical \& Physical Sciences, University of
Toronto Mississauga, Mississauga, ON L5L 1C6, Canada}
\author{Miguel Montarg\`{e}s}
\affil{LESIA, Observatoire de Paris, Universit\'e PSL, CNRS, Sorbonne Universit\'e, Universit\'e de Paris, Meudon, France}
\affil{Institute of Astronomy, KU Leuven, Celestijnenlaan 200D Box 2401, 3001 Leuven, Belgium}

\author{Hilding R. Neilson}
\affil{ David A. Dunlap Department of Astronomy \& Astrophysics, University of
Toronto, 50 St. George Street, Toronto, ON, M5S 3H4, Canada}
\author{Olli Majoinen}
\affil{The CHARA Array of Georgia State University, Mount Wilson Observatory, Mount Wilson, CA 91023, USA}
\author{Ettore Pedretti}
\affil{RAL Space, STFC Rutherford Appleton Laboratory, Harwell, Didcot, OX11 0QX, UK}

\author{Stephen T. Ridgway}
\affil{NSF's NOIRLab, PO Box 26732, Tucson, AZ 85726-6732, USA}

\author{Rachael M. Roettenbacher}
\affil{Yale Center for Astronomy and Astrophysics, Yale University, 46 Hillhouse Avenue, New Haven, CT 06511, USA}

\author{Nicholas J. Scott}
\affil{Space  Science  \&  Astrobiology  Division,  NASA  Ames  Research Center, Moffett Field, CA 94035, USA}

\author{Judit Sturmann} 
\affil{The CHARA Array of Georgia State University, Mount Wilson Observatory, Mount Wilson, CA 91023, USA}
\author{Laszlo Sturmann}
\affil{The CHARA Array of Georgia State University, Mount Wilson Observatory, Mount Wilson, CA 91023, USA}
\author{Nathalie Thureau}
\affil{ Farr Institute Scotland, Nine Edinburgh Bioquarter, Little France, UK}
\author{Norman Vargas}
\affil{The CHARA Array of Georgia State University, Mount Wilson Observatory, Mount Wilson, CA 91023, USA}

\author{Theo A. ten Brummelaar}
\affil{The CHARA Array of Georgia State University, Mount Wilson Observatory, Mount Wilson, CA 91023, USA}

\begin{abstract}

We present \emph {H}-band interferometric observations of the red supergiant (RSG) AZ~Cyg made with the Michigan Infra-Red Combiner (MIRC) at the six-telescope Center for High Angular Resolution Astronomy (CHARA) Array. The observations span 5 years (2011-2016), offering insight into the short and long-term evolution of surface features on RSGs.  Using a spectrum of AZ~Cyg obtained with SpeX on the NASA InfraRed Telescope Facility (IRTF) and synthetic spectra calculated from spherical MARCS, spherical PHOENIX, and SAtlas model atmospheres, we derive $T_{\text{eff}}$ is between $3972 K$ and $4000 K$ and $\log~g$ between $-0.50$ and $0.00$, depending on the stellar model used. Using fits to the squared visibility and Gaia parallaxes we measure its average radius $R=911^{+57}_{-50}~R_{\odot}$. Reconstructions of the stellar surface using our model-independent imaging codes SQUEEZE and OITOOLS.jl show a complex surface with small bright features that appear to vary on a timescale of less than one year and larger features that persist for more than one year. 1D power spectra of these images suggest a characteristic size of $0.52-0.69~R_{\star}$ for the larger, long lived features. This is close to the values of $0.51-0.53~R_{\star}$ derived from 3D RHD models of stellar surfaces. We conclude that interferometric imaging of this star is in line with predictions of 3D RHD models but that short-term imaging is needed to more stringently test predictions of convection in RSGs.

\end{abstract}


\section{Introduction}
 
As stars of mass $9~M_{\odot}$ to $25~ M_{\odot}$ transition to core He burning, they evolve toward the red supergiant (RSG) stage \citep{meynet_evolution}. Reaching radii up to $1500~R_{\odot}$, stars in this evolutionary stage are cool ($3400-4100~K$) and have luminosities  of $20,000$ to $300,000~L_{\odot}$ \citep{levesque_2005}.  RSGs display high mass-loss rates ($10^{-7}$ to $10^{-4}~M_{\odot}~\text{yr}^{-1}$) leading to complex circumstellar environments and a variety of evolutionary outcomes  \citep{vanloon,debeck2010,mauron_josselin}. Stars may remain in this stage until they end in supernovae or they may evolve blueward, possibly to return once more to the RSG stage \citep{meynet_evolution}. Understanding the evolution of RSGs has implications for a variety of astrophysical questions, including the chemical evolution of the Universe and the diversity of supernovae observed.

 RSGs display semi-regular variations comprised of a short period of hundreds of days and a longer period lasting thousands of days \citep{kiss_variability}. Although pulsations have been linked to some of these variations, it is likely that the motions of large surface features also play a role. In the early 1970s, \citet{stothers1972} and \citet{schwarzschild1975} suggested that the observed periodicity was reminiscent of the variability displayed by convection in Sun-like stars but on a larger scale and longer timescale. Using standard mixing length theory, \citet{stothers1972} suggested that RSGs possess a small number of extremely large convection cells ($\sim 0.5-1 R_{\star}$) with lifetimes of 1000s of days. On the other hand, \citet{schwarzschild1975} extrapolated from the behavior of supergranulation in the Sun, predicting a surface occupied by dozens of large granules with lifetimes of 100s of days. Indeed, both types of features are observed in 3D radiative hydrodynamics (RHD) models of RSGs \citep{chiavassa2010,chiavassa2011}. \citet{kravchenko2019} applied tomography to high resolution spectra of the RSG $\mu$~Cep finding that line-of-sight velocity variations exhibited a phase-shift behind temperature and photometric variations, a feature also found in 3D RHD simulations. The authors suggested that this was evidence that the shorter period of hundreds of days was related to convection within the star. The presence of a granulation-related signal in the irregular variability of Betelgeuse was noted by \citet{RSGgranulation2020} who determined a granulation time of 138$^{+8}_{-2}$ days by fitting the posterior of the power density spectrum (PSD) of American Association of Variable Star Observers (AAVSO) observations of the star to a modification of an equation \citet{Harvey} first used to model solar granulation. However, convection is likely not the cause of the dominant short-term periodicity in RSGs. For example, \citet{joyce} found that radial fundamental and first overtone pulsations, originating in the $\kappa$-mechanism, are the cause of Betelgeuse's short-term ($<400$ d) variability.

In 2019, the dramatic dimming of Betelgeuse \citep{dimming} brought further attention to the variability of RSGs and its myriad causes. \citet{betelcoolspot} suggested the dimming was caused by the presence of a large cool spot in the photosphere of the star based on decreased flux at submillimeter wavelengths. On the other hand, other authors have suggested that spectrophotometry \citep{levesequebetel}, polarimetry \citep{polarimetry}, and interferometry \citep{montarges2021} provide evidence that the decrease in visual magnitude was the result of obscuring of the surface, likely due to large-grain dust from mass loss. This claim was bolstered by Dupree  et  al.  (2020) who  reported  the  presence  of  a  large  bright spot  on  Betelguese  prior  to  its  dimming  event.   Because the  location  of  the  spot  was  similar  to  the  location  of  the dimmed  portion  of  the  star,  they  concluded  that  pulsation  and  convection  elevated  material  outward  and  that  this material later cooled, dimming the star. Indeed, the contribution of convection to mass-loss has been a long standing question. Using tomography, \citet{josselin_plez} found velocity gradients indicative of vertical motion on time scales that match the lifetime of convection cells predicted by \citet{schwarzschild1975}. Turbulent pressure from such motions would result in lifting that could play a role in mass loss. However, \citet{torres_paper3} studied the extended atmosphere of three RSGs using spectro-interferometry and found that hydrostatic model atmospheres, 3D RHD simulations of convection, and 1D pulsation models each produced atmospheres that were more compact than observed. Whilst pointing out the limitations of current 3D convection and 1D pulsation models, they noted that these models do not take into account short-timescale velocity gradients, which may result from granulation-like motion in convection cells. 

Because these stars are quite large and because their convection related features are predicted to be large relative to the stellar radius, studying convection in RSGs is amiable to imaging studies. Using aperture masking at the William Herschel Telescope, \citet{BuscherBetelgeuse} and \citet{tuthill1997} found evidence of hotspots via detection of asymmetries in visible wavelength observations of three large RSGs. Because the time scale of the variations matched predictions by \citet{schwarzschild1975}, \citet{tuthill1997} suggested that the hotspots were related to convection. Later observations of Betelgeuse by \citet{youngbetelgeuse} showed that detection of these hotspots is wavelength dependent and that at some wavelengths, RSGs appear featureless. Observations by \citet{haubois} of Betelgeuse in the \textit{H}-band, where H$^{-}$ continuum opacities are at a minimum ($1.6~ \mu$m), suggested that large features do exist in the photosphere, as predicted by 3D RHD models \citep{chiavassaII,montarges2014}. Using the Center for High Angular Resolution Astronomy (CHARA) Array, \citet{baronrsg} imaged the surface of T~Per and RS~Per in the \textit{H}-band, finding more evidence that large surface features are ubiquitous. Looking at short timescale evolution, \citet{montargescetau} monitored the evolving surface features of CE~Tau between November and December 2016, finding evidence of short-term changes. As for longer timescale evolution, \citet{spectropolar} collected spectro-polarimetry of Betelgeuse between 2013 and 2018 and found evidence of warm regions of scale $0.6~R_{\star}$ in the photosphere that persisted for up to 4 years, which the authors tied to giant convection cells. \citet{climent2020} presented reconstructions the surface of V602~Carinae obtained with VLTI-PIONIER. These data, collected in 2016 and 2019, are limited by temporal and angular resolution in their ability to study the size of individual granules. However, they suggest that the surface does not change significantly over a period of at least 70 days and that 3D RHD models are able to reproduce surface features as observed in images of RSGs. Moreover, comparison of continuum images to 3D RHD models suggests that the large features present in these images are indeed due to convection.

 Our work in many ways complements this study of V602~Carinae. In order to better grasp the timescale and spatial scale of surface variations on RSGs, we undertook a 5 year study of the RSG AZ~Cyg using the Michigan Infrared Combiner (MIRC) on the CHARA Array. In section 2 we discuss the observations and data reduction of the interferometric and spectroscopic data we used in this study. In section 3 we present the fundamental stellar parameters of AZ~Cyg. In section 4 we discuss our image reconstruction techniques and present images of AZ~Cyg from the 5 years of observations. In section 5, we use 1D power spectra of the images to compare the reconstructions to predictions of surface feature size by 2D and 3D models. We conclude the paper in section 6 with a summary of the work and discussion about the need for observations at shorter time scales.

\section{Observations}
\subsection{AZ~Cyg}

AZ~Cyg is an M2-4.5 Iab star, with a parallax distance of $d=2090^{+130}_{-115}$ pc based on Gaia data \citep{distgaia}. Using Fourier analysis of AAVSO data, \citet{kiss_variability} determined periods of $495~\pm~40$ days and $3350~\pm~1100$ days for the star. Using self-correlation analysis, \citet{percy1} determined a long secondary period of 2000 days. \citet{rsggaia} found three periods: of 0.2 mag amplitude over 340 days, 0.28 mag over 495 days, and 0.46 over 3350 days.  A study of the star by \citet{kissrsg} with the Michigan InfraRed Combiner (MIRC) using four telescopes on the CHARA Array suggested the presence of complex structures on the surface of the star. These results inspired this more involved study using the updated MIRC after it was upgraded to combine the light from six telescopes, which enabled imaging of the surface structures.

\subsection{CHARA/MIRC Observations}
Over the course of 5 years, we observed AZ~Cyg using MIRC at the Georgia State University's CHARA Array. The CHARA Array is located on Mount Wilson, CA and consists of six 1~m telescopes with 15 baselines ranging from 34~m to 331~m and angular resolutions reaching $\sim$0.5~mas in the \emph {H}-band \citep{ten_brummelaar_first_2005}.

The MIRC instrument, which has since been upgraded to MIRC-X, was a six beam combiner operating in the \emph {H}-band (1.5-1.8 $\mu$m). The instrument was capable of providing 15 visibilities, 10 independent closure phases, and 9 independent closure amplitudes per spectral channel. In the low spectral resolution mode used for our observations (R=30), eight 30~nm wide spectral channels are acquired \citep{monnier_michigan_2004,monnier_resolving_2012}. We recorded 21 data blocks for AZ~Cyg, as listed in Table \ref{tab:obslog}, with each data block corresponding to a continuous observation of a target for 10 minutes. Note that because of  issues with calibration frames taken during the observation run, we were unable to use data from 2012 in our analysis.
\input{Tables/observelog.tex}
\input{Tables/cals.tex}

To reduce data, we used the latest version of the official MIRC reduction pipeline \citep[as of June 2017;][] {monnier_imaging_2007}. In the pipeline, we applied a 17 millisecond coherence time and a cross-talk correction for visibilities less than 0.1.  
Corrections for variations in atmospheric coherence time and optical changes in the beam path were obtained using calibrator stars noted in Table \ref{tab:obslog} and described in Table \ref{tab:cals}.  Following the method of \citet{monnier_resolving_2012}, we also corrected for systematic errors. We used a 6.6\% multiplicative error correction 
and  $2\times10^{-4}$ additive error correction. 
For triple amplitudes, we used a 10\% multiplicative error correction and $1\times10^{-5}$ additive error correction. Following \citet{zhao_toward_2011}, we  applied a $1^{\circ}$ error floor for closure phases.
\input{Figures/uvexample.tex}

Because we do not expect significant brightness variation or rotation over periods of less than one month, we merged calibrated data observed within three weeks of each other into single data files for each epoch prior to analysis and image reconstruction. The purpose of this was to build up (\textit{u},\textit{v}) coverage and increase the number of data to points to assist in image reconstruction. In Figure \ref{fig:uvone}, we display the (\textit{u},\textit{v}) coverage for the 2011 epoch. Figure \ref{fig:uvall} in the Appendix contains plots of (\textit{u},\textit{v}) coverage for all epochs discussed in this paper. Because of observing conditions or ongoing work on different telescopes, it was not always possible to record data with all six telescopes.

\subsection{SpeX IRTF Spectroscopy}
In addition to the interferometric data, we obtained a spectrum of AZ~Cyg on 2016 September 6 using the short cross dispersed (SXD) mode on the SpeX spectrograph on the NASA Infrared Telescope Facility (IRTF) 3~meter telescope atop Mauna Kea \citep{rayner_spex}. SpeX's SXD mode covers 0.7-2.5 $\mu$m at $R\sim2000$ and is matched to a 0.3x15" slit. In order to correct for additive systematic effects such as dark current and sky background, we employed an ``AB'' method in which we observed the target at two different positions along the slit. To build up signal-to-noise (S/N), we collected 5 AB pairs. The total integration time on the target was 13.902 s.

For purposes of flux calibration and correction of telluric features, we obtained 4~AB~pairs of the A0~V standard HD~201320, which we selected using a list provided by the SpeX instrument team. The integration time on each exposure was 29.65~s. We used Spextool, an Interactive Data Language (IDL) package \citep{cushing_spextool} to reduce data. During this process, the standard star spectrum was used to generate a model telluric spectrum by adjusting a convolved model spectrum of Vega to match the standard star, as described in \citet{vacca2003}, and dividing the standard spectrum by this model so that only telluric lines remained. To get a spectrum corrected for telluric lines, we divided the AZ~Cyg spectrum by this derived telluric spectrum. We shifted the telluric spectrum roughly 0.1-0.2 pixels in order to minimize systematic errors around sharp telluric lines. Telluric correction using this method within the SpeX pipeline calibrates flux to 10\% accuracy \citep{irtflibrary}. Calibration frames also included a flat field and Argon arc lamp exposure. In the final step of data reduction, we merged the order separated spectra into a single spectrum. In SpeX spectra, the offset between neighboring orders  is between 1-3\% \citep{irtflibrary}. 

Following reduction, we followed the procedure of \citet{fluxcal} to absolutely flux calibrate our data. In short, we determined a calibration factor, 

\begin{equation}
    \langle C \rangle = \frac{\sum_{X} w_{X}\textrm{Max}[f(C_{X})]}{\sum_{X}w_{X}}
    \label{calibrateflux}
\end{equation}

that we used to scale the entire spectrum, whilst preserving the relative flux calibration between each order. In equation \ref{calibrateflux}, $f(C_{X})=10^{(X_{\textrm{AB}}-\mathcal{N}_{\textrm{obs}})/2.5}$ where $X_{\textrm{AB}}$ is the AB magnitude determined from the observed spectrum, $\mathcal{N}_{\textrm{obs}}$ is a random normal distribution determined using 2MASS magnitudes and errors from \citet{2masscutri}, and $w_{X}$ are weights comprised of $\frac{1}{1~\sigma}$  deviations from $f(C_{X})$.

\section{Stellar Parameters}

\subsection{Angular Diameter}

We determined uniform disk (UD) and power law limb darkened disk (LDD) diameters using $\chi^{2}$ minimization of models to the squared visibilities. The UD complex visibilities are given by the equation $V_{UD}(x)=\frac{2J_{1}(x)}{x}$  where $x=\pi \nu \theta_{UD}$, $J_{1}$ is a Bessel function of the first kind, $\nu$ the spatial frequency at which the visibility function is sampled, and $\theta_{UD}$ the angular diameter being fit. For the limb darkened disk diameter, we used the Hestroffer law \citep{hestlaw}:
\begin{equation}
    I_{\lambda}(\mu)/I(1)=\mu^{\alpha} \, ; \, \alpha \in \mathbb{R}^{+}
    \label{heflaw}
\end{equation}
where $I$ is intensity, $\mu=\sqrt{1-(2r/\theta_{LD})^{2}}$ with $r$ the angular distance from the center of the star, $\theta_{LD}$ the angular diameter of the photosphere, and $\alpha$ is the limb darkening parameter. The complex visibility function corresponding to this law, as a function of $x=\pi \nu \theta_{LD}$, is:
\begin{equation}
    V(x)=\sum_{k\geq 0} \frac{ (-1)^k \Gamma(\alpha/2+2)}{\Gamma(\alpha/2+k+2)\Gamma(k+1)}\left(\frac{x}{2}\right)^{2k}
    \label{visld}
\end{equation}
where $\Gamma$ is the Gamma function \citep{arcturusiota}.

\input{Tables/diameters.tex}
We present the results of our fits in Table \ref{tab:diameters}. Note that we did not observe any significant dependence on wavelength in the angular diameters when fitting using individual channels. 
The angular diameter and limb darkening parameter of the star seem to vary temporally. Because we did not limit our fits to the first lobe, the reduced $\chi^2$ values for fits to the uniform disk are high. This is to be expected for an object which deviates from a uniform disk, such as a star with spots on it and a darkened limb. Although lower, reduced $\chi^2$ values for fits to the limb darkened disk are likewise impacted by the presence of such features. In Figure \ref{fig:temporal} we plot the visibility curve of each epoch, denoted by different colors. Looking at the first two lobes, which provide the diameter information, one can see that the differences in diameter arise from the second lobe and lower visibilities of the first lobe. This suggests that the variation in measured diameters comes from difficulties in fitting to the limb of the star, which would be sensitive to the presence of bright features.


\input{Figures/temporalfig.tex}

\subsection{Fundamental Stellar Parameters}

\input{Figures/phoenixfit.tex}
To determine the fundamental stellar parameters of AZ~Cyg, we made use of atlases of synthetic spectra based on spherical MARCS models \citep{gustafsson_grid_2008-1}, SAtlas models \citep{SAtlas}, and synthetic spectra from the \citet{phoenixrsg} spectral library, which is based on spherical PHOENIX models. Within the \citet{phoenixrsg} grid, we selected the spectra with solar metallicity but $^{4}$He, $^{12}$C,$^{14}$N, and $^{16}$O abundances modified to match those of a $M=20~M_{\odot}$ RSG near the end of its life. The MARCS grid consisted of 280 models, covering  $T_{\text{eff}}=3300-4500~K$, $\log~g=-0.5 \pm 1.0$, and [Fe/H]$=0.0 \pm 1.0$. All models were calculated for $M=15~M_{\odot}$. To generate the model spectra from MARCS models we used TURBOSPECTRUM~v19.1, a 1D~LTE spectral synthesis code that incorporates numerous atomic and molecular features \citep{turbospectrum,turbspectrumcode}. We calculated  spectra with  $v_{\text{\rm turb}} = 2~\text{km}~\text{s}^{-1}$ and 0.1~\AA~steps from 650-5500~nm, in order to cover the complete range of our SpeX observations.

To generate the grid of SAtlas model atmospheres we first calculated a grid of plane parallel model atmospheres, starting with the initial grid of model atmospheres from Fiorella Castelli's website\footnote{\url{http://wwwuser.oats.inaf.it/castelli/}}  which covered the parameter space $T_{\text{eff}}=[3500,3750,4000,4250,4500]~K$, $\log~g=[0,0.5,1.0]$, [Fe/H]=[-5.5:0.5 in 0.5 steps], and v$_{\text{\rm turb}}=2~\text{km}~\text{s}^{-1}$ \citep{castelli_atlas09}. We used the $\kappa_{Rosseland}$ coefficients, the ``new'' opacity distribution functions recomputed with an updated H$_{2}$O line list, and Kurucz's molecular and atomic line lists from Castelli's website. Using this grid, we calculated an expanded grid covering $T_{\text{eff}}=3000-4500$ in $100~K$ steps and $\log~g=-1.0\pm1.0$ in 0.25 steps at [Fe/H]=0.0 and v$_{\text{\rm turb}}=2~\text{km}~\text{s}^{-1}$. Using this grid of plane parallel models, we then calculated spherical models with radius $R=100-900~R_{\odot}$, mass $M=8,10,15,20,30~M_{\odot}$, and luminosity determined by using the corresponding radius and temperatures from $3000-4300 ~K$ in $100~K$ steps. Prior to calculation we checked whether the L, R, and M fell into RSG parameter space, which we defined as $10000~L_{\odot}<L<500000~L_{\odot}$ and $-1.0 < \log~g< 1.0$. This resulted in a grid of 542 model atmospheres.  We then used the SYNTHE suite\footnote{Available in a Linux port at https://wwwuser.oats.inaf.it/castelli/sources.html} to generate model spectra at resolution R=500,000 from 680 to 2700~nm using the Kurucz linelists as inputs \citep{syntheorg,synthelinux}.

We obtained stellar parameters by minimizing the $\chi^{2}$ between  the observed spectra and our model spectra (after convolving to the resolution of SpeX) using the Amoeba function within IDL, which uses the Nelder-Mead downhill simplex method for multi-dimensional minimization. We included a radial velocity correction of $-3.88~ \text{km}~\text{s}^{-1}$ based on the measurement of the radial velocity from Gaia Data release 2 and corrected for reddening using a modified version of the Cardelli reddening law \citep{cardelli} as proposed by \citet{reddening}, setting $R_{v}=4.2$ rather than the normal 3.1 in order to account for circumstellar dust. We allowed the color excess $E(B-V)$ to vary as a parameter in our fits. We limited our fit to $850-2290~\mu\text{m}$ in order to avoid strong TiO and CN lines which emerge further in the atmosphere of the star and would drive the temperature lower \citep{davies_temperatures}.  We report the results of our fitting in Table \ref{tab:parameters} and present an example of the fits in Figure \ref{fig:phoenixfit}, with the rest in Figure \ref{fig:spectra} in the Appendix. Keep in mind that the grids and input parameters for each model atmosphere atlas are somewhat different. Thus, in Table \ref{tab:parameters} luminosities and radii for the MARCS and PHOENIX rows are derived from the effective temperature, $\log~g$, and mass of the best fitting model. For the SAtlas row, effective temperature and $\log~g$ are  derived from the luminosity, radius, and mass of the best fitting model. Note also that the luminosities for the SAtlas models were originally derived from a grid of radii and effective temperatures; as a result, the luminosity grid does not correspond to rounded numbers. 

\input{Tables/parameters.tex}


We find that $T_{\text{eff}}$ ranges from 3972 to 4000 $K$ and $\log~g$ from -0.5 to 0. The color excess we find, ranges from $E(B-V) = 0.54 $ to 0.56.  Gaia \citep{gaia2,gaia2cat} reports $T_{\text{eff}}=3294~K$ for AZ~Cyg, which is close to the value of $T_{\text{eff}}=3300~K$ given by \citet{oldazcyg}. The temperature scale of \citet{levesque_2005}, based on fits of TiO rich optical spectra to spherical MARCS models, gives $T_{\text{eff}}=3535-3660~K$ for spectral types M2-4.5. On the other hand, \citet{vanbelle} gives $T_{\text{eff}}=3513\pm168$~K to $3755\pm194~K$ for spectral types M2-M4.7, based on angular diameters derived from the Palomar Testbest Interferometer data and fits of SEDs to templates from \citet{pickles}. The cooler temperatures reported previously likely result from the inclusion of regions of cool molecules further in the atmosphere of the star. \citet{davies_temperatures} found that measurements of RSG temperatures using spectral regions with strong molecular lines resulted in cooler temperatures. Using the limb-darkened diameters and Gaia distances from \citet{distgaia} ($ d=2090^{+130}_{-115}$ pc) we find an average radius of $R=911^{+57}_{-50}~R_{\odot}$.
 
\section{Image Reconstruction}

\subsection{Reconstruction codes} 

Interferometric imaging constitutes an ill-posed inverse problem, both due to sparse (\textit{u},\textit{v}) coverage and due to the incompleteness of the phase information carried by closure phases. The problem is typically solved by regularized maximum likelihood \citep{thiebaut2017}. The reconstructed image $\V{x}^\text{opt}\in\mathbb{R}_{\ge 0}^{N\times N}$ is defined as the most probable $N \times N$-pixel image given the data, under a set of prior assumptions about the image (such as a given level of smoothness or sparsity). The data are assumed to be normally distributed, following the OIFITS standard \citep{pauls2005}.
 The priors are then imposed both by the optimization engine for image positivity, and by a regularization function $R(\V{x})$, so that $\V{x}^\text{opt}$ is given by:
\begin{equation}
\V{x}^\text{opt} = \underset{ \V{x} \in \mathbb{R}_{\ge 0}^{N\times N}}{\operatorname{argmax}} \left( \chi^2(\V{x}) + \mu R(\V{x}) \right). \label{eq:regularized_imaging}
\end{equation}
where $\mu$ is a hyperparameter that sets the relative strength of the regularization with respect to the $\chi^2$. In this paper, we use two very different imaging codes to produce the reconstructions. Both exemplify two different optimization techniques to solve Eq.~\ref{eq:regularized_imaging}. 

OITOOLS.jl\footnote{\url{https://github.com/fabienbaron/OITOOLS.jl}} uses a quasi-Newton method with positivity constraint \citep{thiebaut2002}, a fast minimization technique which makes use of the analytic expressions of the gradients $\frac{\partial \chi^2}{\partial \V{x}}$ and $\frac{\partial R(\V{x})}{\partial \V{x}}$. A major drawback of OITOOLS.jl is consequently its restricted choice of regularization function, which needs to be both differentiable and convex. 

In contrast, SQUEEZE\footnote{\url{https://github.com/fabienbaron/squeeze}} \citep{baron2010,baron2012} employs slower, gradient-less, stochastic methods. In this paper we used its default simulated annealing mode. The image $\V{x}$ is modeled as a superposition of large number of flux elements (typically 1000-10000 elements). These elements are free to randomly move within the image, increasing or decreasing the $\chi^2 + \mu R$ functional as they do so and as the posterior distribution is explored. The evolution of the image as the elements move constitutes a Markov Chain. After a period of burn-in where elements are settling, the chain stabilizes: the posterior distribution is still being explored, but only close to the optimal solution. Averaging the chain over the last hundreds of iterations then provides a mean image that, while it does not strictly minimize Eq.~\ref{eq:regularized_imaging}, still characterizes the posterior around the solution. A standard deviation is also produced, that quantifies the spread around the mean and thus reflect the posterior sharpness.

\subsection{Regularization} 

The best non-committal regularization in optical interferometry is generally admitted to be total variation \citep{renard2011}, which acts on the image $\V{x}$ by imposing greater sparsity on its spatial gradient $\nabla \V{x}$. The total variation functional $R_{\text{TV}}$ is the $\ell_1$ norm of the spatial gradient, and can be approximated by:
\begin{equation}
R_{\text{TV}}(\V{x}) = || \nabla \V{x} ||_1 \simeq \sum\limits_{i,j} \sqrt{(x_{i,j} - x_{i-1,j})^2 + (x_{i,j} - x_{i,j-1})^2  + \epsilon}  \label{eq:totvar}
\end{equation}
where $\epsilon$ is chosen to be a small number close to numerical precision. This approximation is differentiable and can be readily used in OITOOLS.jl. 

Total variation's effectiveness can be traced to two main effects. First, it is very effective in regularizing the outside of the star, i.e., pixels with zero flux. As such it acts as a soft-support constraint, as opposed to a hard-support constraint enforced e.g., by a mask. Second, total variation prevents high frequency noise.
Total variation, by design, also tends to favor zones of uniform flux separated by sharp boundaries. Since 3D simulations of stellar convection do show convection cells separated by black outline, total variation may be a good choice as long as $\mu$ remains reasonable.

In addition to total variation, we also employ a Laplacian regularizer and $\ell_{0}$ regularizer in this study. The Laplacian regularizer is the $\ell_{1}$ norm of the Laplacian: $R_{\text{LA}}(\V{x}) = \norm{ \nabla^{2} \V{x} }_1$. Because it favors contours, it is particularly useful in imaging stellar surfaces with non-uniform flux. The $\ell_{0}$ regularizer is a pseudo-norm, which is not a norm since it is non-convex: 


\begin{equation}
\ell_0 (\V{x}) = \sum_{i}^{N} \mathbb{1}_{\mathbb{R}}(x_i) , 
\end{equation}
where $\mathbb{1}_{\mathbb{R}_{\ge 0}}$ is the indicator function of $\mathbb{R}_{\ge 0}$, the set of real non-negative numbers. The $\ell_0$ pseudo-norm counts the number of non-zero elements in the vector it is applied to and hereby heavily penalizes stray flux outside of the stellar disk.


\subsection{Results}
Prior to reconstructing images of AZ~Cyg, we ran an experiment to test the impact of different combinations of regularizers and hyperparameters on reconstructions of RSGs. To start, we selected a snapshot of a 3D RHD simulation from \citet{chiavassa2011}, scaled the snapshot to the angular diameter of AZ~Cyg, and copied the (\textit{u}, \textit{v}) coverage from each epoch to make four OIFITS files of simulated observations of a RSG. Using each simulated observation, we ran a large stack of reconstructions with SQUEEZE using different regularizers and hyperparameters. In each reconstruction, we used a mask of a 4-4.25 mas disk and an initial image corresponding to the best fitting limb darkened disk for a given epoch. Each reconstruction ran on 5 independent chains and we aligned and averaged the results of all the chains to produce a single final mean image. We used the $\ell_{1}$ norm to compare each mean image to the source snapshot, convolved to the resolution of the reconstruction and rebinned to match a 4.0~mas star. \citet{gomes2017} found the $\ell_{1}$ norm to be the optimal metric for comparing reconstructions to source images. 

\begin{equation}
    \ell_{1}(\V{x},\V{y})=||\V{x}-\V{y}||_1=\sum_{i}|x_{i}-y_{i}| .
    \label{eq:l1norm}
\end{equation}
For the reconstruction of images using real data from each epoch, we selected the regularizers and hyperparameters used in the reconstruction with the lowest $\ell_{1}$ norm of the simulated data corresponding to that epoch. 
In Figure \ref{fig:experiment}, we depict the source image and the best fitting reconstruction for the simulated observation based on the 2011 observations. In Table \ref{tab:recon}  we present the reconstruction parameters we used for each epoch, based on the results of this experiment.
\input{Figures/experiment.tex}

To produce our images, we ran SQUEEZE using the parameters from Table \ref{tab:recon} on 10 chains for 500 iterations using 4500 elements on a 64x64 grid at a scale of 0.125 mas/pixel, which is roughly $1/8$\textsuperscript{th} the resolution of our data. In each reconstruction we used an initial image of a 4.0-4.25 mas disk, which we smoothed with a Gaussian in order to enforce artificially sharp boundaries at the edges, and used as a prior image the best fitting limb darkened disk for a given epoch. 

\input{Tables/reconstructiontab.tex}

The use of multiple chains offers a chance to investigate the impact of variations in the reconstruction process on the final image. For each epoch, we determined mean and standard deviation images from the result of all chains. We found that the standard deviation image was not very informative, likely because of the use of a mask and initial image, and that the result from each chain was only marginally different.  

We also tested our images for artifacts by simulating an observation of a uniform disk using the (\textit{u}, \textit{v}) coverage of our observations and reconstructing an image using exactly the same parameters. We depict the result of that experiment in Figure \ref{fig:azcygall}. Although the reconstructed images do have some noisy features, we note that in every case the patterns observed in the reconstructions of the true data are not present and that the pixel intensities are much lower than found in the images from the real data.

In the second row of Figure \ref{fig:azcygall}  we present SQUEEZE reconstructed images of AZ~Cyg from 2011 to 2016. In Figure \ref{fig:compare2011}, we present the squared visibilities and closure phases of the 2011 data compared to synthetic squared visibilities and closure phases generated from the mean reconstructions. A figure presenting a comparison of data from all epochs is available in the Appendix in Figure \ref{fig:compare}. The reduced $\chi^{2}$ of the reconstructions, based on comparison of the squared visibilities, triple amplitude, and closure phases of the data, are listed in Table \ref{tab:recon}. The reduced $\chi^{2}$ in Table \ref{tab:recon} and residuals in the bottom panels of Figure \ref{fig:compare} indicate a close match between reconstructed images and data and low frequencies, but that there are notable deviations at higher frequencies and in datasets with longer temporal coverage.

Although the spectral resolution of the observations is rather low, we also reconstructed images for each wavelength channel in order to test for wavelength dependence. These images were produced using the same reconstruction parameters as used in reconstructing images using data from all wavelength channels. We present these images in Figure \ref{fig:azcygallwave} in the Appendix.

\input{Figures/allimage.tex}
\input{Figures/compare2011.tex}

In order to test the results of our reconstructions, we also used OITOOLS.jl to reconstruct the data using total variation and a centering prior. We used the same masks and initial images as used with SQUEEZE. We display the reconstructions made using OITOOLS.jl in the top row of Figure \ref{fig:azcygall}. We note that prominent features found in the images produced using SQUEEZE are present in the reconstructions made with OITOOLS.jl, which suggests that these features are not the result of a bias in the reconstruction process. In the following analysis section, we will use only the images produced using SQUEEZE.

As further tests of our images, we split up the observations from the 2016 epoch into two parts and reconstructed images with these split data. We find that the prominent features in the image produced with the complete dataset remain in the split data sets. We depict the result of this experiment in Figure \ref{fig:split2016}.

\section{Discussion}

In Figure \ref{fig:azcygall} we can see that the photosphere of AZ~Cyg is non-uniform and consists of features of varying size and intensity. The accompanying reconstructions of uniform disks based on simulated observations using the same (\textit{u},\textit{v}) coverage do not show similar features as the reconstructions of AZ~Cyg. Thus we posit that the features in the reconstructions of the star are not artifacts of observational coverage.  

\input{Figures/split2016.tex}

 In Figure \ref{fig:azcygallwave}, it is clear that the images have a wavelength dependence. The spectrum of AZ~Cyg is dominated by CN lines from roughly 1.450 to 1.560~$\mu$m and between 1.560 and 1.640~$\mu$m are several CO bands. Thus, it is likely that reconstructions in these channels display the appearance of the star at higher levels in the atmosphere where these lines form. On the other hand, the channels between 1.670 and 1.760 $\mu$m cover a continuum region of the spectrum. Thus these reconstructions in these channels present the surface of the star.

In each epoch, there are a number of relatively small ($<30$\% R$_{*}$), bright features of various intensities, as well as larger medium intensity regions and smaller dark regions. The size of the small, bright features appears to be similar in every year, but as these features are less than the nominal resolution of the CHARA Array, it could be that this is the result of a limitation in our data. On the other hand, the medium intensity regions seem to be consistent in their size, occupying $>0.5~R_{\star}$, as was found in \citet{chiavassaII}. The smaller, dark features are of comparable size to the bright regions.

3D RHD models, as well as spectroscopic and photometric evidence suggest that the surface of RSGs is occupied by large, long lived convection cells with spatial extent $\ge0.5~R_{\star}$. Within these cells there are expected to be smaller granules with a lifetime on the order of months. In the reconstruction of the 3D RHD model presented in Figure \ref{fig:experiment}, we see that the large convection cells in the source image appear as bright to medium intensity regions in the reconstruction, with the bright regions being granules of rising hot gas. The darker regions in the reconstruction correspond to the thin lanes of infalling gas between the large convection cells. Although the source image has four or five large cells, in the reconstruction, these regions blur together and it is difficult to distinguish more than three separate regions. 

Carrying this same analysis to the reconstructions of AZ~Cyg, we see a strikingly similar surface pattern. In 2011, the surface of AZ~Cyg exhibited several distinct regions that may be convection cells bounded by dark lanes. One of these is in the northwest quadrant of the star, another is in the northeastern quadrant and one is in the southeastern quadrant. In addition, there are several bright regions that may be granules. The size of these granule-like features is at the limit of the interferometer, so these images provide an upper limit of $0.25~R_{\star}$ on the granules.

The image from 2011 is separated too far in time from the 2014 image to make any temporal comparison. However, comparing the 2014-2016 images, allows us to make a rough estimate of the lifetime of these features. In 2014, there is a region in the northwest that may be a convection cell, a feature in the south that may be a separate cell, and a feature in the east that may be another. In 2015, the bright granules are no longer in the same location, as expected, but there remain a medium intensity region in the northwest and another in the south. The similarities between 2015 and 2016 are even more striking. The pattern in the southwest in the 2015 image remains in the 2016 image. This suggests that the lifetime of the larger features is at least one year and possibly two. The rotation periods of most RSGs are on the order of 20 years. Thus, it is not likely that yearly changes on the surface are the result of long lived features moving out of view. To better understand the lifetime of surface features will require data obtained at a shorter time interval, perhaps monthly.


In order to quantitatively analyze our images we first determined the root-mean-square (rms) relative intensity contrast (which we will call contrast throughout the rest this paper) of the surface granulation using the definition of \citet{tremblay2013} which has found use by \citet{wittkowski2017}, \citet{paladiniagb} and \citet{montargescetau} in studies of evolved stars.

\begin{equation}
    \frac{\delta~I_{\text{rms}}}{<I>}=\bigg \langle \frac{\sqrt{<I(x,y,t)^{2}>_{x,y}-<I(x,y,t)>^{2}_{x,y}}}{<I(x,y,t)>_{x,y}}\bigg \rangle_{t}.
    \label{contrast}
\end{equation}

As noted in \citet{tremblay2013}, this quantity is a measure of the deviation of a star from plane-parallel approximation and 3D models with the highest contrast are those with the least efficient convection \citep{trempedach2013}. The contrast is correlated with the Mach number, which is the ratio of the flow and sound speeds, such that stars with high contrast ratios will have larger Mach numbers. This is explained by mixing length theory, which suggests that larger convective velocities correspond to larger temperature fluctuations between convective features and their surroundings. The intensity contrast ratio is correlated with density, with lower densities corresponding to higher Mach numbers. Because lower temperature stars have lower densities, there also exists a correlation between temperature and intensity contrast, with lower temperatures corresponding to larger intensity contrast ratios. This relation is expressed quantitatively (for giant stars) by \citet{trempedach2013}:

\begin{equation}
    I_{\text{rms}}/\% =(20.81\pm1.81) \times \text{log}\text{ } T_{\text{eff}}-(1.11\pm0.14)\times \text{log}\text{ }g-55.46\pm0.29 .
    \label{eq:contrastquant}
\end{equation}

We present the average intensity contrasts in Table~\ref{tab:contrasts} and Figure~\ref{fig:contrasts}. To get these values, we took the average of the contrast measured on the mean image of each chain. This permits our measured values to incorporate the spread resulting from the reconstruction process. We also performed this measurement on the OITOOLS.jl reconstruction, which does not have an error value because the reconstruction process results in only one image. 

We find that our contrast ranges from roughly 5.0\% to 14.5\% in the continuum wavelength channels. Contrasts in observations from 2014 are notably lower that other years. It is unclear if this is the result of an actual difference in the star, or the lower quality of that year's data. Using Equation~\ref{eq:contrastquant} with the parameters from Table \ref{tab:parameters} gives 19.35$\pm$6.52\% through 20.05$\pm$6.53. \citet{paladiniagb} found contrasts of  $11.9\pm0.4$\% to $13.1\pm0.2$\% for an AGB of $T_{\text{eff}}=3200~K$ and $\log~g=-0.4$. \citet{montargescetau} found contrasts of 5\% to 6\% for the RSG CE Tau with $T_\text{eff}=3820\pm135~K$ and $\log~g=0.05^{+0.11}_{-0.17}$. \citet{v776cen} found a contrast of 10$\pm$4\% for the RSG V776 Cen with   $T_\text{eff}=4290\pm760~K$. Thus, our contrasts fall within the range of those found around similar stars.
Based on the models described by \citet{tremblay2013}, which we note do not cover RSG parameter space, these intensity contrasts correspond roughly to Mach numbers between 0.25 and 0.75 at unity Rosseland optical depth ($<\tau_{R}>_{x,y}=1$). In models of red giant stars, \citet{tremblay2013} found relative intensity contrasts of roughly 18\% and 22\%. Though these models have similar temperature to AZ~Cyg, they are of higher gravity ($\log~g=1.0$ and $1.5$). A lower gravity would suggest a lower density and thus larger Mach number and larger intensity contrast, a trend also described by Equation~\ref{eq:contrastquant}. That we find a lower contrast rather than higher suggests some difference between red giants and red supergiants. If we focus on the Mach number as a guide, one explanation is that the convective flow velocity behaves differently. A lower contrast suggests a lower Mach number and thus a lower convective flow velocity, which could result if the depth of the cells was larger in RSGs than in red giants. A lower convective velocity would also suggest a larger advective time scale for the same pressure scale height and thus longer cell lifetimes.
\input{Tables/contrasts.tex}

\input{Figures/contrastfigs.tex}

\input{Tables/sizes.tex}

\subsection{Spatial scales}

\citet{tremblay2013} and \citet{trempedach2013} also offer a way of measuring the characteristic size of features using the power spectra of the intensity maps they generated. This method was adapted by \citet{paladiniagb} to measure the characteristic size of granulation on the AGB $\pi^{1}$ Gruis using images generated from Very Large Telescope Interferometer (VLTI) data. In order to derive the size of features on AZ~Cyg, we followed a similar method. We first took the 1D spatial power spectra of each mean image. We then filtered the disk out of the images by setting the pixel values of an image from the disk boundary outwards equal to the median flux of the pixels within the star. This had the effect of smoothing out the sharp boundary at the edge of the disk and enabled detection of prominent power carrying features within the disk. To make identification of the peak easier, we padded this filtered image by 200 pixels.  We identified the characteristic size of granulation as the remaining peak in a power spectrum. 
We present the power spectrum of each epoch in Figure \ref{fig:powerspectra}. We also present a comparison of power spectra of images produced with SQUEEZE and OITOOLS.jl in Figure \ref{fig:powerspectraoitools}. We note the characteristic size of granulation in Table 
\ref{tab:features}.

 Using mixing length theory with turbulence, \citet{antia1984} predicted a small number of large granules on the surface of cool, evolved stars such as red giants and red supergiants. This was in line with the qualitative arguments by \citet{schwarzschild1975} which  extrapolated from solar data to suggest that the atmospheric depth of the maximum of the superadiabatic temperature gradient was related to the vertical extent of a granule and thus also related to the horizontal size of such a feature. 2D and 3D models \citep{freytag1,trempedach2013,tremblay2013} have found that granule size, $x_{\text{gran}}$ is proportional to the scale pressure height, $H_{p}$, a result of the fact that the mixing length, the length at which a convective element dissipates into its environment, is proportional to $H_{p}$ in current theories of convection. Interestingly, \citet{freytag1} found that the superadiabatic temperature gradient was less useful as a descriptor of $x_{\text{gran}}$ because it deviates from a linear relation at lower gravities. Instead, they proposed the relation $H_{p}=RT_{\text{eff}}/g$ with $R$ the universal gas constant. This can be rewritten \citep{paladiniagb} as 

\begin{equation}
    \frac{x_{\text{gran}}}{R_{\star}}\approx~0.0025~\frac{R_{\star}}{R_{\odot}}\frac{T_{\text{eff},*}}{T_{\text{eff},\odot}}\frac{M_{\odot}}{M_{*}}.
    \label{eq:freytag}
\end{equation}

 \citet{tremblay2013} found that the relation between $x_{\text{gran}}$ and $H_{p}$ was not linear but varied based on Mach number. They noted that $x_{\text{gran}}/H_{p}$ might be related to the ratio between horizontal and vertical velocity, with larger velocity ratios corresponding to larger ratios of $x_{\text{gran}}/H_{p}$. To describe the relationship between stellar parameter and $x_{\text{gran}}$ they determined a parametrization using least squares fits to the granule size described by a power spectrum:

\begin{equation}
    \frac{\text{Char.size}}{[\text{km}]}=13.5 g^{-1}[T_{\text{eff}}-300 \log~g]^{1.75} 10^{0.05[Fe/H]},
    \label{eq:tremeq}
\end{equation}
with $g$ in cgs and $T_{\text{eff}}$ in $K$.
Also using least square fits to characteristic sizes determined from power spectra of 3D models, \citet{trempedach2013} found that the granulation size $A_{\text{gran}}$ was described by the relation 

\begin{equation}
 \label{eq:trempsize}
    \text{log}\frac{A_{\text{gran}}}{[Mm]}\simeq (1.310\pm0.0038)\times \text{log}\text{} T_{\text{eff}}
    -(1.0970\pm0.003)\times \text{log}\text{ }g+0.0306\pm0.0359.
\end{equation}
Finally, using 3D RHD models of red supergiants, \citet{chiavassa1} found the equation of \citet{freytag1} needed to account for turbulent pressure. Including this factor, they used

\begin{equation}
    H_{p}=\frac{kT_{\text{eff}}}{g\mu m_{H}}\Big(1+\beta \gamma \big(\frac{v_{\text{\rm turb}}}{c_{s}}\big)^{2}\Big)
    \label{eq:chiavassa}
\end{equation}
with $\beta$ a parameter close to one, $\gamma$ the adiabatic exponent, $c_{s}$ the sound speed, $\mu$ the average particle mass, $m_{H}$ the atomic mass of hydrogen, and $v_{\rm turb}$ the turbulent velocity.  
Using this equation and models of RSGs, \citet{chiavassa1} found that convective feature size could be obtained by multiplying Equation \ref{eq:freytag} by five. The feature size predicted by this equation matched that of the large convection features in their 3D RHD models. \citet{chiavassa2011} conducted further analysis of the relation between pressure scale height and large surface features, this time as suggested by the standard deviation of photocenter displacement. They found that both 3D RHD models and interferometric observations of RSGs suggests the relation between convection cell size and pressure scale height is different in RSGs than it is in giant stars.
\input{Figures/powerspec.tex}
\input{Figures/powerspecoitools.tex}

\input{Tables/azcygcalc.tex}

In Table \ref{tab:calcazcyg}, we display the parameters of the surface of AZ~Cyg derived from Equations \ref{eq:freytag}-\ref{eq:chiavassa}, which we determined using values from Table \ref{tab:parameters}. We report ranges in calculated granule size for Equations \ref{eq:tremeq} and \ref{eq:trempsize} because these equations depend on $\log~g$, for which we found a wide-range of possible values, depending on the stellar atmosphere code used in the fit (see Table \ref{tab:parameters}). We did not use averages of the parameters in Table \ref{tab:parameters} because they resulted from very different stellar atmosphere codes. Based on these results, it appears that the features measured in the images of AZ~Cyg presented in this paper are the same as those calculated by Equation \ref{eq:chiavassa}. These features are likely the longer lived convection cells. If we assume that these cells last 3350 days, the length of the longest period observed by \citet{kiss_variability} and \citet{rsggaia}, the same cell could be present during the entirety of the observing cycle.  The similarity in measured granule size between 2015 and 2016 suggests that features of that scale do last longer than a year. The difference between those years and the measured size in 2011 may indicate the emergence of a new cell, or it could indicate that there was a change in the size of a cell during its lifetime. Further, long term observations of the star, which we are pursuing, will be necessary to image the full life-time of a convection cell. On the other hand, the features derived using Equations \ref{eq:freytag}, \ref{eq:tremeq} and \ref{eq:trempsize} could correspond to the smaller bright features visible in the images in Figure \ref{fig:azcygall}, which may represent smaller scale granulation with a shorter life time. These features may last on the order months, indicating a need for short-term observations as well as long term study.

\section{Conclusion}

The observations presented here have allowed us to derive the fundamental stellar parameters of the RSG AZ~Cyg. In addition, we have reconstructed \textit{H}-band images of the star, finding that the surface varies from year to year with large features of scale $\sim0.5~R_{\star}$, as predicted by 3D RHD simulations. We also find small bright spots on the surface which are perhaps akin to the granulation predicted by such models, but these features are of a scale which is below the resolution of the CHARA Array and thus analysis of their size beyond the scope of this work. Although it seems like the larger features last for longer than one year, the surface pattern of the small features varies substantially. Observations on a shorter time-scale, on the order of months, will be necessary to get a better understanding of the lifetime of these features. Overall, it seems like the surface of this RSG, and likely others, is dominated by a pattern of large, long lived convection cells and smaller features consisting of hot granules of rising gas which persist for months, but less than one year.

\acknowledgments

RN would like to thank Dr. Carlos Lopez Carrillo for the suggestion to use mean images from each chain when measuring contrast and feature size in order to obtain an understanding of the spread in measurements.

FB and RN would like to acknowledge funding of this research through NSF Awards \#1616483 and \#1814777. RN would also like to acknowledge start-up funding from the Office for Research at New Mexico Tech. This work is based upon observations obtained with the Georgia State University Center for High Angular Resolution Astronomy Array at Mount Wilson Observatory. The CHARA Array is supported by the National Science Foundation under Grant No. AST-1636624 and AST-1715788. Institutional support has been provided from the GSU College of Arts and Sciences and the GSU Office of the Vice President for Research and Economic Development.

This work is based upon observations obtained with the Infrared Telescope Facility, which is operated by the University of Hawaii under contract NNH14CK55B with the National Aeronautics and Space Administration.

This research has made use of the Jean-Marie Mariotti Center \texttt{OIFits Explorer}\footnote{Available at \url{http://www.jmmc.fr/oifitsexplorer}} and  \texttt{Aspro}\footnote{Available at \url{http://www.jmmc.fr/aspro}} services. 

This research has also made use of the SIMBAD database and the VizieR catalogue access tool, CDS, Strasbourg, France (DOI: 10.26093/cds/vizier). The original description of the VizieR service was published in \citet{vizier}.

This publication makes use of data products from the Two Micron All Sky Survey, which is a joint project of the University of Massachusetts and the Infrared Processing and Analysis Center/California Institute of Technology, funded by the National Aeronautics and Space Administration and the National Science Foundation. 

This work has made use of data from the European Space Agency (ESA) mission {\it Gaia} (\url{https://www.cosmos.esa.int/gaia}), processed by
the {\it Gaia} Data Processing and Analysis Consortium (DPAC,
\url{https://www.cosmos.esa.int/web/gaia/dpac/consortium}). Funding for the DPAC has been provided by national institutions, in particular
the institutions participating in the {\it Gaia} Multilateral Agreement.

\bibliographystyle{aasjournal}
%
%
%
\bibliography{azcyg.bib}

\appendix
\section{Additional Material}
\input{Figures/uvall.tex}
\input{Figures/spectra.tex}
\input{Figures/compare.tex}

\input{Figures/wavelength_1.tex}

\end{document}

%% file: Tables/observelog.tex
\begin{deluxetable}{lccc}[h]
\tablewidth{0pt}
\tablecaption{Observing Log for AZ~Cyg \label{tab:obslog}}
\tablehead{
\colhead{Date (UT)} & \colhead{N$_{telescopes}$} & \colhead{N$_{block}$} & \colhead{Calibrators}}
\startdata
2011 Jul 21 & 6 & 1 & 7~And \\
2012 Jun 06 & 5 & 1 & 7~And \\
2012 Jun 11 & 6 & 1 & $\sigma$~Cyg \\
2014 Jul 16 & 5 & 1 & $\sigma$~Cyg, 7~And \\
2014 Jul 17 & 5 & 1 & $\sigma$~Cyg, 7~And \\
2015 Aug 22 & 5 & 2 & $\sigma$~Cyg \\
2015 Aug 23 & 6 & 2 & $\sigma$~Cyg \\
2016 Aug 29 & 6 & 1 & $\sigma$~Cyg
\\
2016 Aug 30 & 6 & 1 & $\sigma$~Cyg
\\
2016 Sep 03 & 6 & 1 & $\sigma$~Cyg
\\
2016 Sep 06 & 5 & 1 & $\sigma$~Cyg
\\
2016 Sep 08 & 5 & 1 & $\sigma$~Cyg, 7~And \\
2016 Sep 09 & 6 & 1 & $\sigma$~Cyg, 7~And \\
2016 Sep 10 & 6 & 4 & $\sigma$~Cyg, 7~And \\
2016 Sep 11 & 6 & 2 & $\sigma$~Cyg, 7~And \\
\enddata

\end{deluxetable}

%% file: Tables/cals.tex
\begin{deluxetable}{ccc}[h!]
\tablewidth{0pt}
\tablecaption{Calibrators used in this study \label{tab:cals}}
\tablehead{
\colhead{Star Identifier} & \colhead{UD (\textit{H}-Band) (mas)} & \colhead{Source}}
\startdata
7 And & 0.65~$\pm$~0.03 & \citet{2015mourard} \\
$\sigma$~Cyg & 0.54~$\pm$~0.02 & \citet{zhao2008} \\
\enddata

\end{deluxetable}

%% file: Figures/uvexample.tex
\begin{figure}[h!]
\begin{center}
\includegraphics[scale=0.25,angle=0]{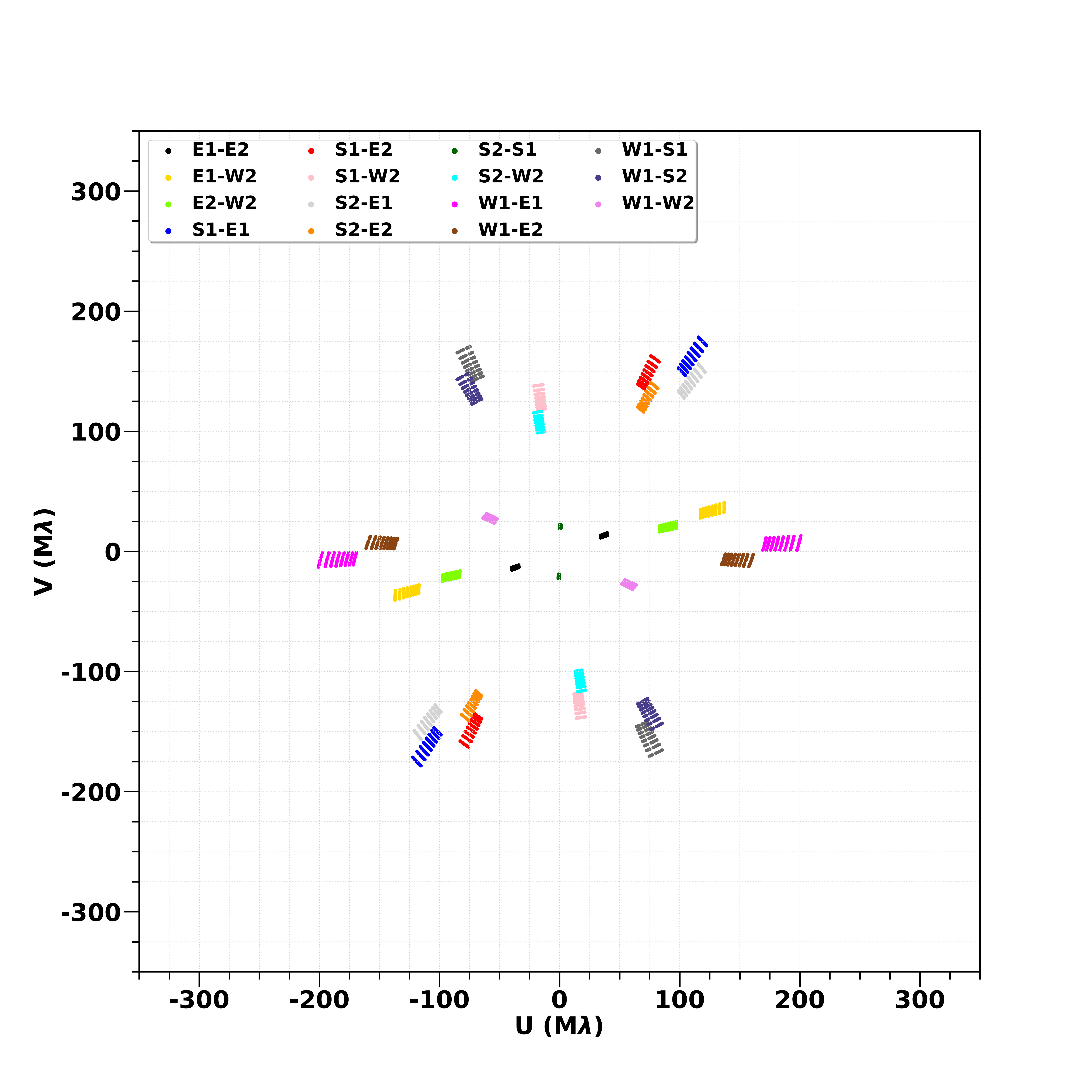}
\caption{(\textit{u},\textit{v}) coverage of AZ Cyg 2011 observations \label{fig:uvone}}
\end{center}
\end{figure}

%% file: Tables/diameters.tex
\begin{deluxetable}{ccccccc}[b]
\tablewidth{0pt}
\tablecaption{Best fit UD and power-law LDD for AZ~Cyg at 1.50-1.75 $\mu$m. Note that fits were made using only $V^{2}$ and not the other interferometric observables. $\chi^2$ are reduced  $\chi^2$. \label{tab:diameters}}
\tablehead{
\colhead{Parameter} & \colhead{2011} & \colhead{2014} & \colhead{2015} & \colhead{2016} &\colhead{All years}}
\startdata
$\theta_{UD}$ (mas) & 3.82$\pm$0.01 & 3.81$\pm$0.01 & 3.90$\pm$0.01 & 3.99$\pm$0.07 & 3.85$\pm$0.01 \\
$\chi^{2}_{UD}$ & 16.08 & 41.05 & 23.23 & 22.49 & 24.68\\
$\theta_{LD}$ (mas) & 3.93$\pm$0.01 & 4.09$\pm$0.01 & 4.11$\pm$0.01 &4.09$\pm$0.01 &4.05$\pm$0.01\\
$\alpha$ & 0.42$\pm$0.01 & 0.59$\pm$0.02 & 0.62$\pm$0.01 &0.72$\pm$0.02 & 0.57$\pm$0.01 \\
$\chi^{2}_{LD}$ & 5.24 & 8.07 & 3.40 & 3.44 & 5.51 \\
\enddata

\end{deluxetable}

%% file: Figures/temporalfig.tex
\begin{figure}[ht]
\begin{center}
\includegraphics[scale=0.6,angle=0]{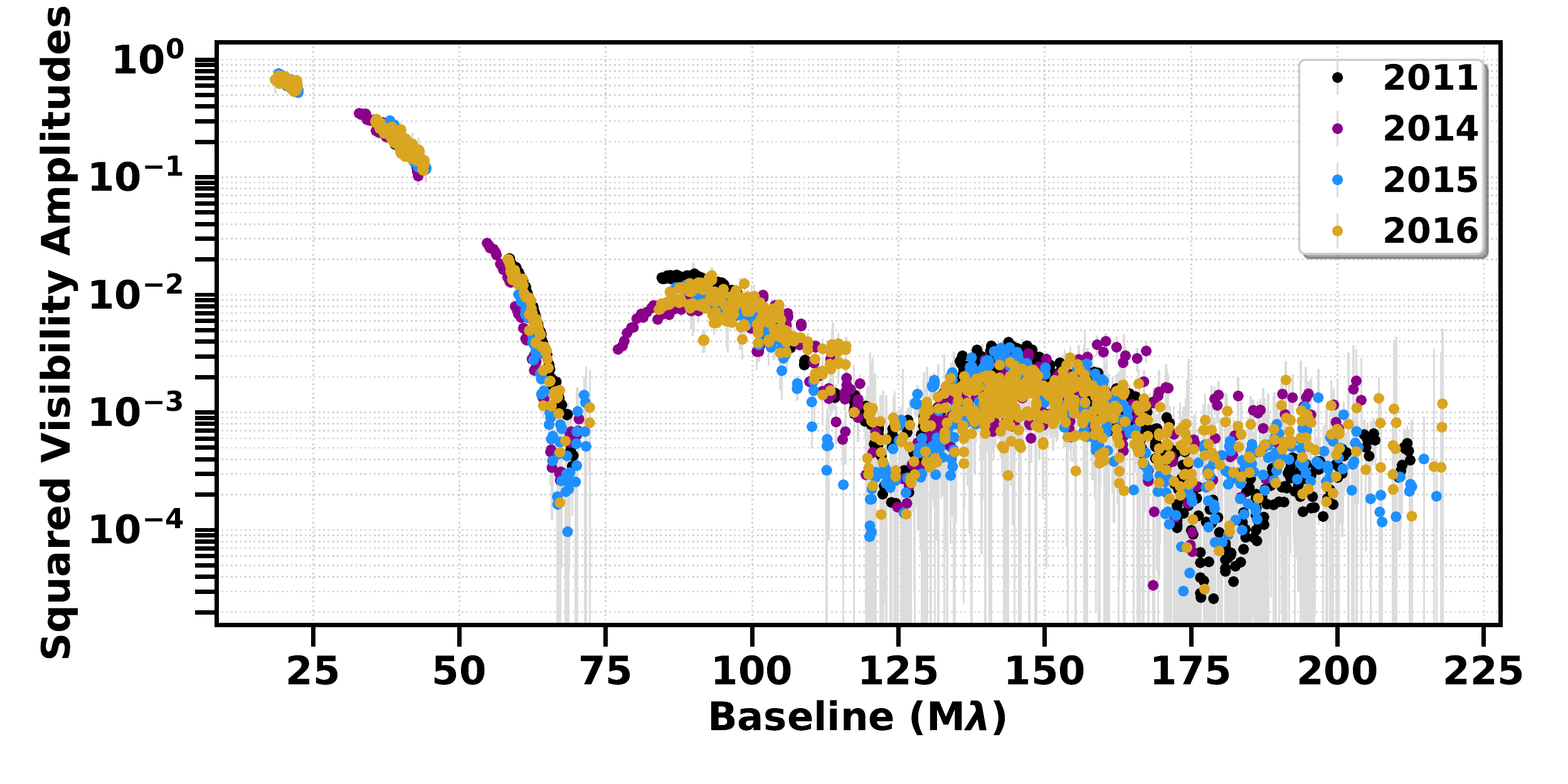}
\caption{Squared Visibilities for each epoch, denoted by different colors \label{fig:temporal}}
\end{center}
\end{figure}

%% file: Figures/phoenixfit.tex
\begin{figure}[ht]
\begin{center}
\includegraphics[scale=0.5,angle=0]{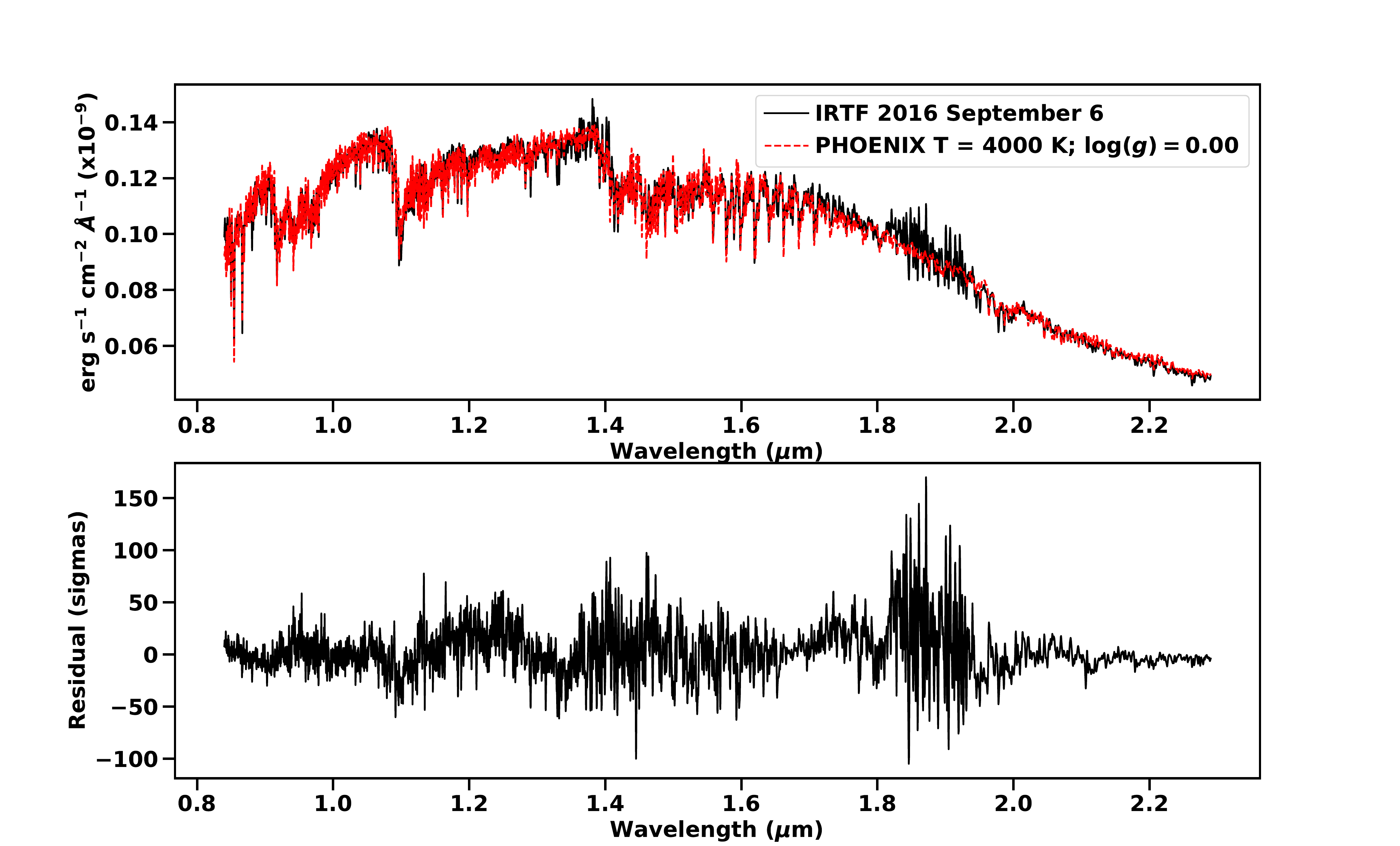}
\caption{ Upper: Best fitting PHOENIX model spectrum (red) compared to observed spectrum. Lower: Residuals (in sigmas) of the observed and model spectrum. \label{fig:phoenixfit}}
\end{center}
\end{figure}

%% file: Tables/parameters.tex
\begin{deluxetable*}{cccccccc}[!h]
\tablewidth{0.2\textwidth}
\scriptsize
\tablecaption{Stellar Parameters of AZ~Cyg corresponding to the best fitting model atmospheres \label{tab:parameters}}
\tablehead{
\colhead{Model} & \colhead{T$_{\text{eff}}$} & \colhead{log \textit{g}} & \colhead{Radius} &  \colhead{Luminosity} & \colhead{Mass} & [Fe/H] & $E(B-V)$\\
\colhead{} & \colhead{(K)} & \colhead{} & \colhead{(R$_{\odot})$} & \colhead{(L$_{\odot})$} &\colhead{(M$_{\odot})$}
&\colhead{(dex)} & \colhead{(mag)}
}
\startdata
MARCS & 4000 & -0.50 & 1040 & 249443 & 15 & 0.0 & 0.56\\  
  PHOENIX & 4000 & 0.00  & 641 & 94759 & 15 & 0.0 & 0.55\\   
 SATLAS & 3972 & -0.07  & 700 & 109828 & 15 & 0.0 & 0.54\\   
\enddata

\end{deluxetable*}

%% file: Figures/experiment.tex
\begin{figure*}
\gridline{\fig{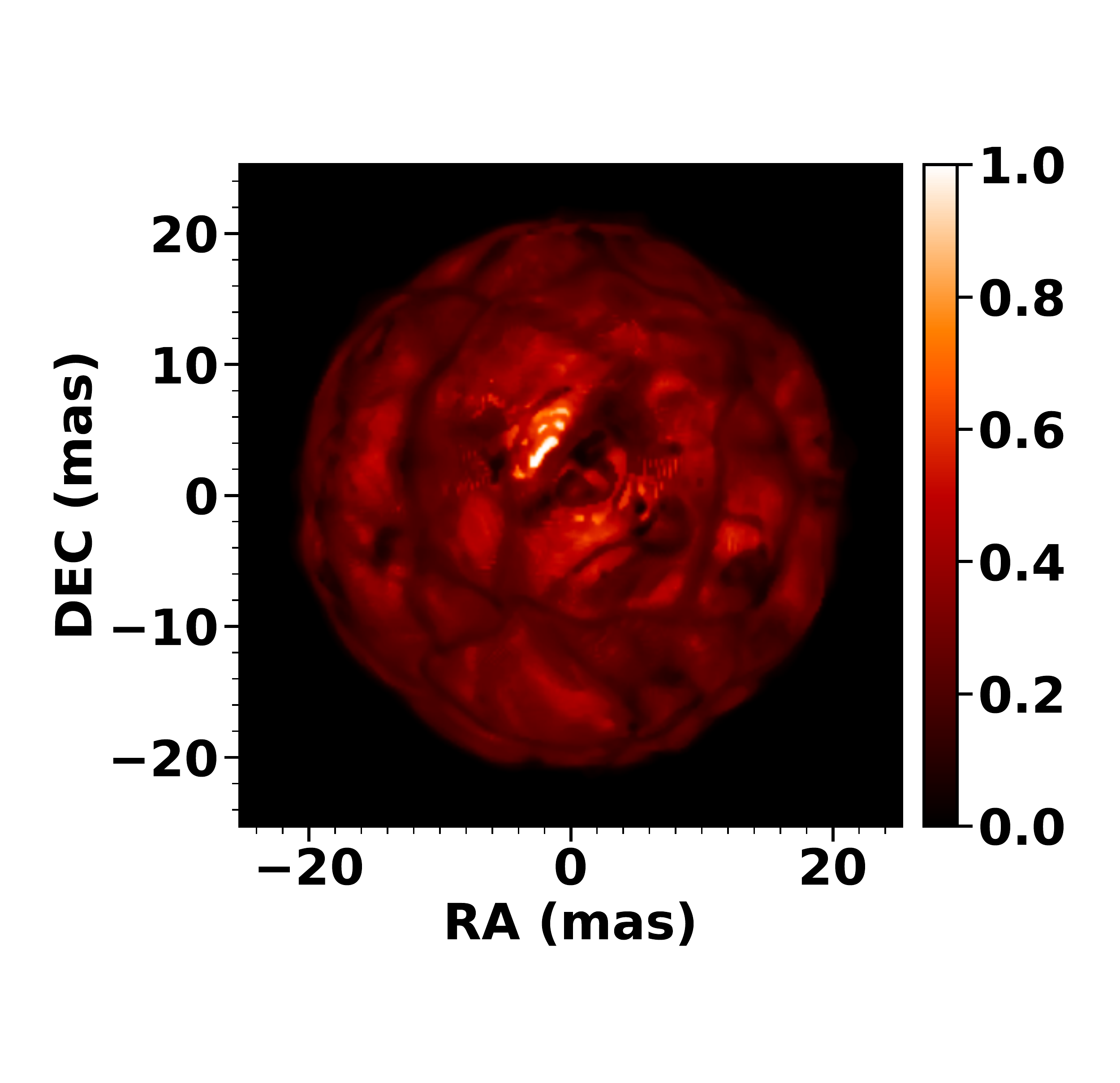}{0.3\textwidth}{Unconvolved Source Image}\\
     \fig{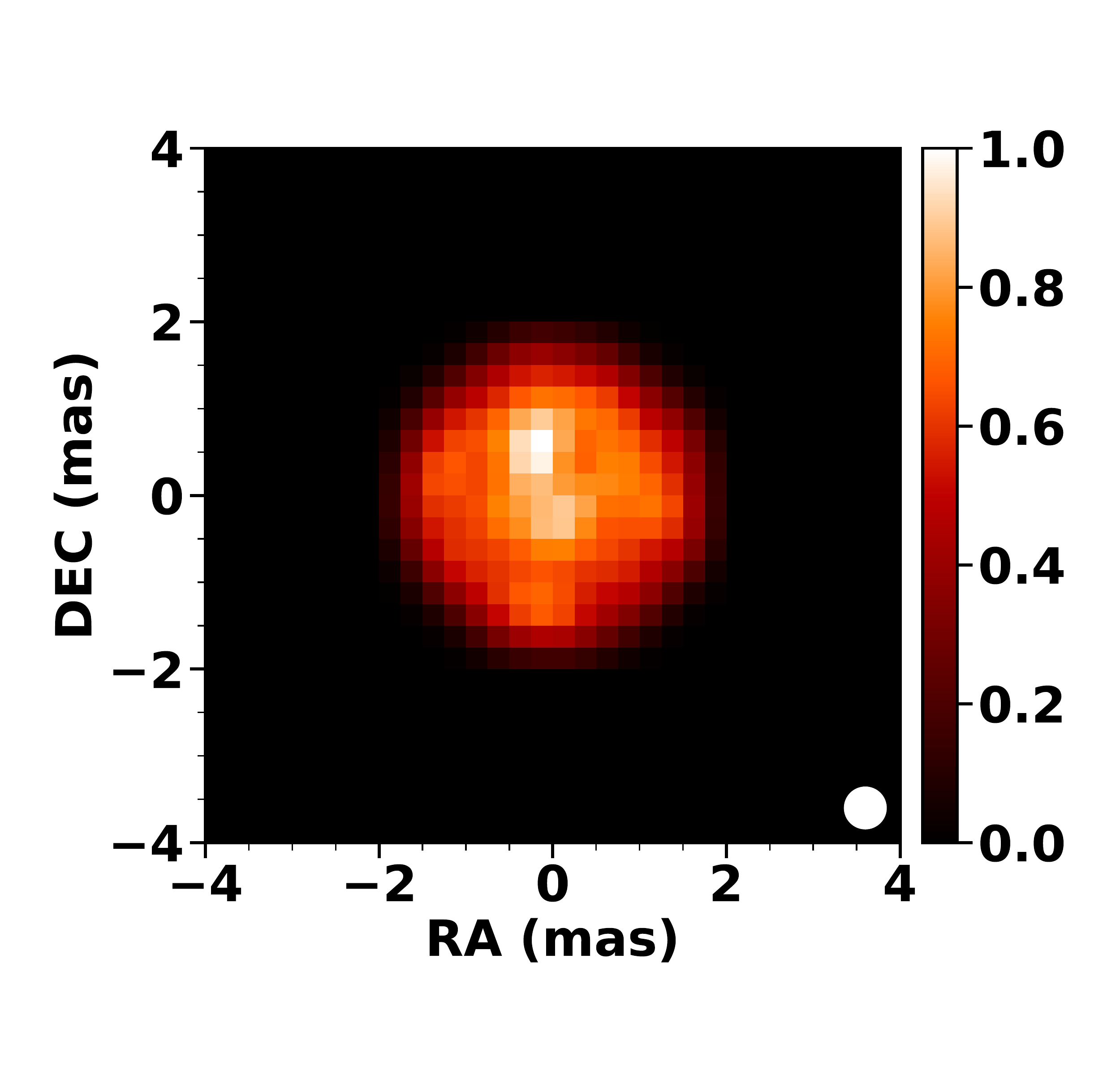}{0.3\textwidth}{Scaled down, convolved Source Image}
     \fig{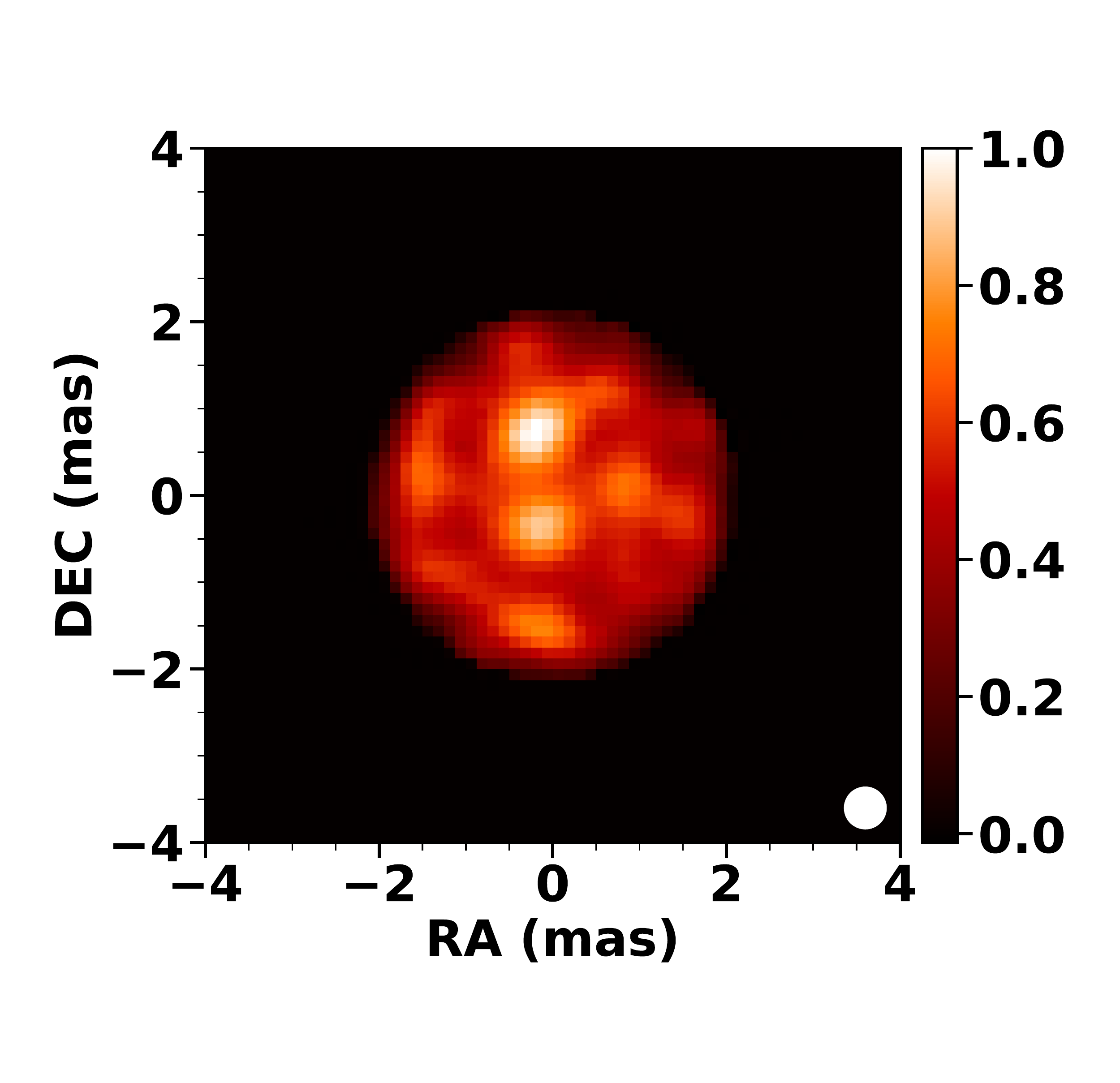}{0.3\textwidth}{Best fitting reconstruction}
          }
\caption{Reconstruction of a simulated observation of the 3D RHD simulation of Betelgeuse from \citet{chiavassa1}, using the (\textit{u},\textit{v}) coverage of the AZ~Cyg 2011 observation. The left-most panel presents an image of the source simulation, the middle panel the source image scaled down and convolved to the resolution of CHARA, and the right most the reconstructed image using the simulated observation. In each image, the beamsize in the lower right corner is the reconstruction resolution. The reconstructed image shows it is possible to reconstruct surface features predicted by 3D RHD models at the resolution of CHARA using typical observational coverage.
\label{fig:experiment}}
\end{figure*}

%% file: Tables/reconstructiontab.tex
\begin{deluxetable}{cccc}
\tablewidth{0.5\textwidth}
\scriptsize
\tablecaption{Regularization parameters used for reconstructing the images in Figures \ref{fig:azcygall},\ref{fig:split2016}, and \ref{fig:azcygallwave}. Note that the $\chi^{2}$ describes the reduced $\chi^2$ of fits of the reconstructions to all the interferometric observables.\label{tab:recon}}
\tablehead{
\colhead{Year} & \colhead{Regularizers} & \colhead{Hyperparameters} &
\colhead{$\chi^{2}$}}
\startdata
 2011 & Total Variation (TV), $\ell_{0}$ & 1000,0.3 & 2.06 \\
 2014 & Laplacian(LA), TV & 2500,500 & 2.36\\
 2015 & LA, TV & 2500,1500 & 4.09\\
 2016 & LA, TV & 300,2000 & 4.80\\
\enddata

\end{deluxetable}

%% file: Figures/allimage.tex
\begin{figure*}

\gridline{\fig{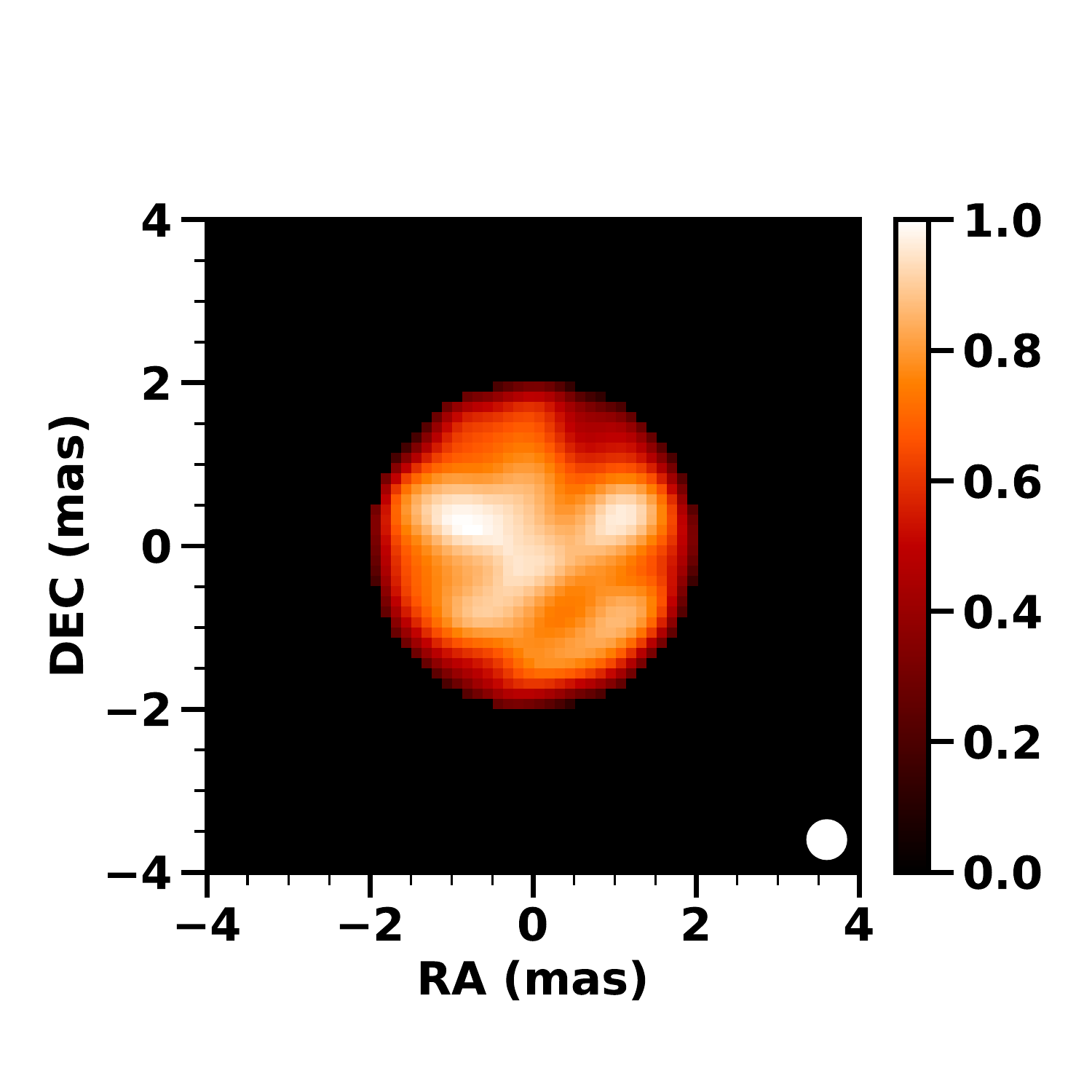}{0.25\textwidth}{AZ~Cyg (OITOOLS.jl) 2011}
          \fig{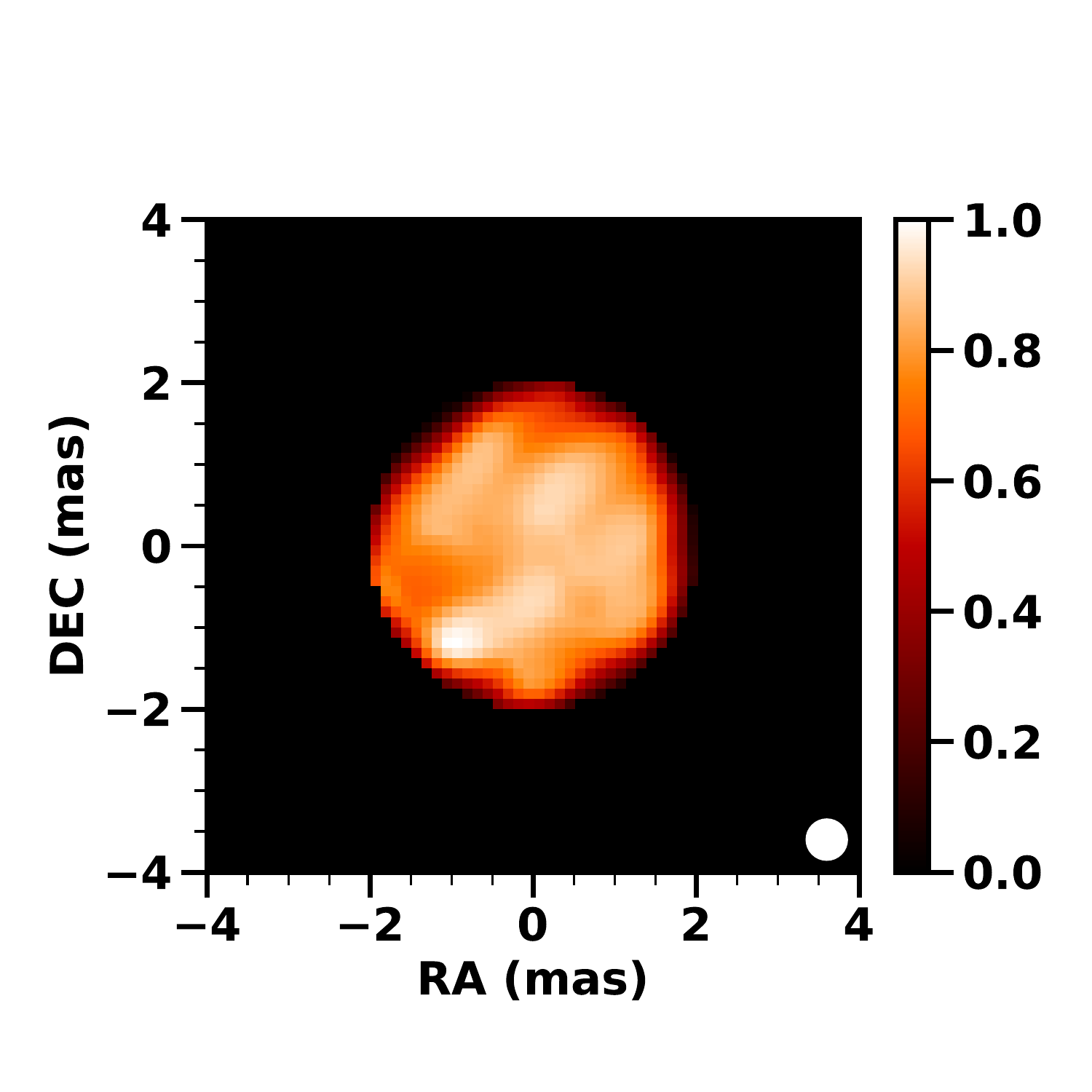}{0.25\textwidth}{AZ~Cyg (OITOOLS.jl) 2014}
          \fig{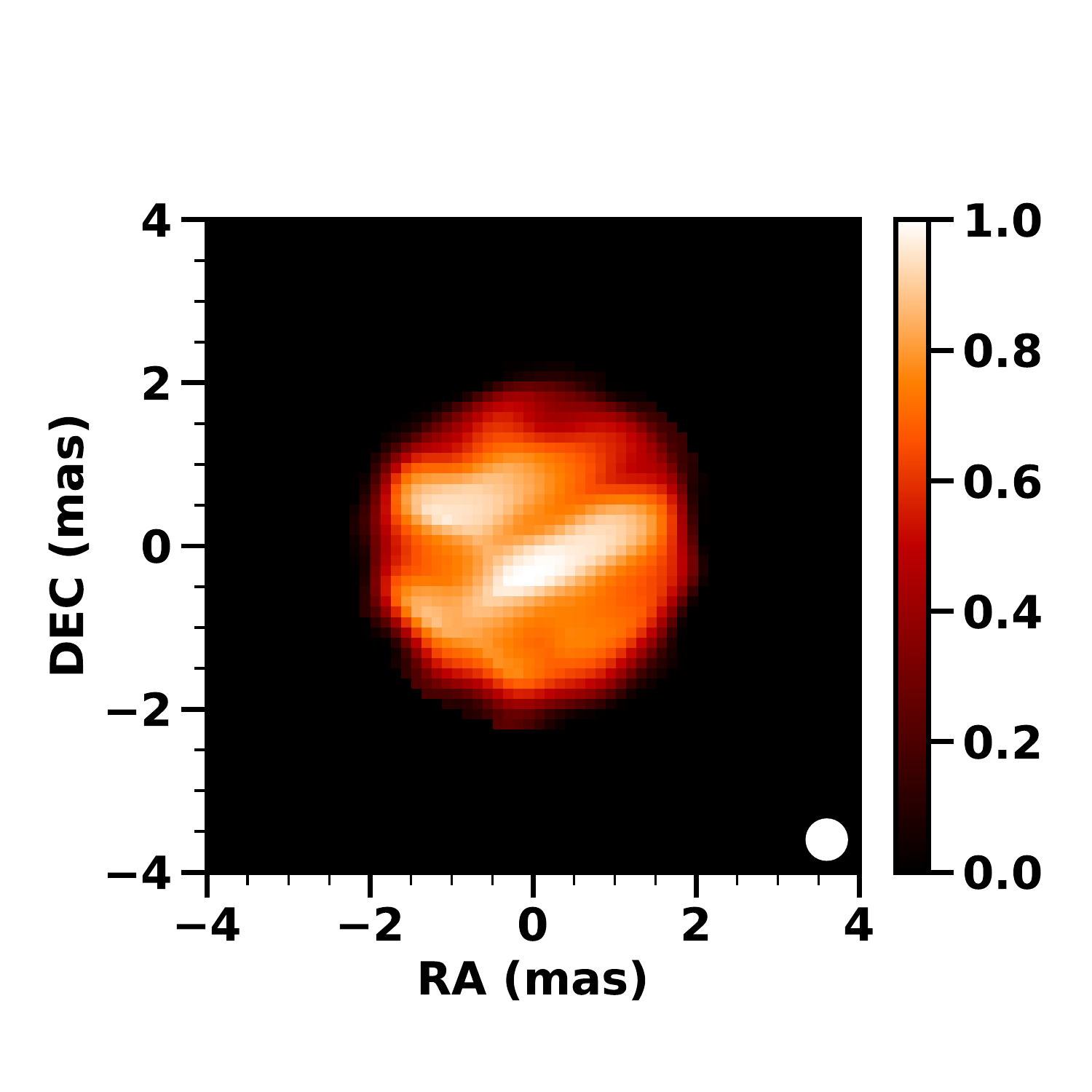}{0.25\textwidth}{AZ~Cyg (OITOOLS.jl) 2015}
          \fig{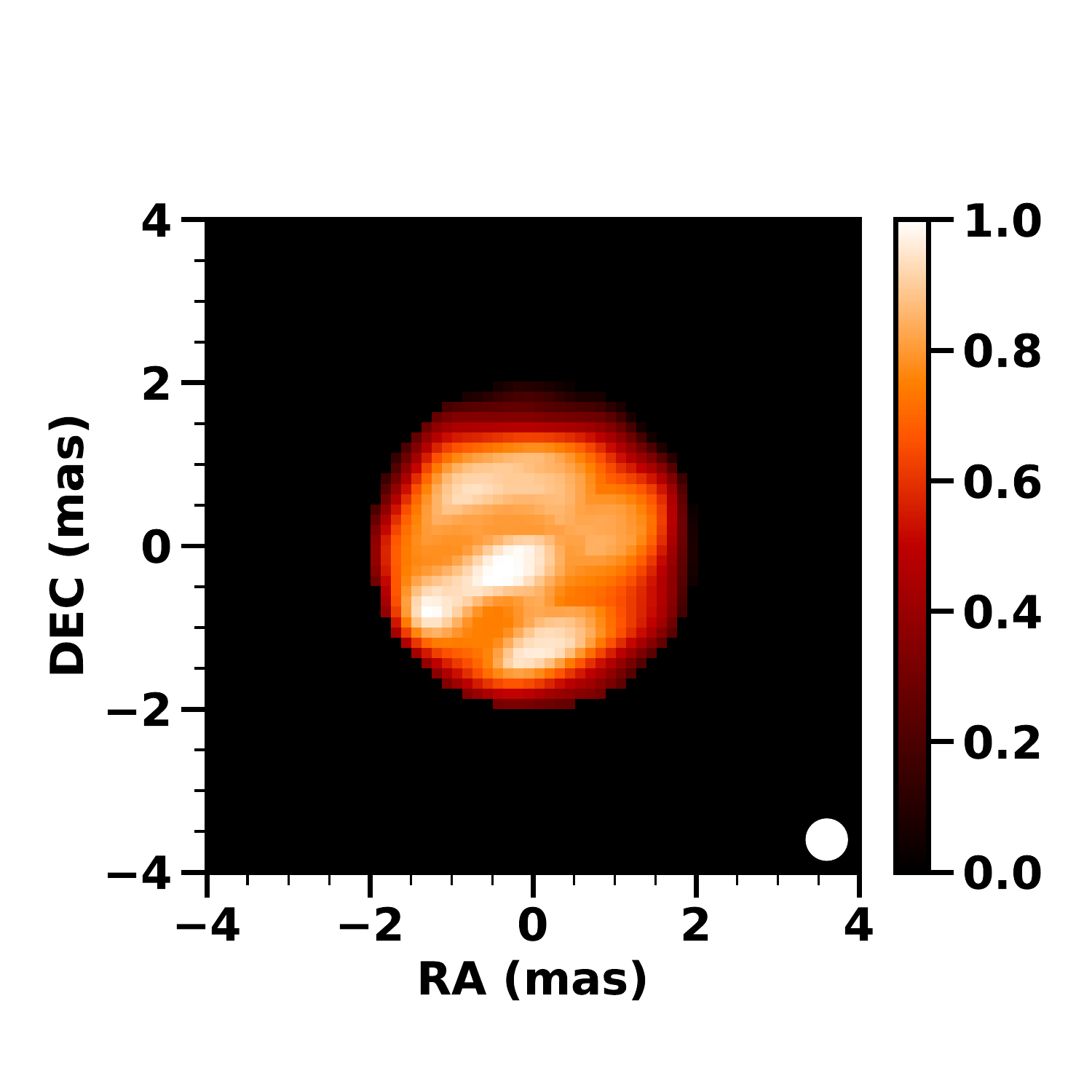}{0.25\textwidth}{AZ~Cyg (OITOOLS.jl) 2016}
          }

\gridline{\fig{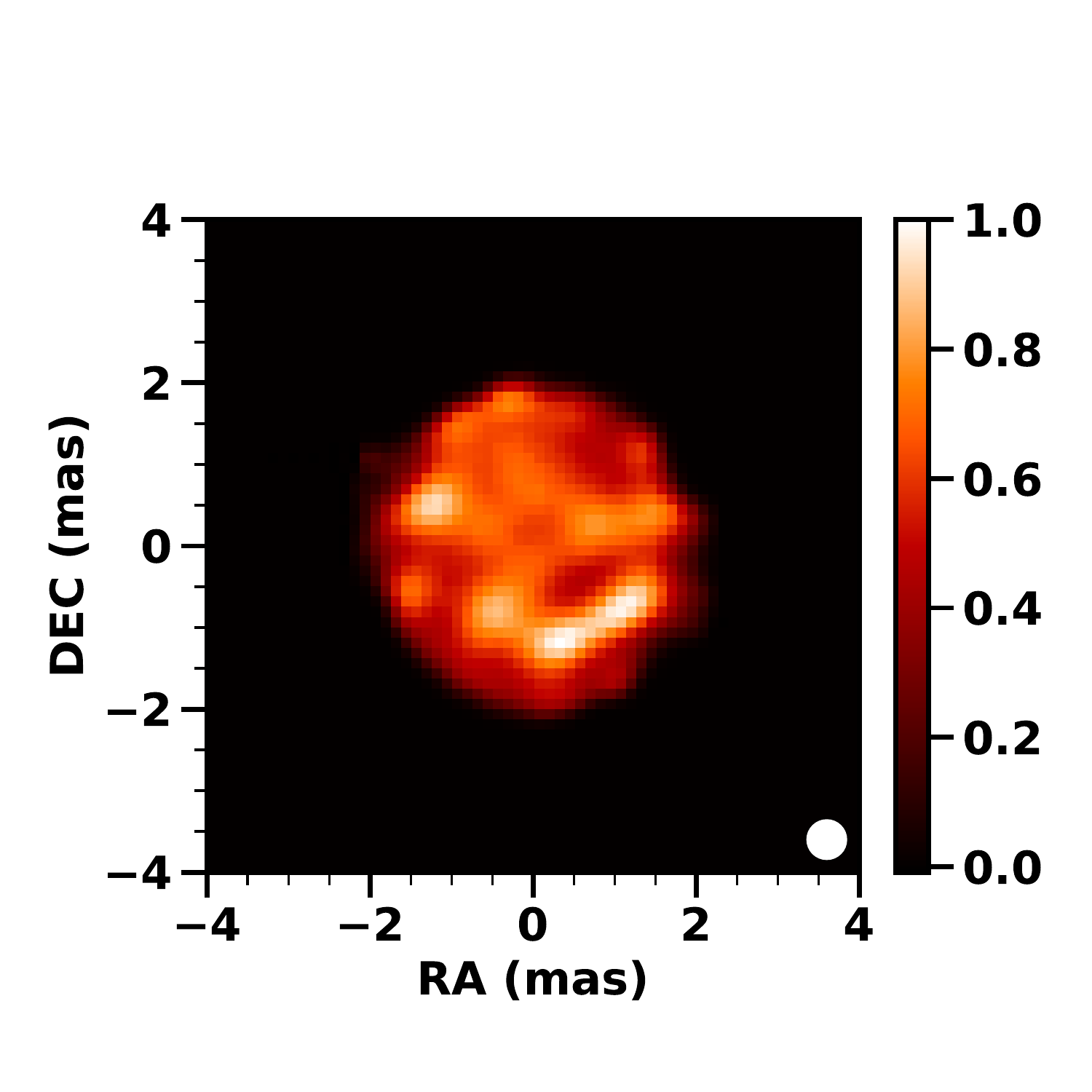}{0.25\textwidth}{AZ~Cyg (SQUEEZE) 2011}
          \fig{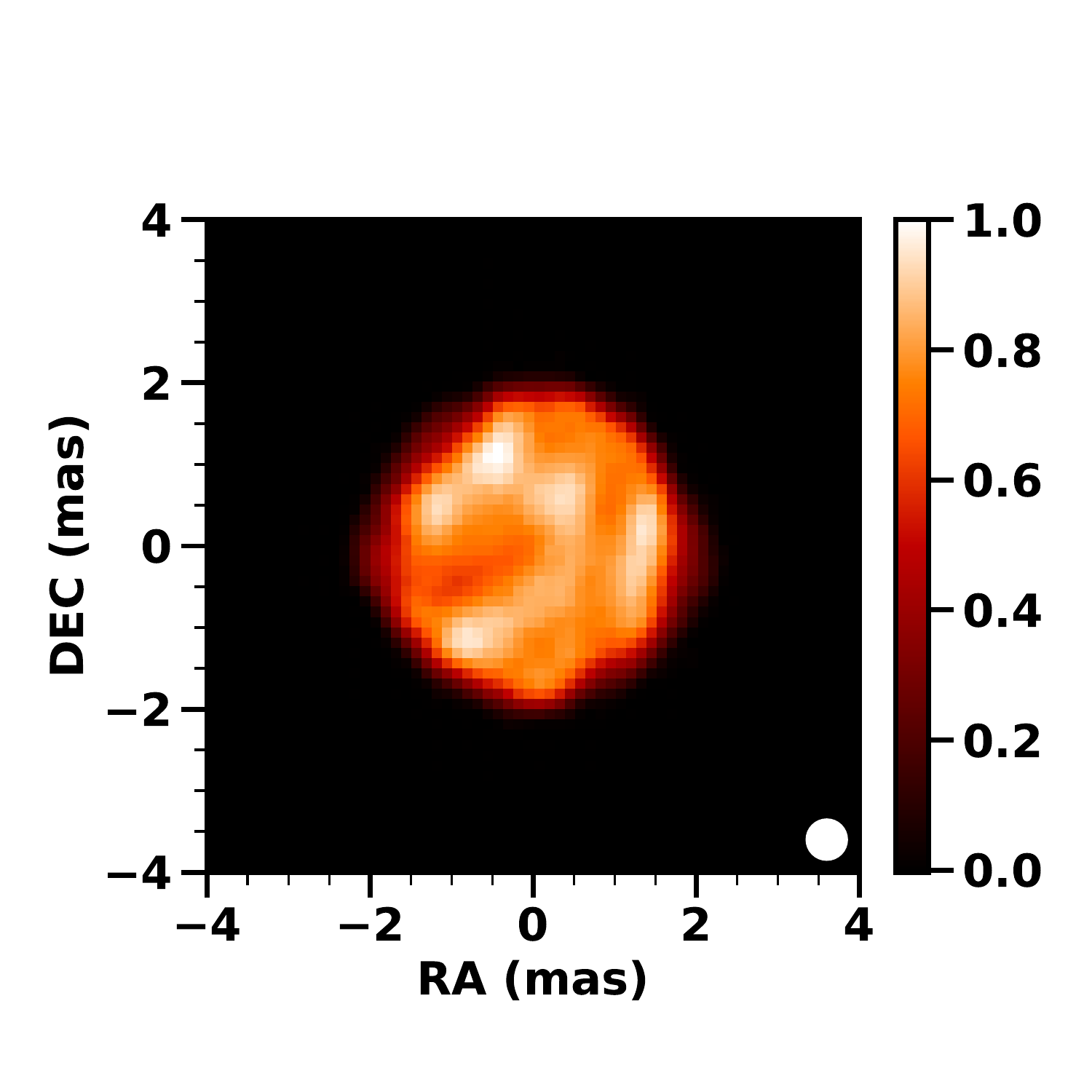}{0.25\textwidth}{AZ~Cyg (SQUEEZE) 2014}
          \fig{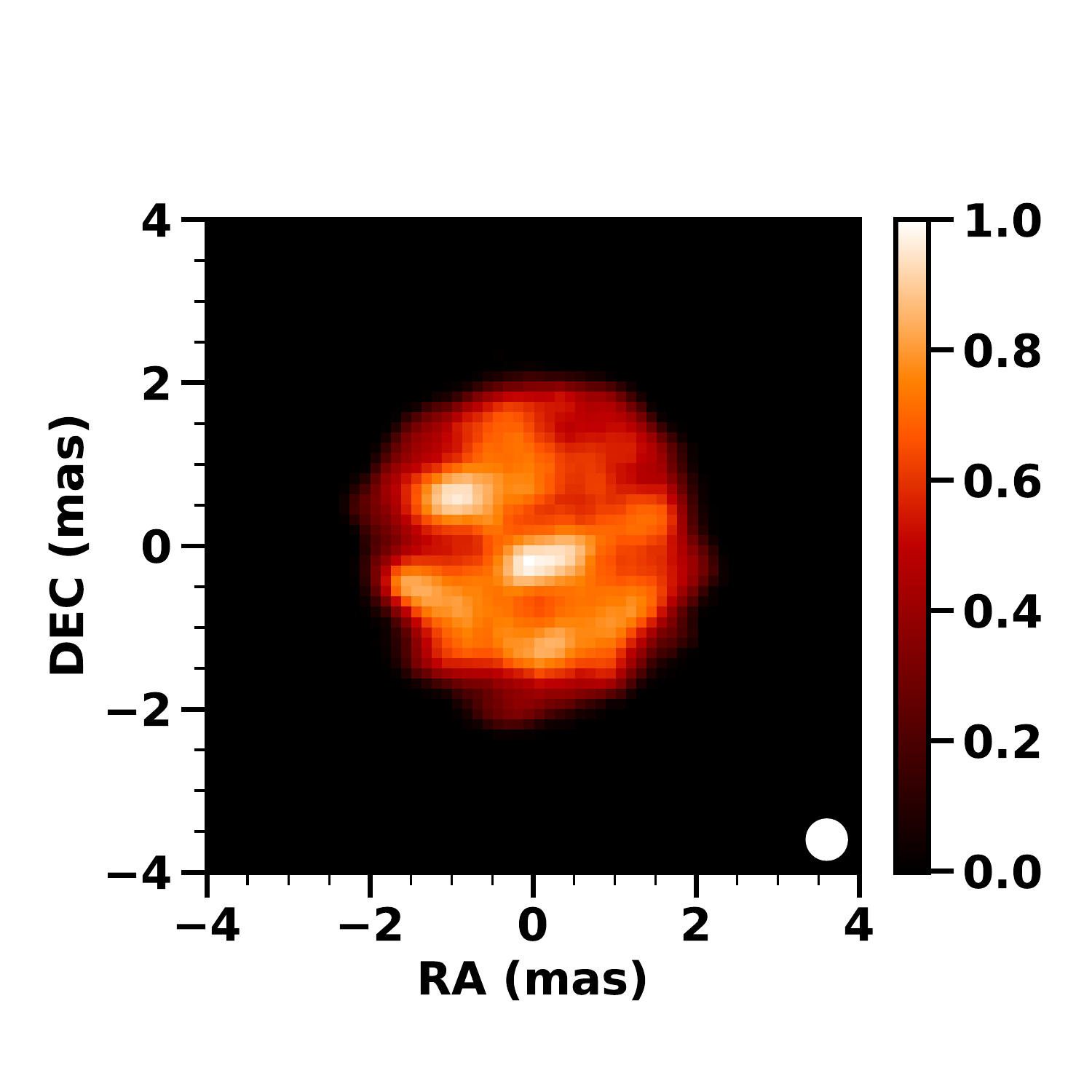}{0.255\textwidth}{AZ~Cyg (SQUEEZE)
          2015}
          \fig{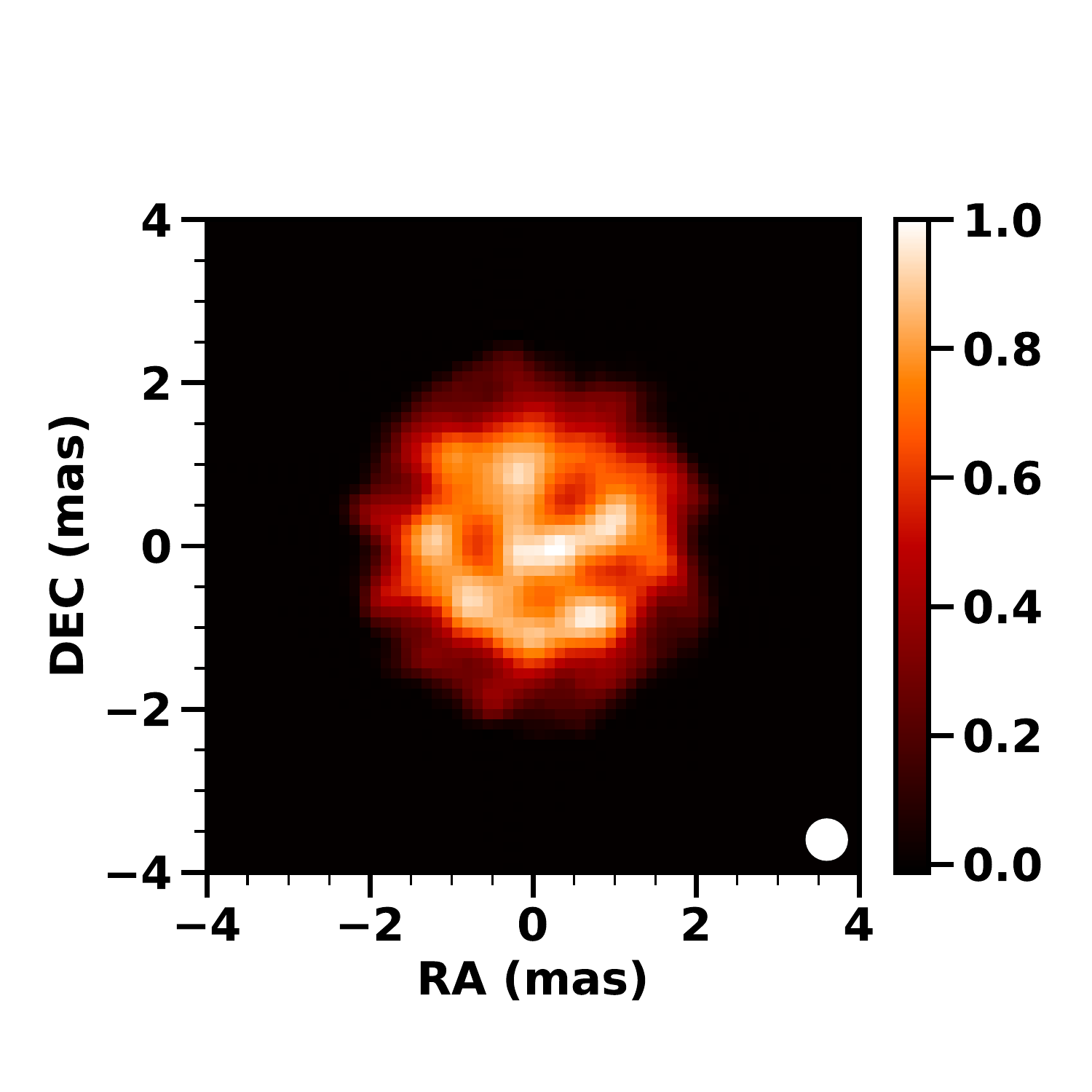}{0.25\textwidth}{AZ~Cyg (SQUEEZE) 2016}
          }
          \gridline{\fig{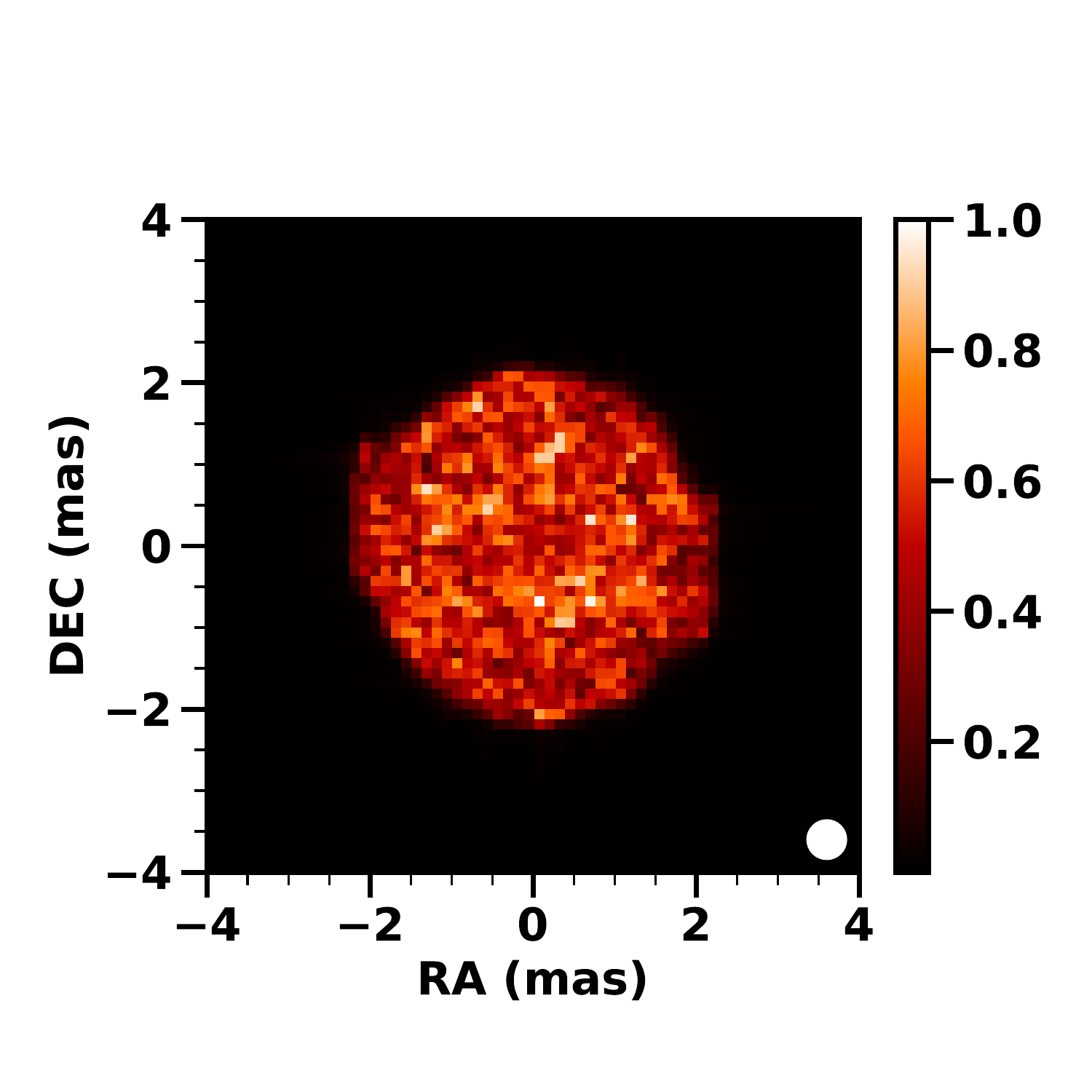}{0.25\textwidth}{AZ~Cyg 2011}
          \fig{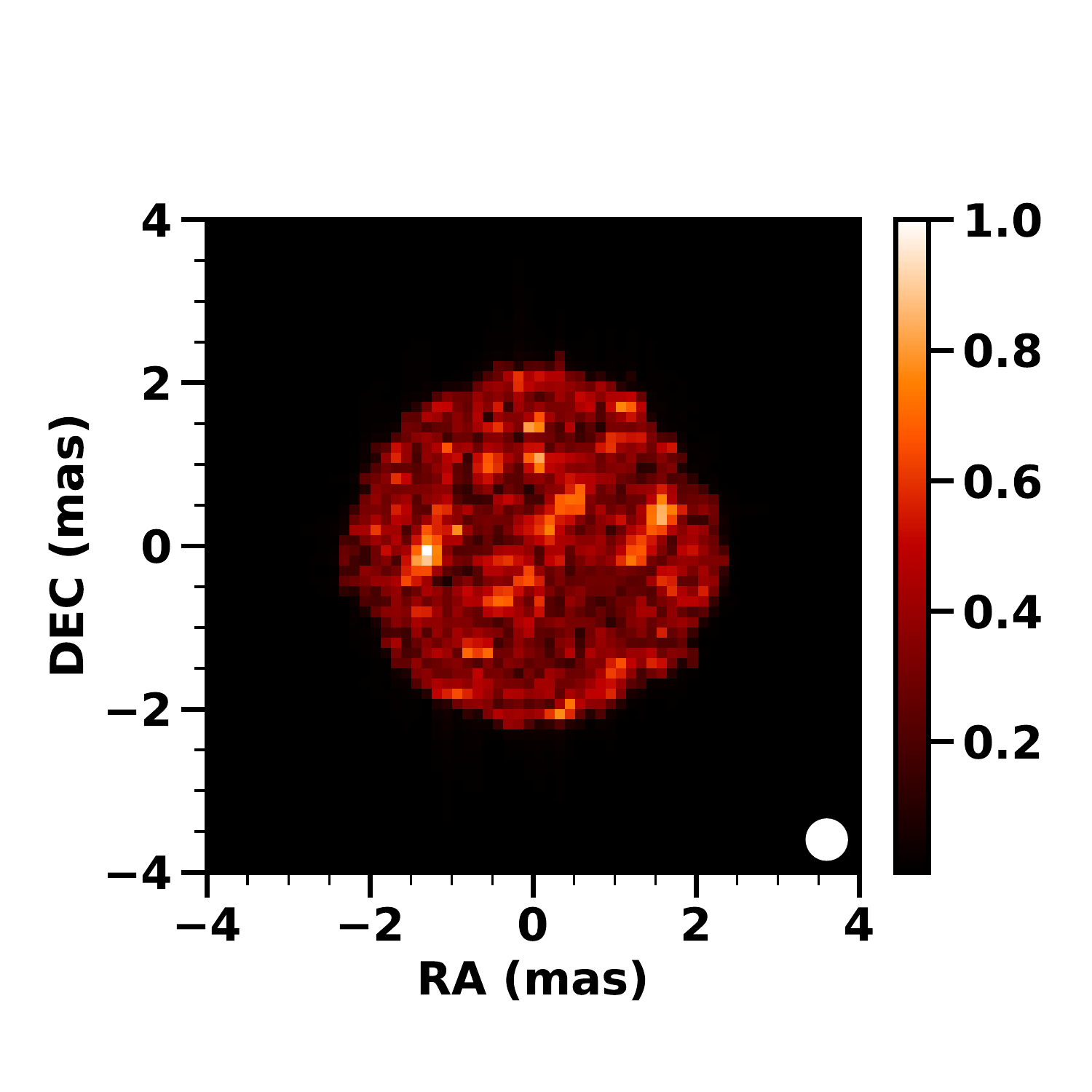}{0.25\textwidth}{AZ~Cyg 2014}
          \fig{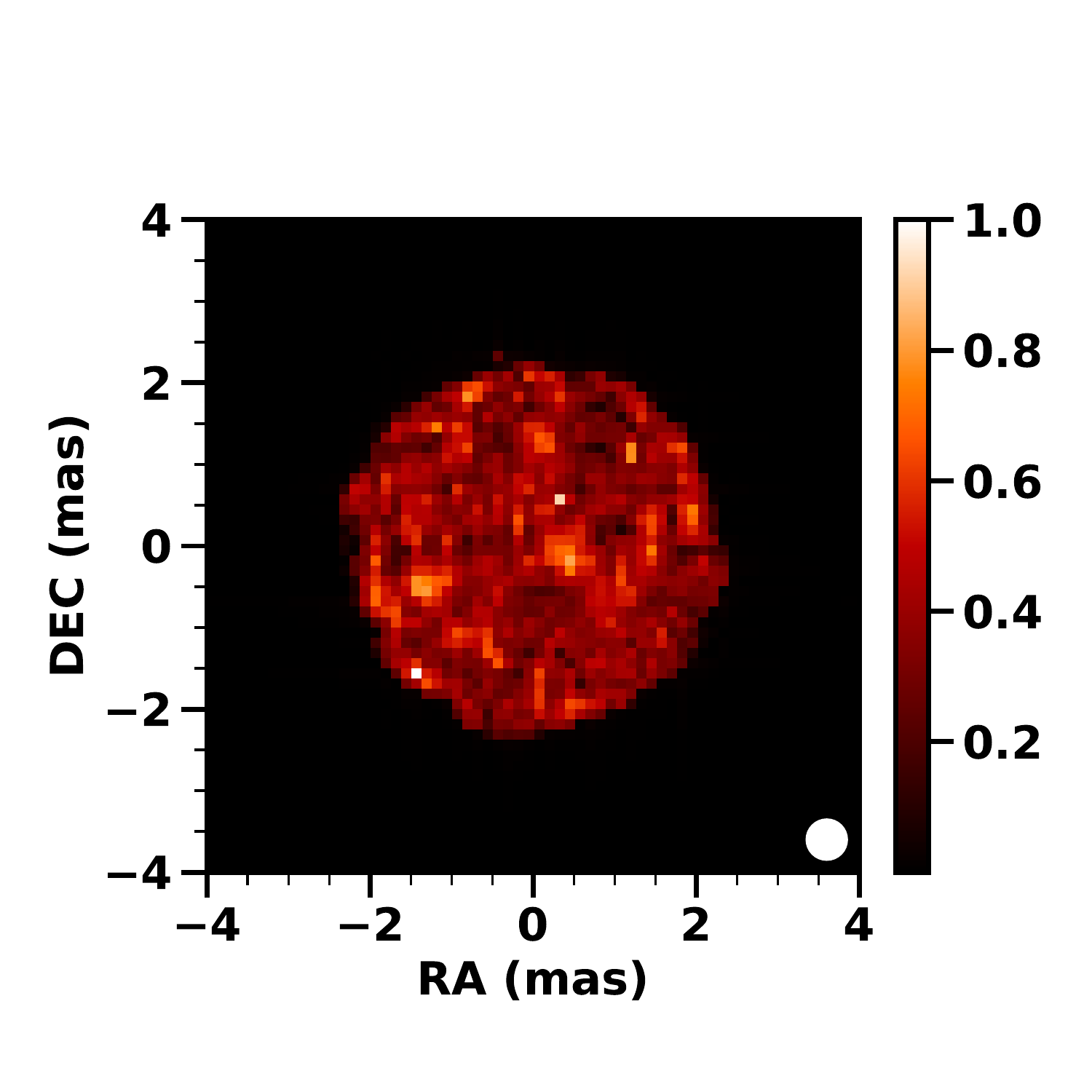}{0.25\textwidth}{AZ~Cyg
          2015}
          \fig{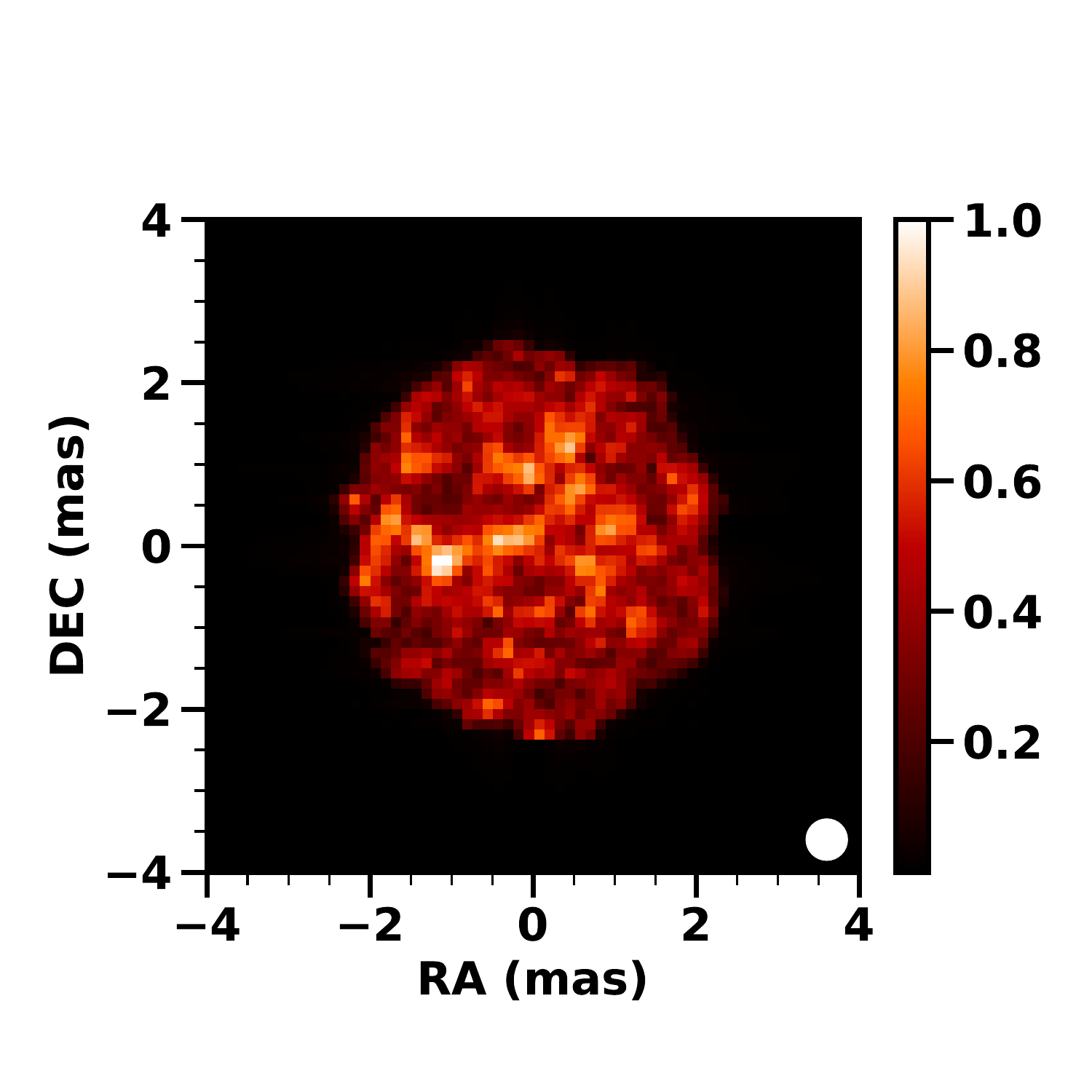}{0.25\textwidth}{AZ~Cyg 2016}
          }
          
          \gridline{
         \fig{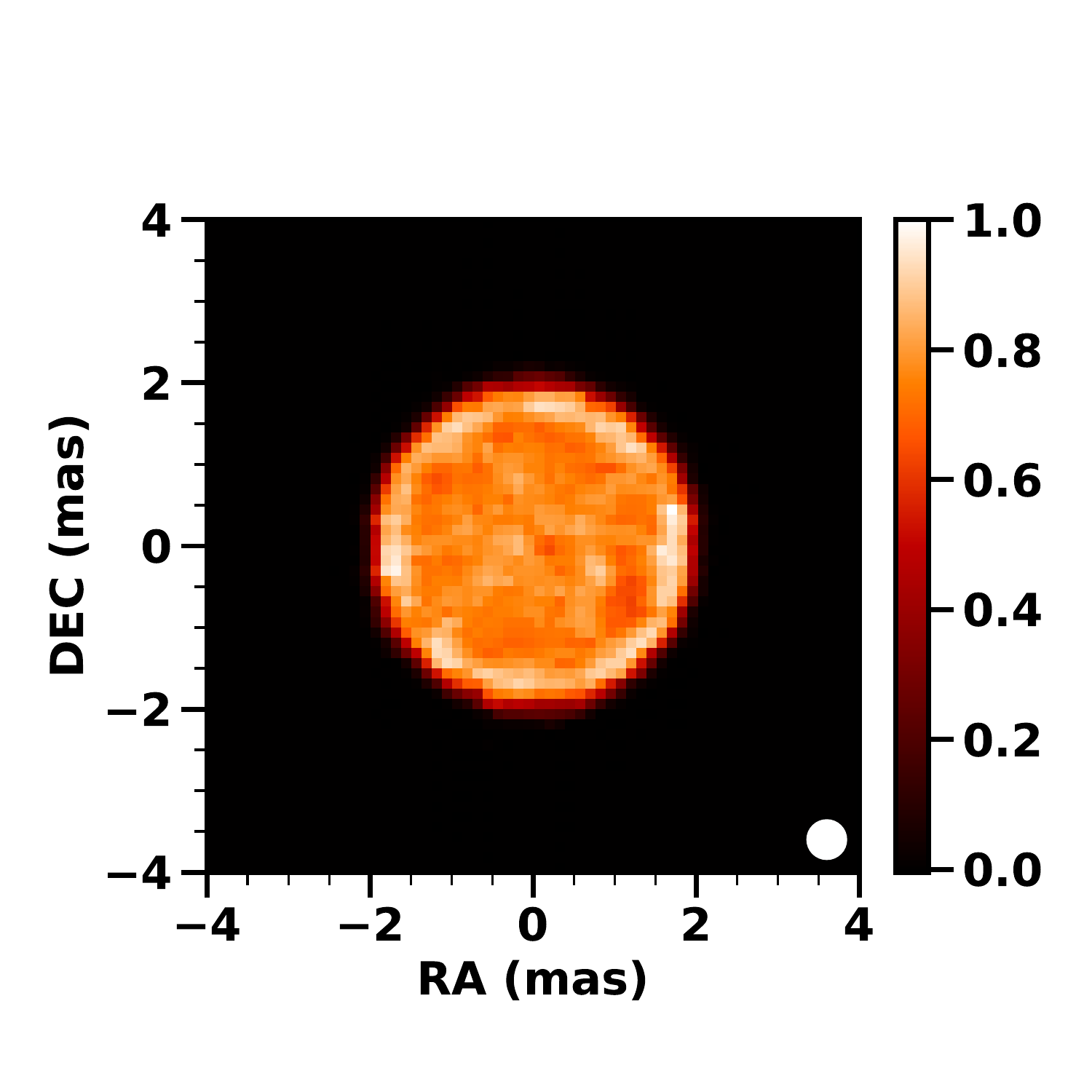}{0.25\textwidth}{UD 2011}
          \fig{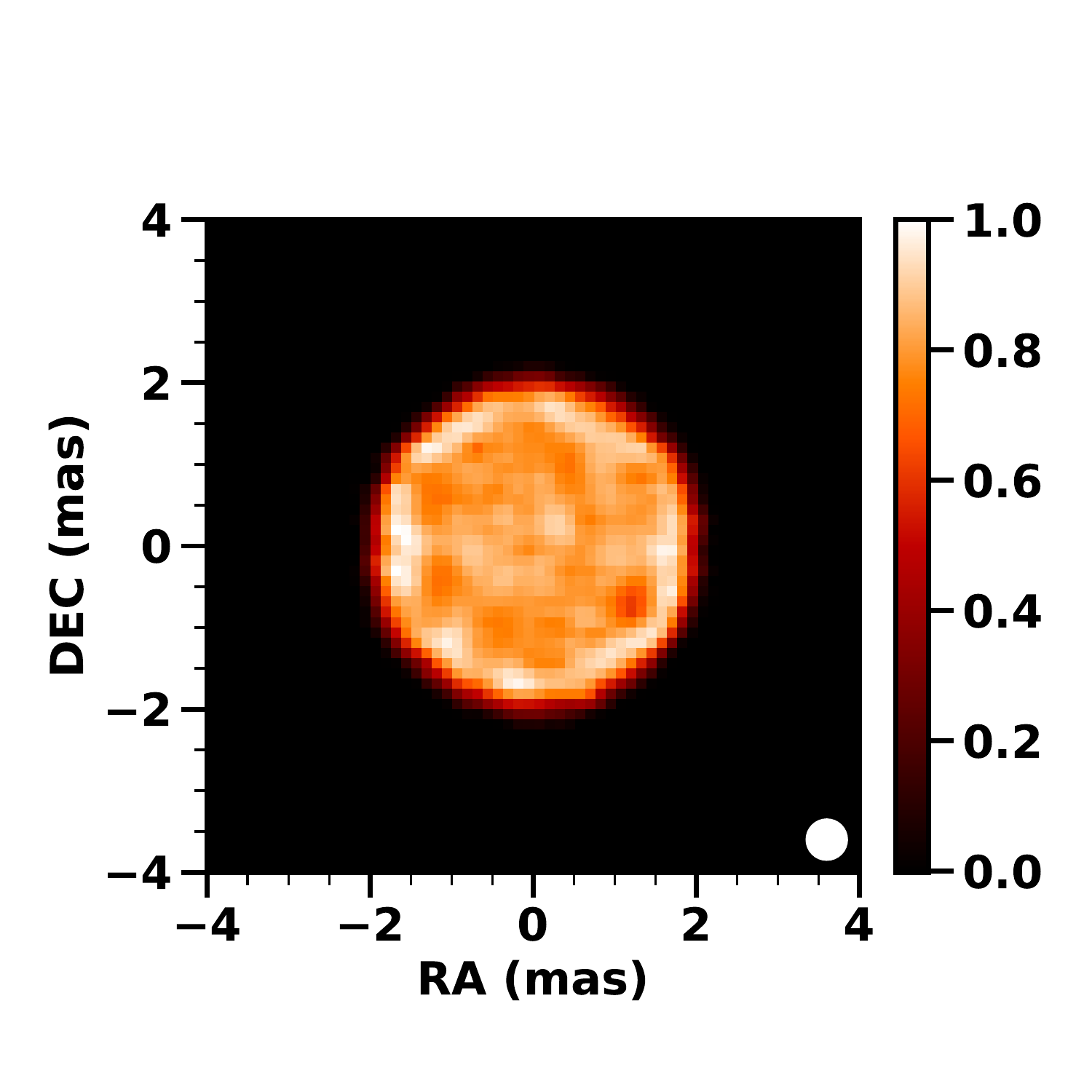}{0.25\textwidth}{UD 2014}
          \fig{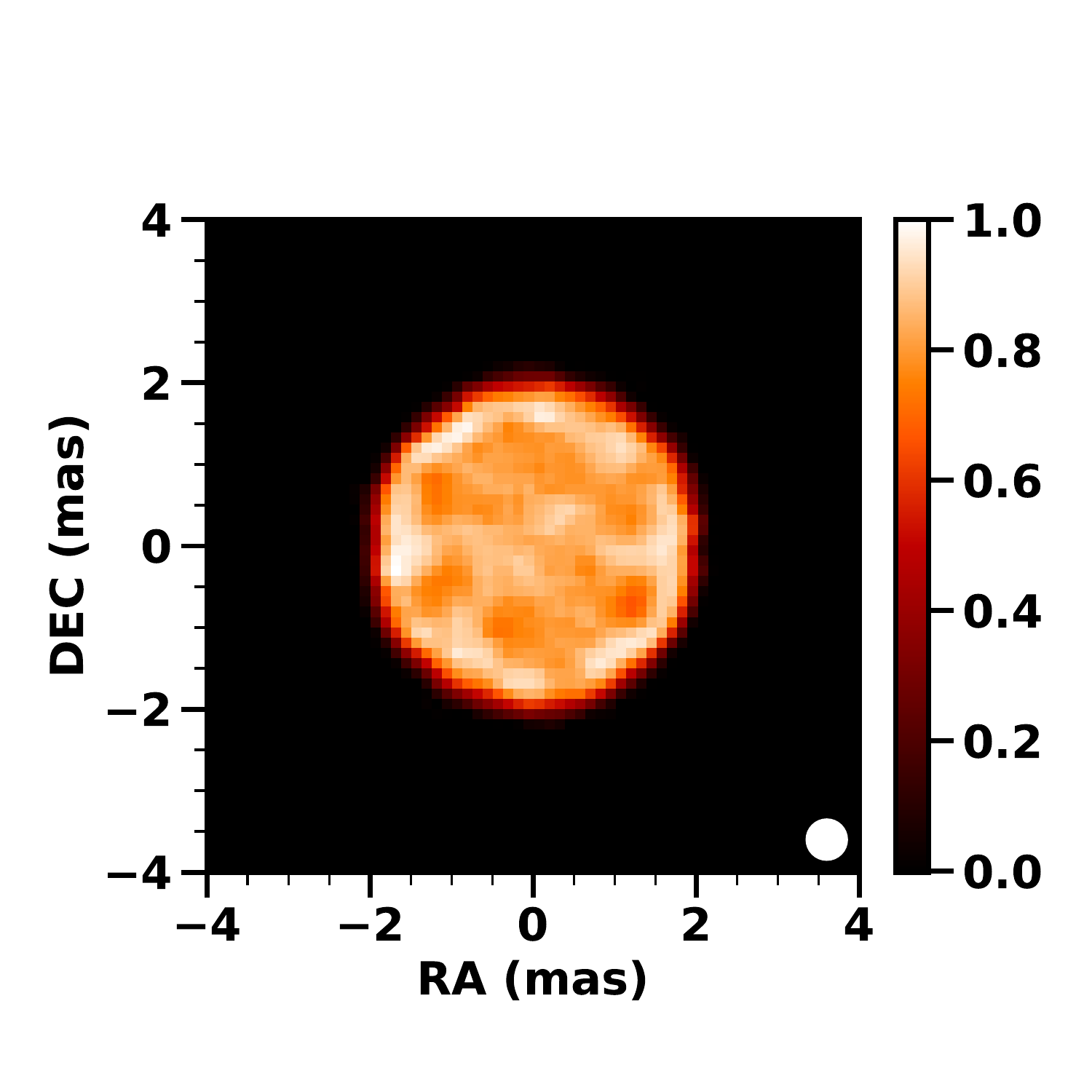}{0.25\textwidth}{UD 2015}
          \fig{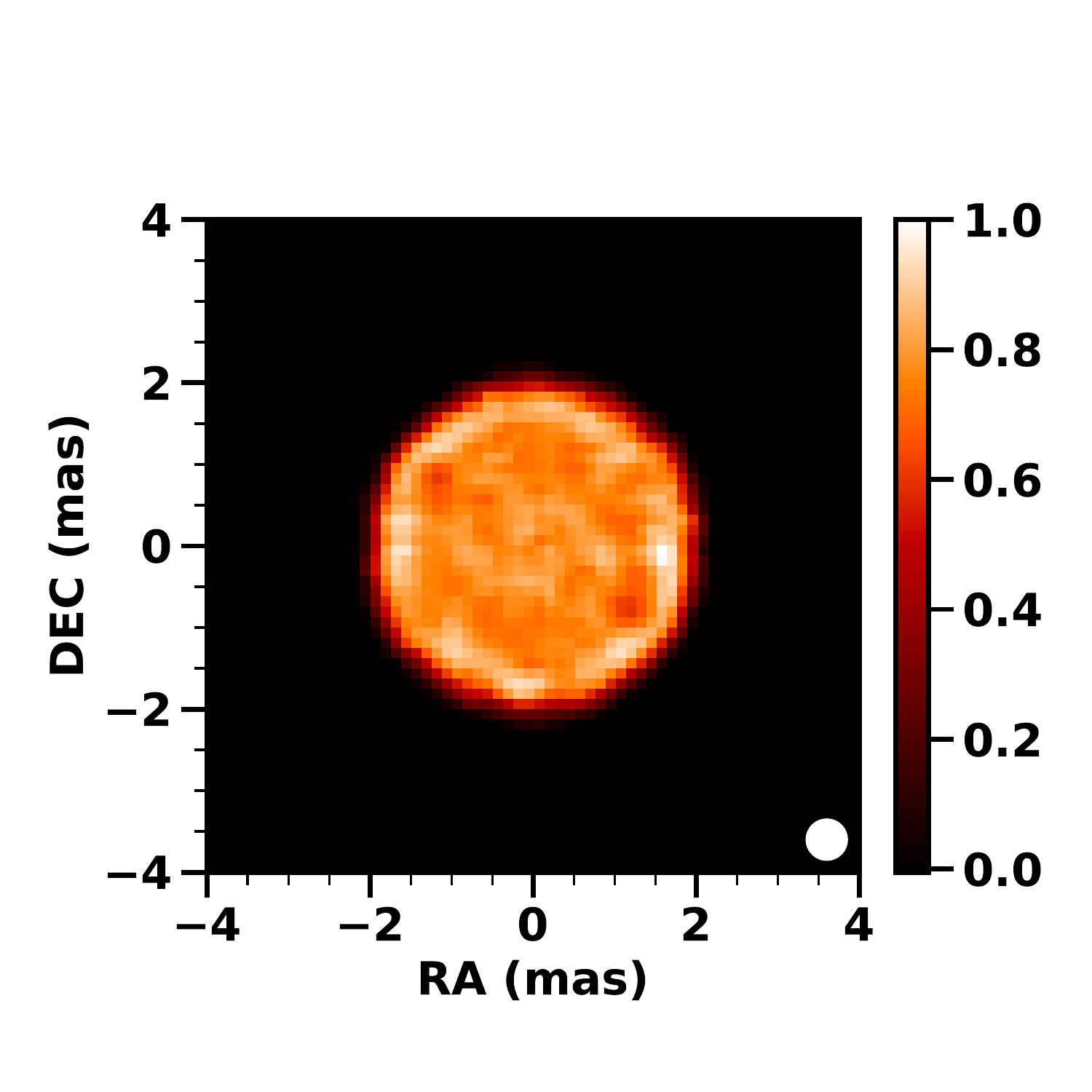}{0.25\textwidth}{UD 2016}
          }
          
\caption{First row: Images of AZ Cyg from 2011-2016 produced with OITOOLS.jl. Second row: Mean images of AZ Cyg from 2011-2016 produced with SQUEEZE. 
Third row: Standard deviation of the reconstructed images in row two.
Fourth row: Corresponding reconstruction of a simulated uniform disk (UD). Intensities of images in each column are scaled to the maximum pixel value of the corresponding mean image in row one. The beamsize in the right hand corner corresponds to resolution given by the maximum projected baseline of that observation.
\label{fig:azcygall}}
\end{figure*}

%% file: Figures/compare2011.tex
\begin{figure*}[!h]
\gridline{
\fig{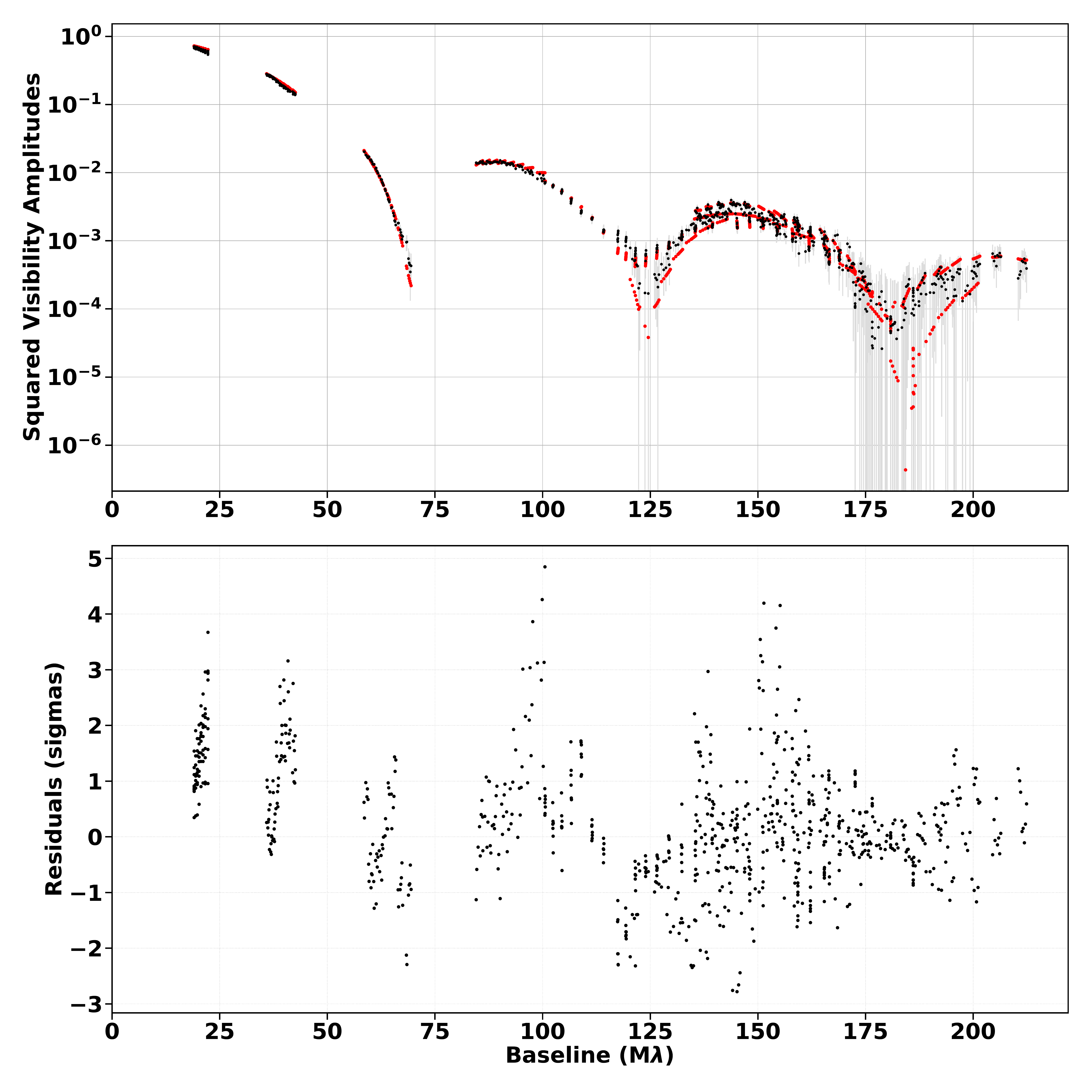}{0.45\textwidth}{AZ~Cyg 2011 Squared Visibilities}
\fig{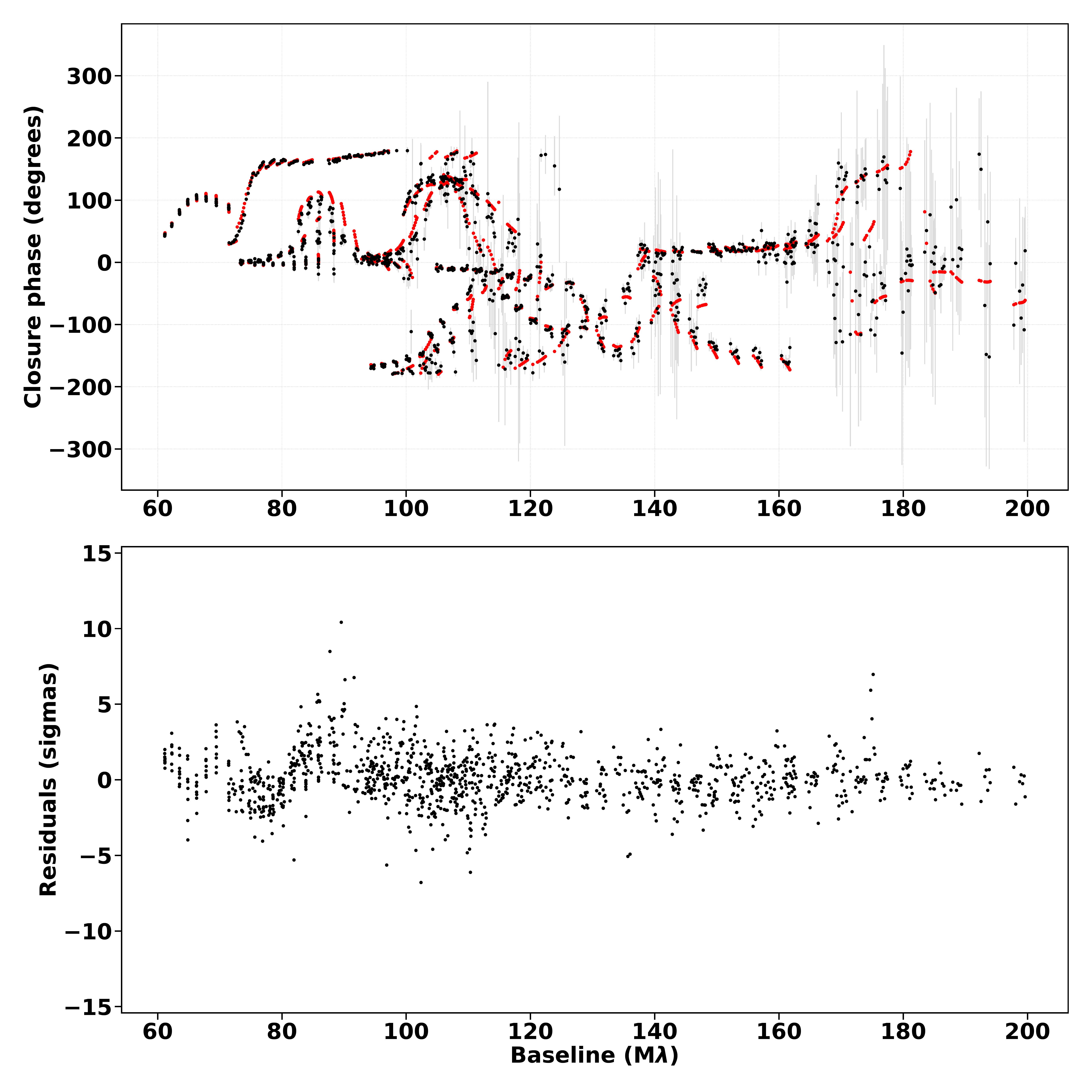}{0.45\textwidth}{AZ~Cyg 2011 Closure Phases}
          }
          \caption{Upper: Comparison of AZ~Cyg observations  2011 (black) to squared visibilities and closure phases calculated from the mean SQUEEZE image for the same epoch (red). Lower: Residuals (in sigmas) between the quantities in each upper graph. 
\label{fig:compare2011}}
\end{figure*}

%% file: Figures/split2016.tex
\begin{figure*}
\gridline{\fig{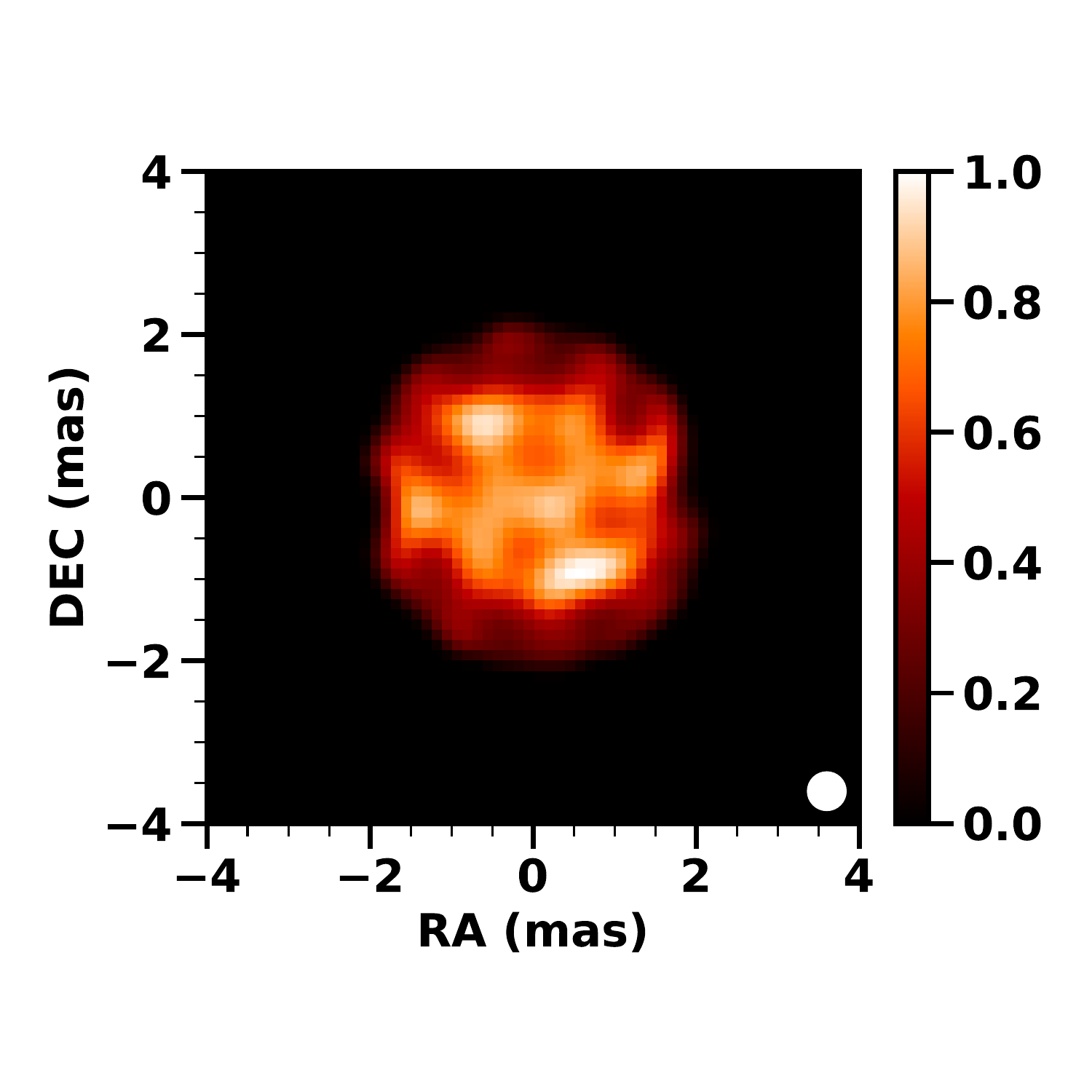}{0.3\textwidth}{2016 Aug 28-Sep 06}
          \fig{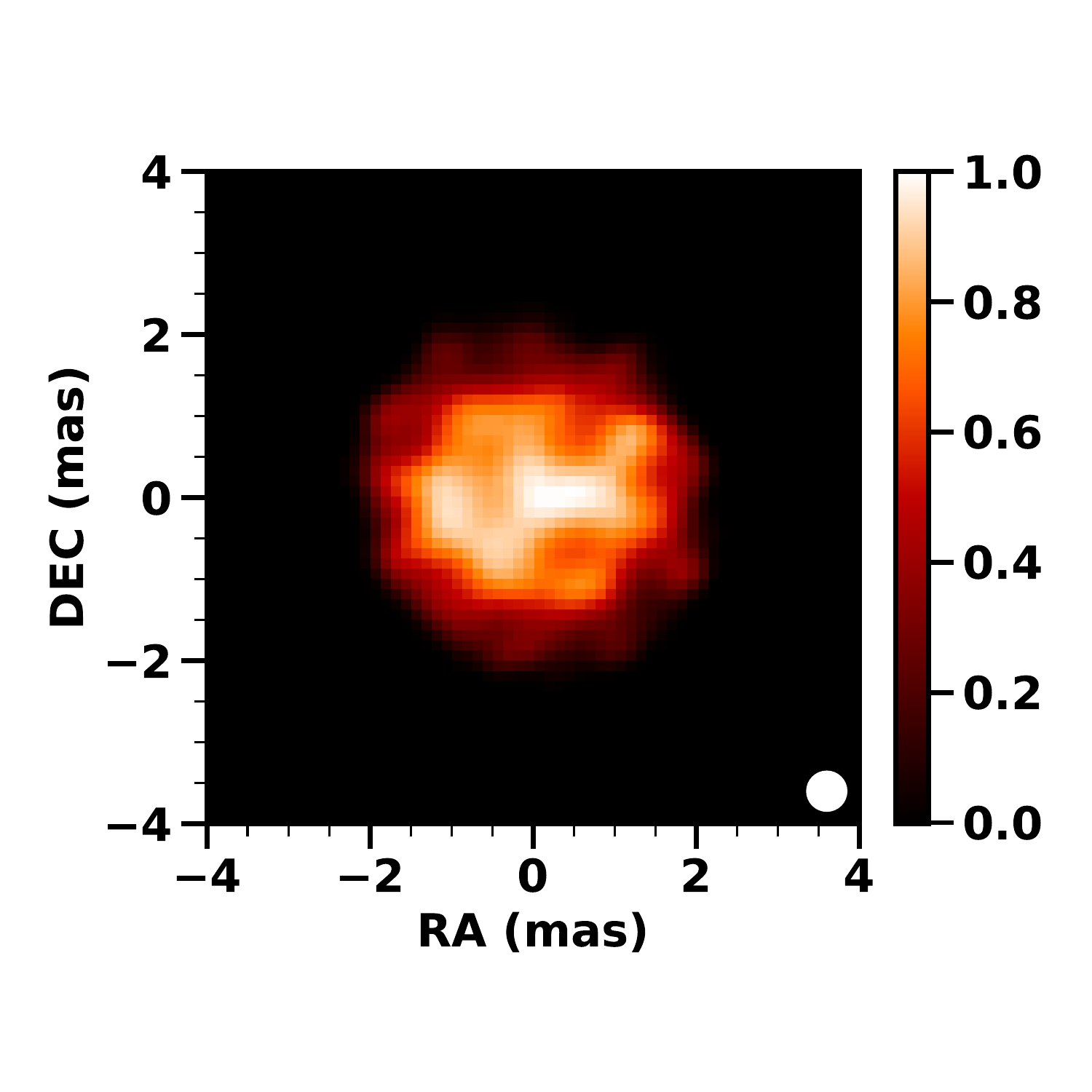}{0.3\textwidth}{ 2016 Sep 8-11}
          \fig{Files/azcyg2016meanupdated.pdf}{0.3\textwidth}{2016 Aug 28-Sep 11}
          }
\caption{SQUEEZE reconstructions of AZ~Cyg in 2016 using data from August 28-September 6, September 8-11, and entirety of the AZ~Cyg 2016 observations. The beamsize in the right hand corner corresponds to resolution given by the maximum projected baseline of that observation. \label{fig:split2016}}
\end{figure*}

%% file: Tables/contrasts.tex
\begin{splitdeluxetable*}{ccccccBcccccc}
\tablewidth{0pt}
\tablecaption{Surface Contrast AZ~Cyg \label{tab:contrasts}}
\tablehead{
\colhead{Year} & \colhead{1.47-1.52 $\mu~m$}  & \colhead{1.52-1.56 $\mu~m$} & \colhead{1.56-1.60 $\mu~m$} & \colhead{1.60-1.63 $\mu~m$} & \colhead{1.63-1.67 $\mu~m$} & \colhead{Year} &
\colhead{1.67-1.70 $\mu~m$} & \colhead{1.70-1.73 $\mu~m$} &
\colhead{1.73-1.76 $\mu~m$} &
\colhead{1.47-1.76 $\mu~m$}&
\colhead{OITOOLS (1.47-1.76 $\mu~m$)}}
\startdata
2011 & 20.69$\pm$1.20 & 16.41$\pm$1.07 & 15.07$\pm$1.85 & 12.84$\pm$0.08 & 12.81$\pm$0.08 & 2011 & 12.37$\pm$1.99 & 13.33$\pm$3.22 & 11.93$\pm$0.96 & 13.89$\pm$0.18 &
8.84\\
2014 & 6.27$\pm$0.92 & 6.27 $\pm$0.92 & 6.27$\pm$1.30 &  6.86$\pm$0.52 
& 4.83$\pm$0.35 &  2014 & 6.42$\pm$1.77 & 6.69$\pm$2.10 & 5.17$\pm$1.10 & 9.59$\pm$1.65 &
4.39\\
2015 & 10.50$\pm$2.86 & 11.95$\pm$1.09 & 11.11$\pm$0.79 & 11.36$\pm$0.70 &  12.35$\pm$0.68 & 2015 & 10.13$\pm$0.53 & 10.56$\pm$1.04 & 9.81$\pm$1.54 & 12.16$\pm$0.79 
& 9.40\\
2016 & 18.27$\pm$2.25 & 18.13$\pm$2.74 & 14.23$\pm$1.06 & 13.57$\pm$2.79 & 14.55$\pm$2.00 & 2016 & 14.26$\pm$1.22 & 11.72$\pm$1.90 & 10.81$\pm$2.10 & 14.46$\pm$1.30 &
6.85\\
\enddata

\end{splitdeluxetable*}

%% file: Figures/contrastfigs.tex
\begin{figure*}[!h]
\includegraphics[scale=0.5,angle=0]{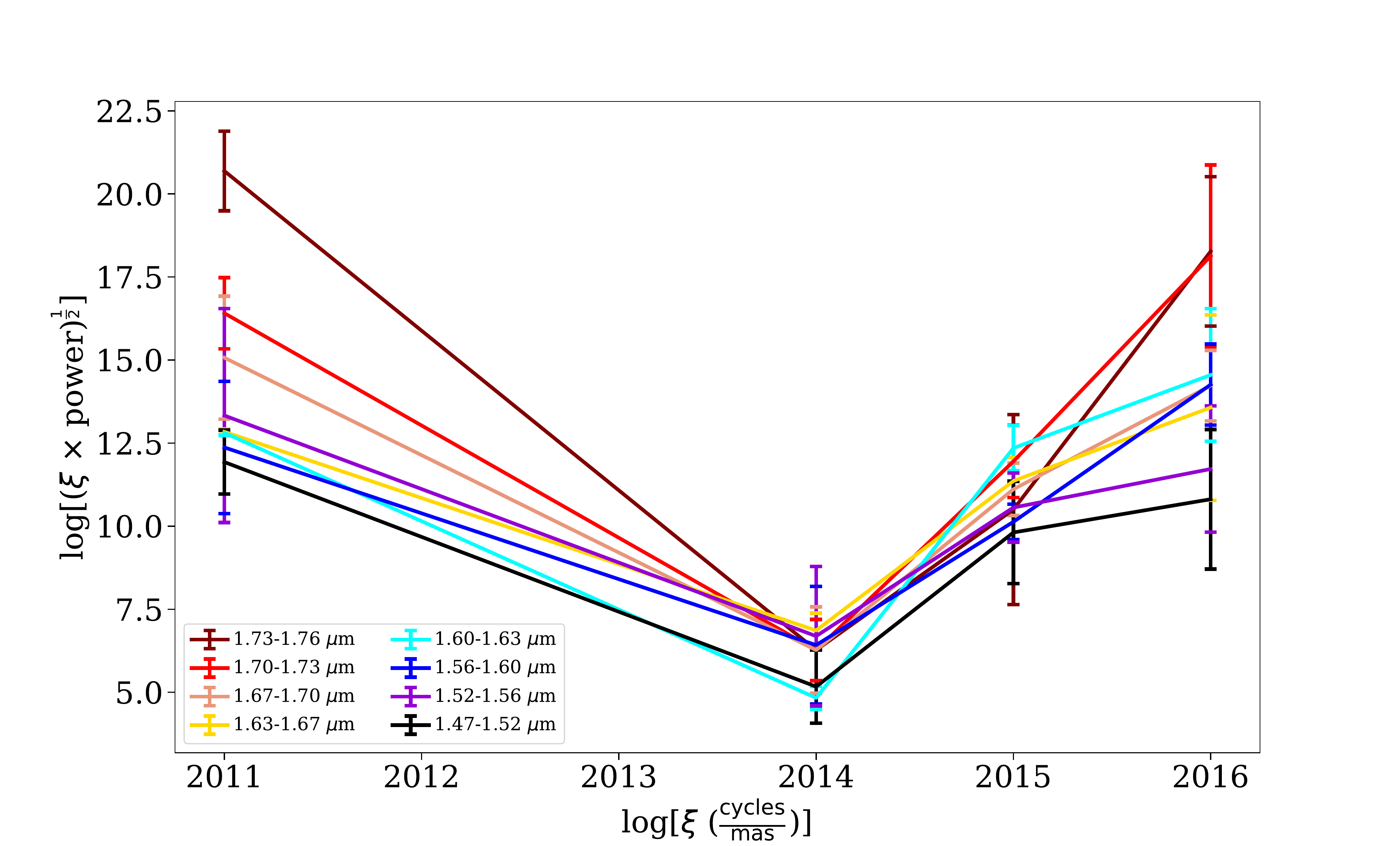}
\caption{Surface contrasts of AZ Cyg for different wavelength channels in each epoch of observation as determined using the mean reconstructed images in Figure \ref{fig:azcygall}.
\label{fig:contrasts}}
\end{figure*}

%% file: Tables/sizes.tex
\begin{splitdeluxetable*}{cccccccBcccccc}
\tablewidth{0pt}
\tablecaption{Characteristic granulation size for AZ~Cyg (R$_{*}$) \label{tab:features}}
\tablehead{\colhead{Year} &
\colhead{1.47-1.52 $\mu~m$} &
\colhead{1.52-1.56 $\mu~m$} &
\colhead{1.56-1.60 $\mu~m$} &
\colhead{1.60-1.63 $\mu~m$} &
\colhead{1.63-1.67 $\mu~m$} &
\colhead{1.67-1.70 $\mu~m$} & 
\colhead{Year} &
\colhead{1.70-1.73 $\mu~m$} &
\colhead{1.73-1.76 $\mu~m$} &
\colhead{1.47-1.76 $\mu~m$}&
\colhead{OITOOLS (1.47-1.76 $\mu~m$)}}
\startdata
2011 &
0.74$\pm$0.06 &
0.52$\pm$0.04 &
0.67$\pm$0.16 &
0.51$\pm$0.04 &
0.48$\pm$0.03 &
0.56$\pm$0.08 &
2011 &
0.70$\pm$0.18 &
0.59$\pm$0.04 &
0.69$\pm$0.01 &
1.06$\pm$0.01 &
\\
2014 &
0.52$\pm$0.09 &
0.46$\pm$0.11 &
0.79$\pm$0.30 &
0.71$\pm$0.14 &
0.77$\pm$0.06 &
1.13$\pm$0.03 &
2014 & 
0.94$\pm$0.28 &
0.90$\pm$0.25 &
0.67$\pm$0.19 &
0.8$\pm$0.01 &
\\
2015 &
0.66$\pm$0.11 &
0.78$\pm$0.12 &
0.83$\pm$0.12 &
0.80$\pm$0.24 &
1.00$\pm$0.34 &
1.14$\pm$0.16 &
2015 &
1.11$\pm$0.06 &
1.16$\pm$0.17 &
0.52$\pm$0.03 &
0.51$\pm$0.02 &
\\
2016 &
0.67$\pm$0.17 &
0.57$\pm$0.05 &
0.54$\pm$0.04 &
0.53$\pm$0.14 &
0.57$\pm$0.06 &
0.63$\pm$0.07 &
2016 &
0.90$\pm$0.23 &
0.78$\pm$0.26 &
0.54$\pm$0.02 &
0.51$\pm$0.02 &
\\
\enddata
\end{splitdeluxetable*}

%% file: Figures/powerspec.tex
\begin{figure*}[!h]
\gridline{\fig{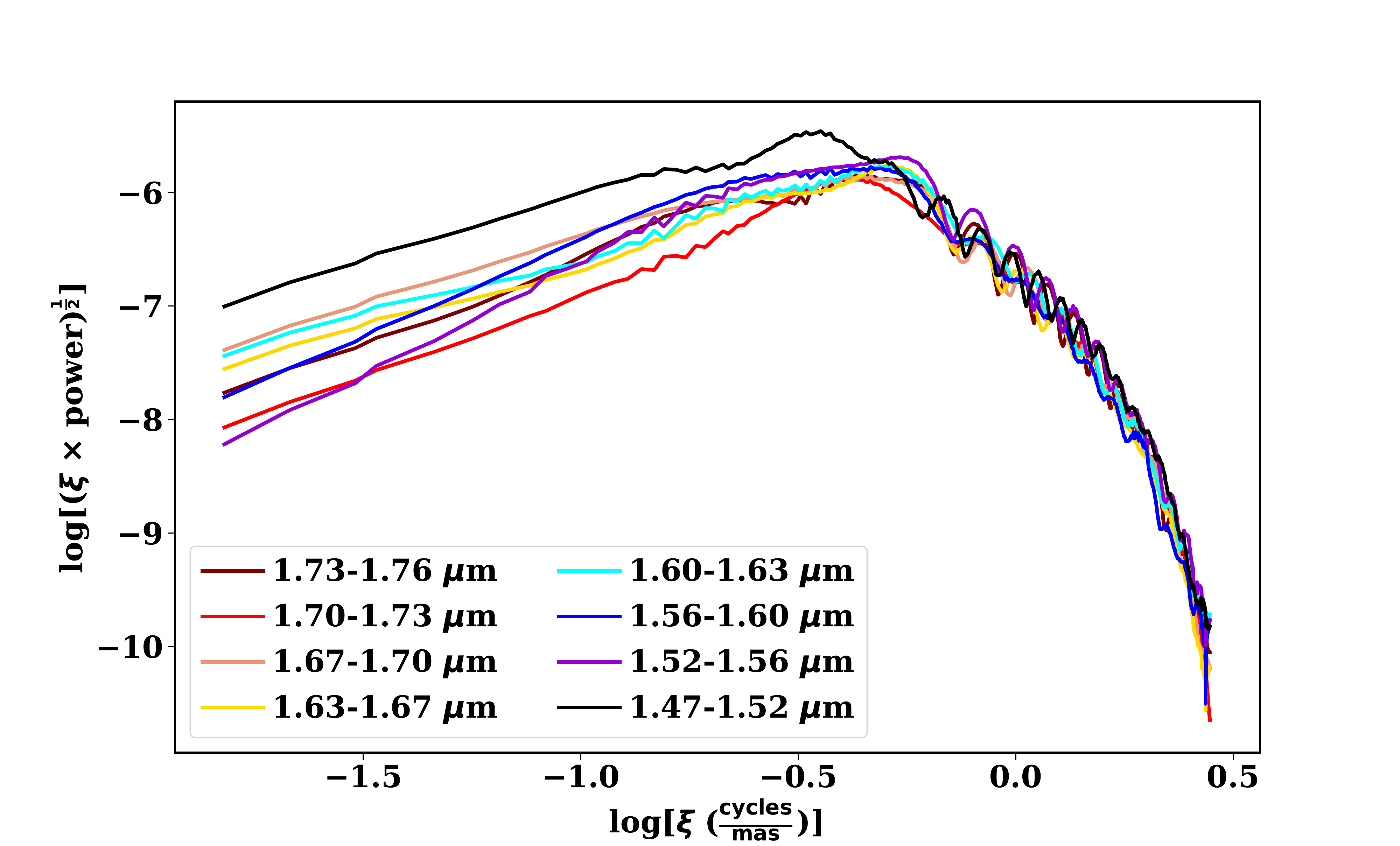}{0.5\textwidth}{2011}\\
     \fig{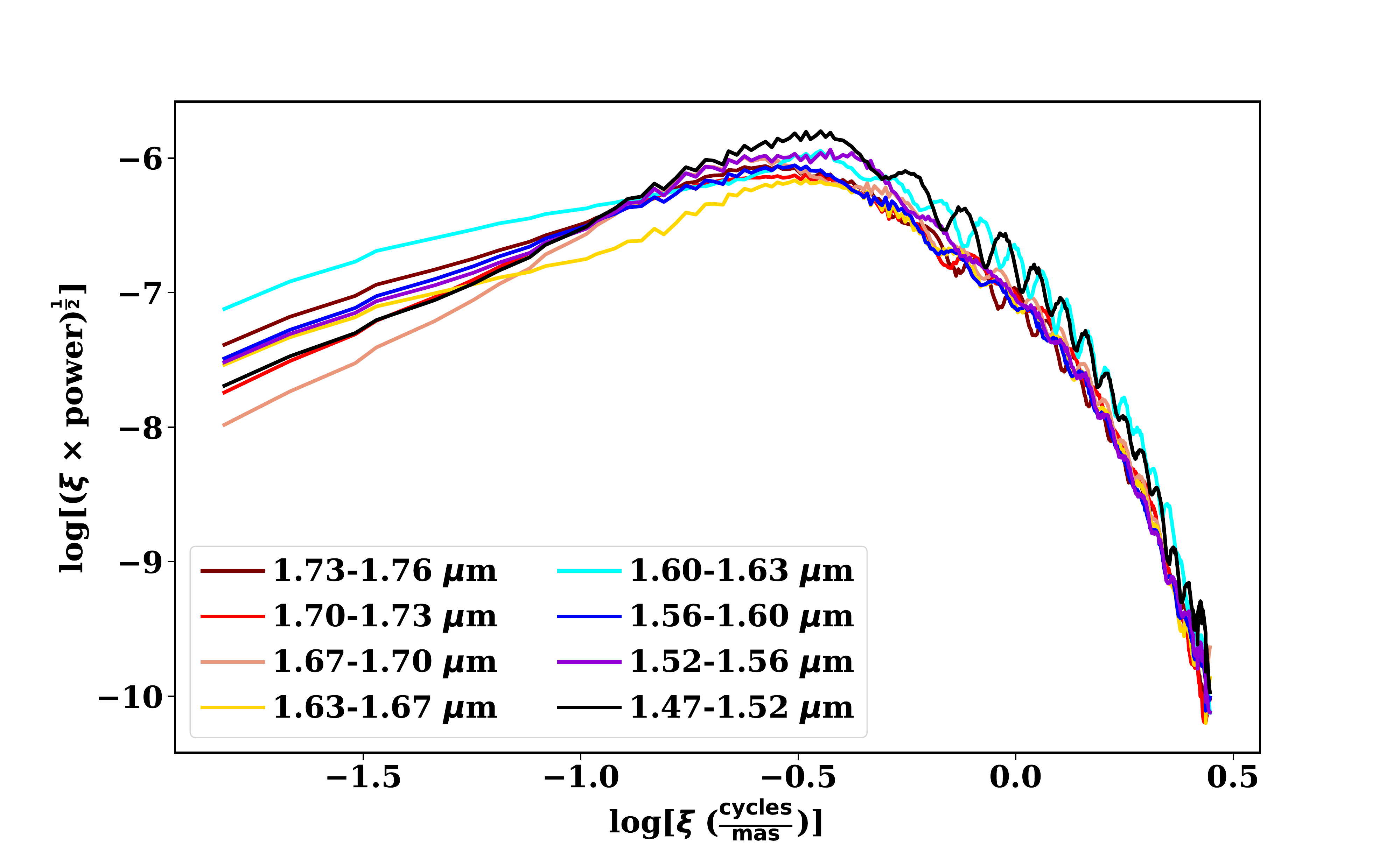}{0.5\textwidth}{2014}
         }
        \gridline{ \fig{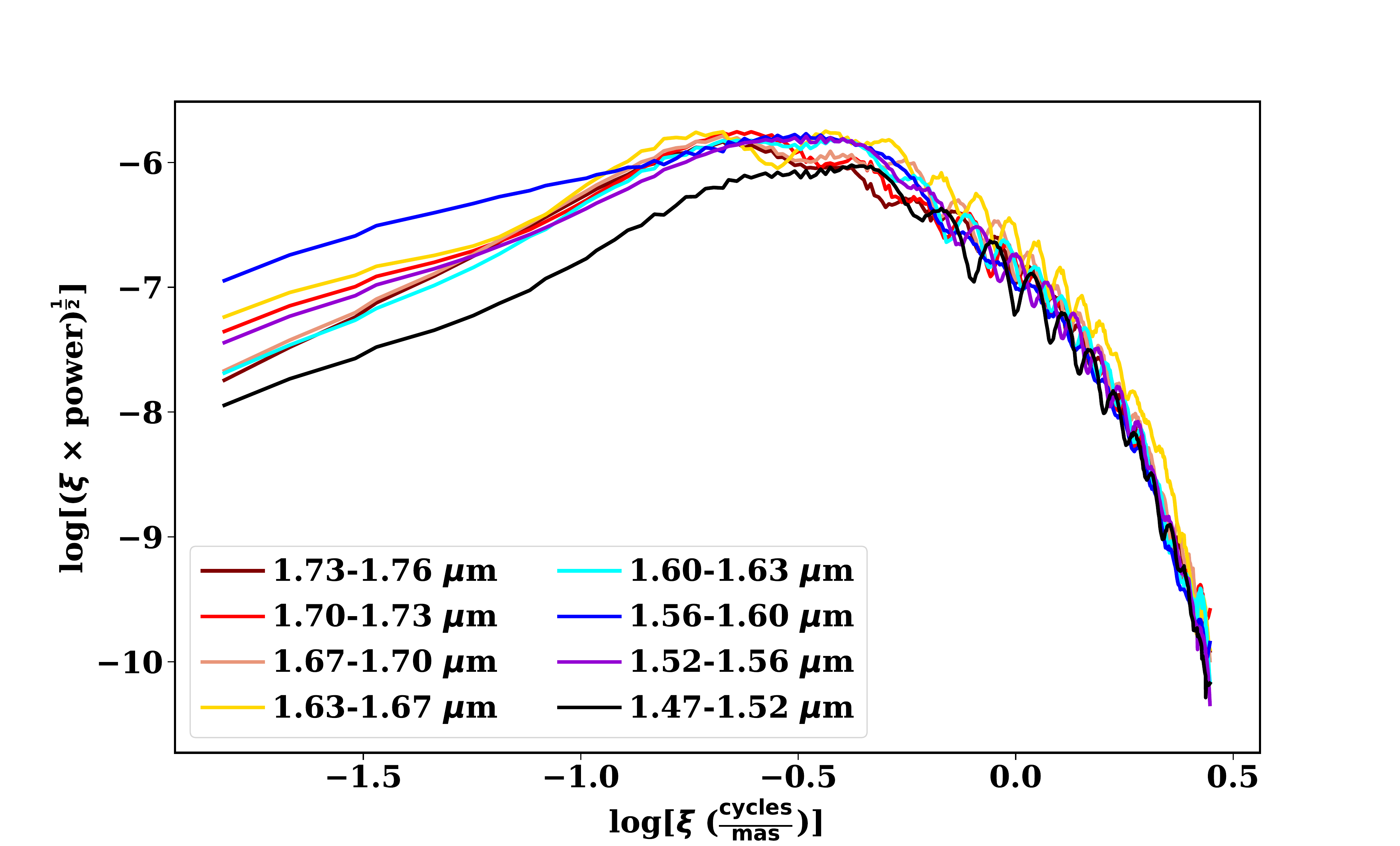}{0.5\textwidth}{2015}
        \fig{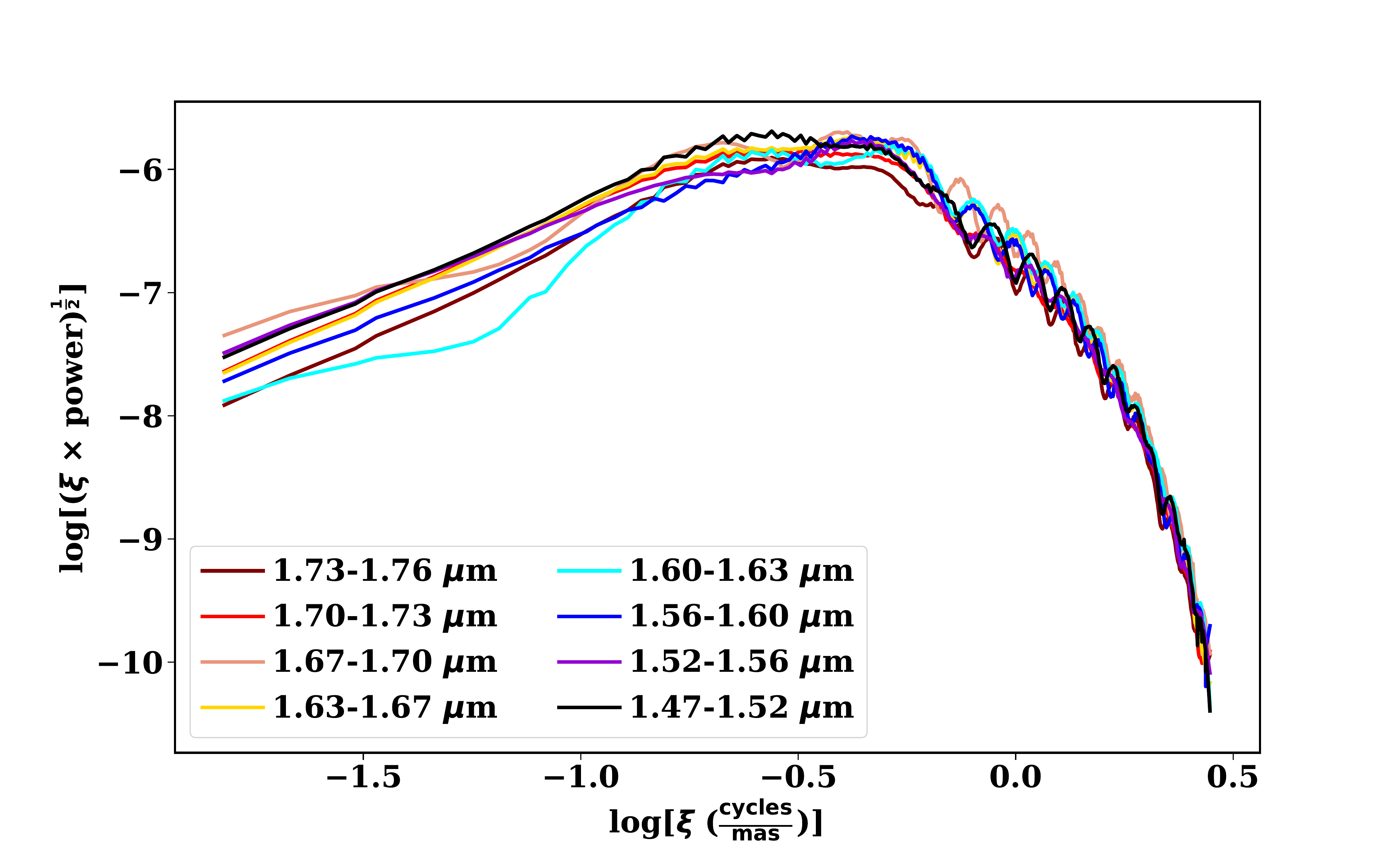}{0.5\textwidth}{2016}
          }
\caption{Power spectra of the mean reconstructed images in Figure 5, with the limb darkened disk edge filtered out.
\label{fig:powerspectra}}
\end{figure*}

%% file: Figures/powerspecoitools.tex
\begin{figure*}[!h]
\gridline{\fig{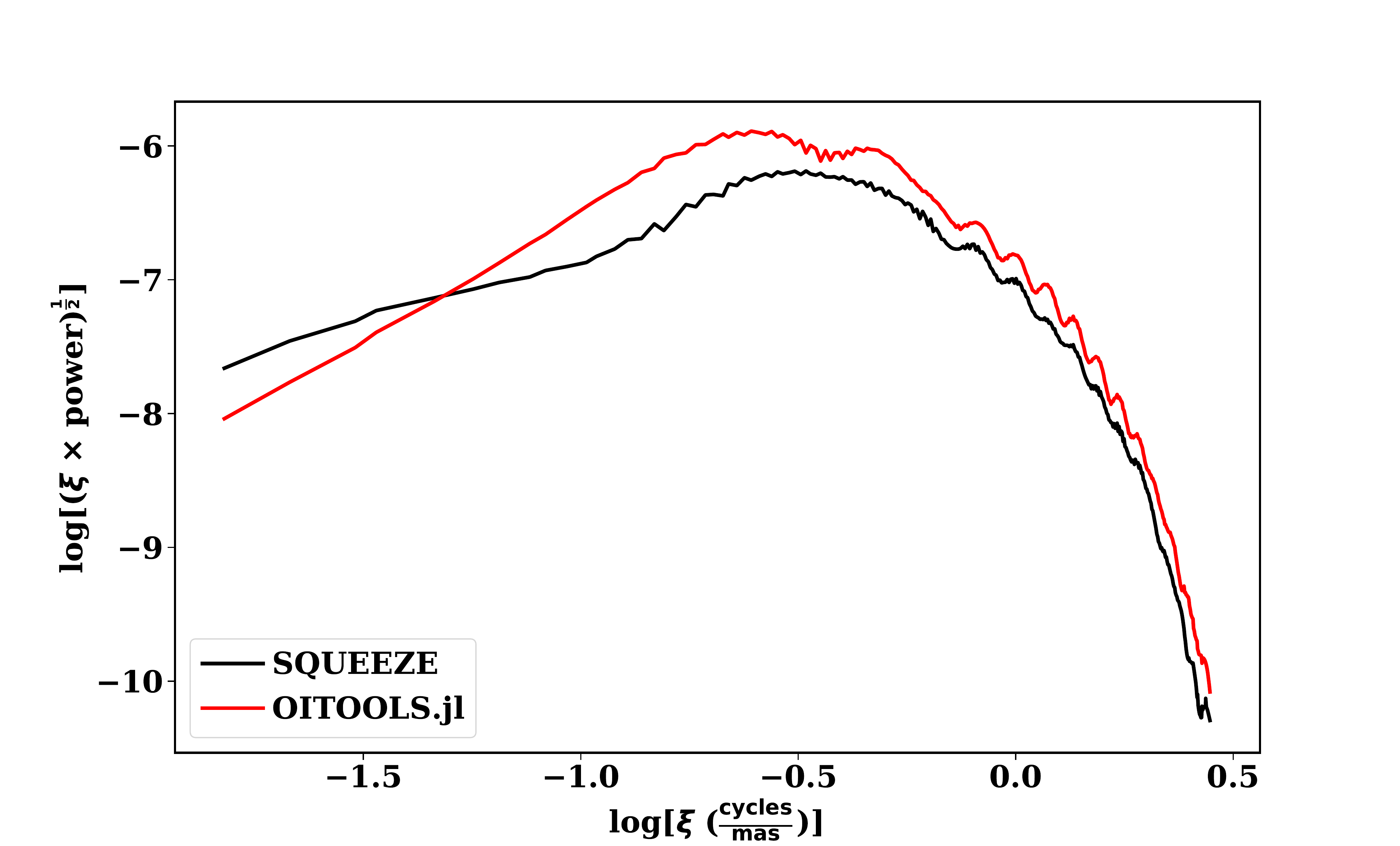}{0.5\textwidth}{2011}\\
     \fig{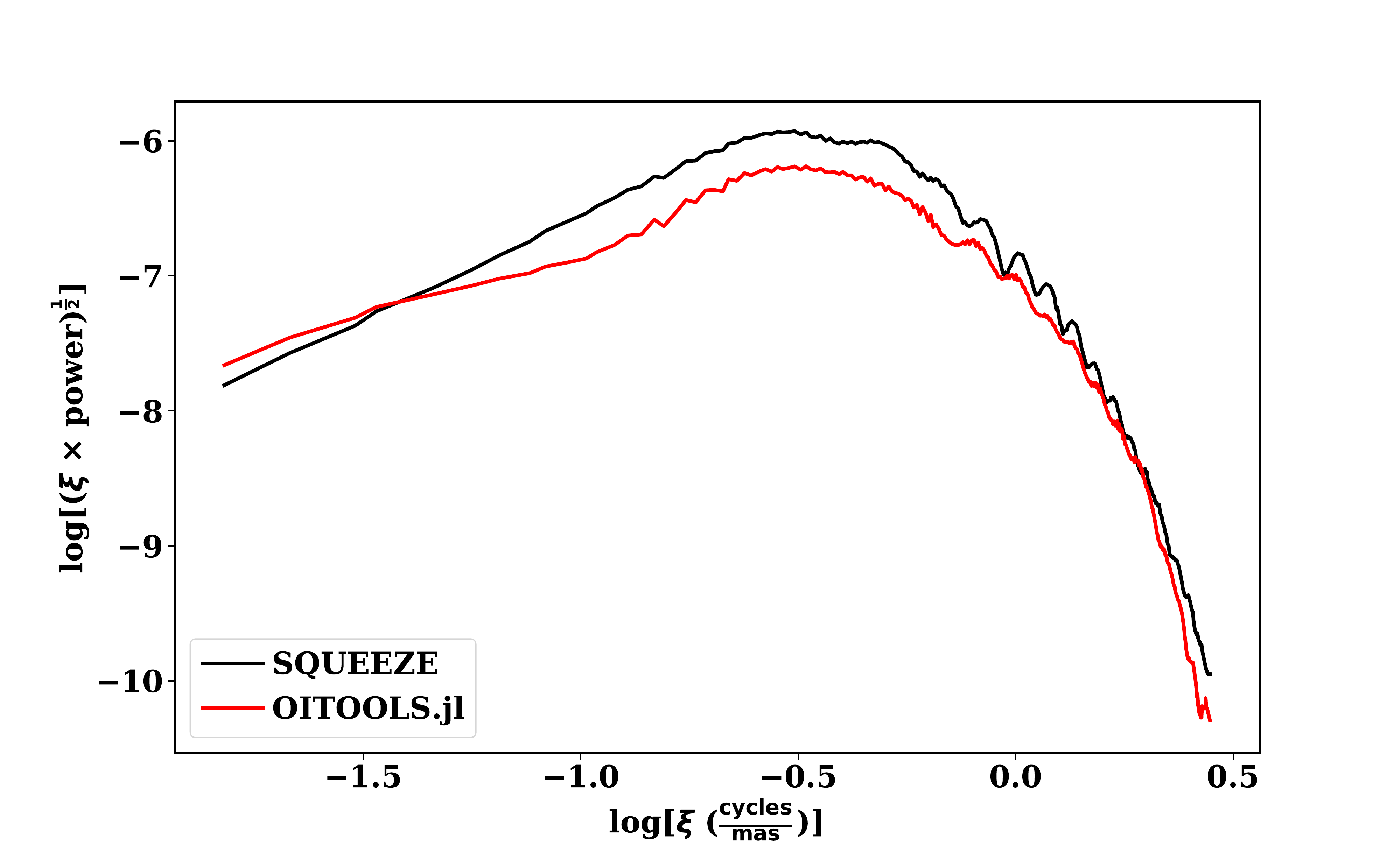}{0.5\textwidth}{2014}
         }
        \gridline{ \fig{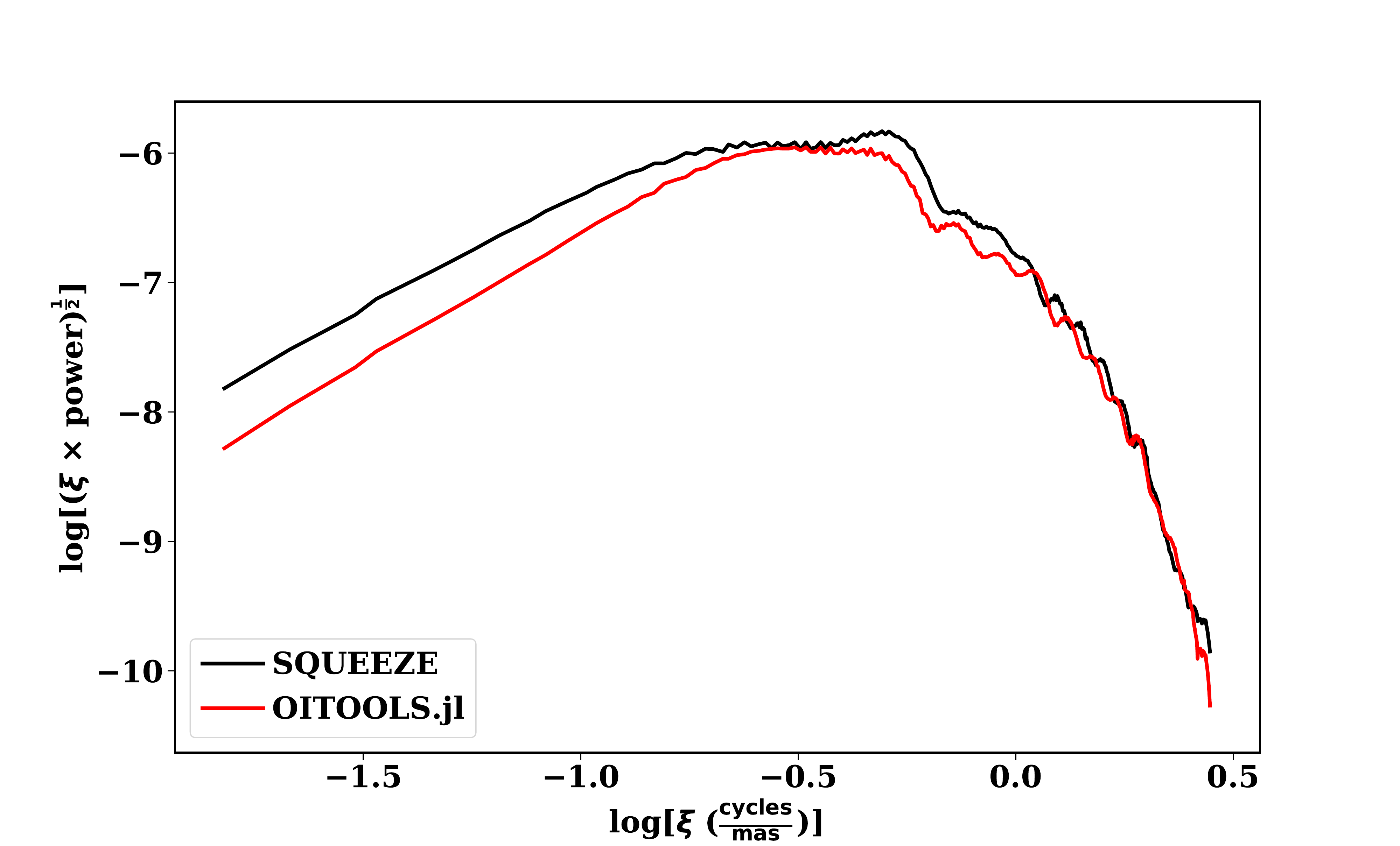}{0.5\textwidth}{2015}
        \fig{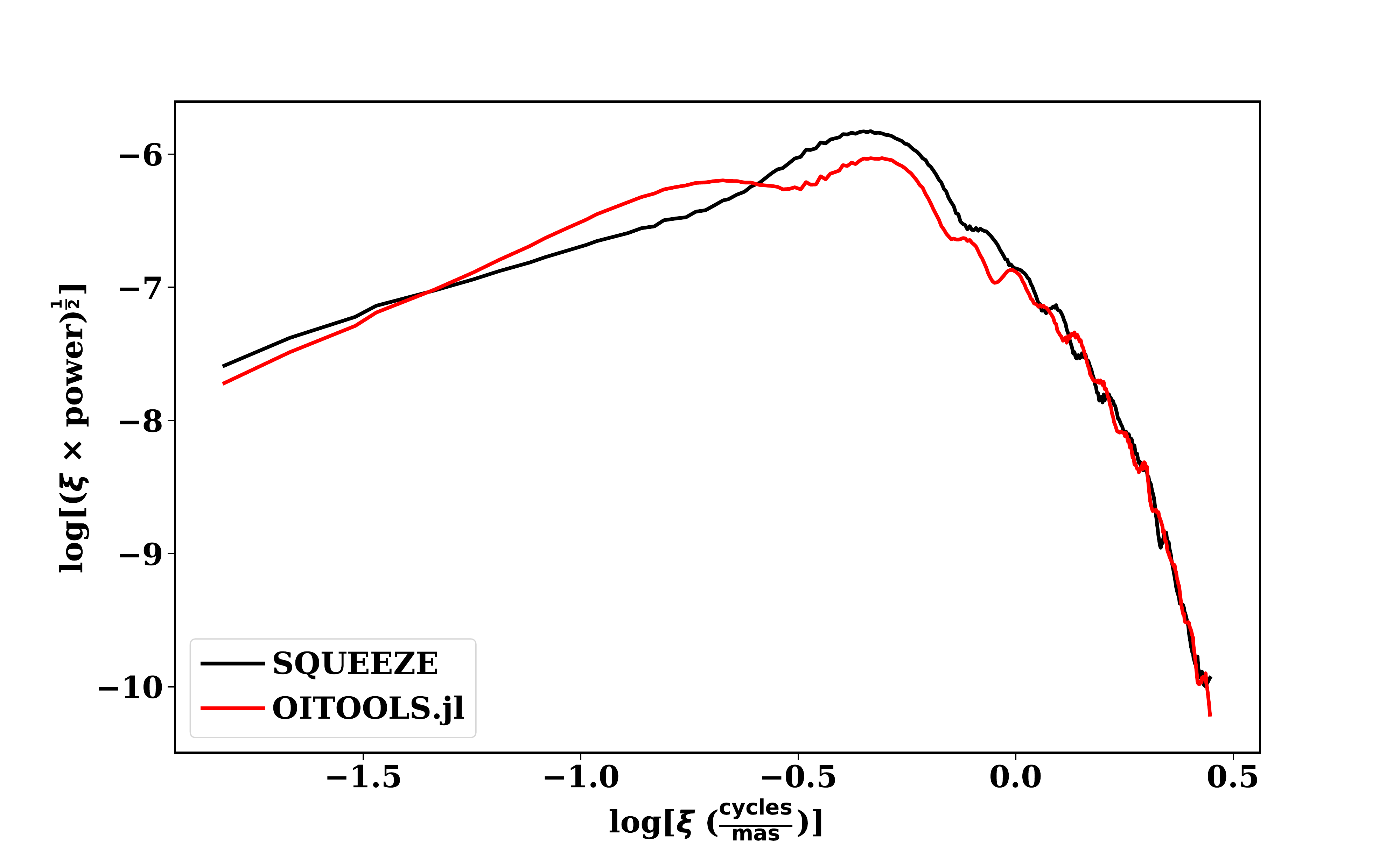}{0.5\textwidth}{2016}
          }
\caption{Power spectra of the mean reconstructed images from OITOOLS in Figure 7 compared to the reconstructions from SQUEEZE in Figure 5, both with the limb darkened disk edge filtered out.
\label{fig:powerspectraoitools}}
\end{figure*}

%% file: Tables/azcygcalc.tex
\begin{deluxetable*}{cccccc}
\tablewidth{0.2\textwidth}
\scriptsize
\tablecaption{Calculated Surface Properties of AZ~Cyg. When ranges are reported they represent the lower and upper values determined using the parameters in Table \ref{tab:parameters}\label{tab:calcazcyg}}
\tablehead{
\colhead{Year} & \colhead{$x{_\text{gran, obsv}}$} & \colhead{$x{_\text{gran, Freytag}}$} & \colhead{$x{_\text{gran, Tremblay}}$} &  \colhead{ $x{_\text{gran, Trampedach}}$} & \colhead{$x{_\text{gran, Chiavassa}}$}\\
\colhead{} & \colhead{(R$_{*}$)} & \colhead{(R$_{*}$)} & \colhead{(R$_{*}$)} & \colhead{(R$_{*}$)} & \colhead{(R$_{*}$)}
}
\startdata
2011 & $0.69\pm 0.01$ & 0.10$^{+0.01}_{-0.01}$ & 0.04-0.19 & 0.09-0.38 & 0.51$^{+0.03}_{-0.03}$\\
2014 & $0.67\pm0.19$ & 0.11$^{+0.01}_{-0.01}$ & 0.04-0.19 & 0.09-0.37 & 0.53$^{+0.04}_{-0.03}$\\
2015 & $0.52\pm0.03$ & 0.11$^{+0.01}_{-0.01}$ &0.04-0.19& 0.08-0.37 & 0.53$^{+0.04}_{-0.03}$\\
2016 & 0.54$\pm0.02$ & 0.11$^{+0.01}_{-0.01}$ & 0.03-0.18 & 0.08-0.35 & 0.53$^{+0.04}_{-0.03}$\\
\enddata
\end{deluxetable*}

%% file: Figures/uvall.tex
\begin{figure}[ht]
\gridline{\fig{Files/UV2011_update4.pdf}{0.46\textwidth}{2011}
\fig{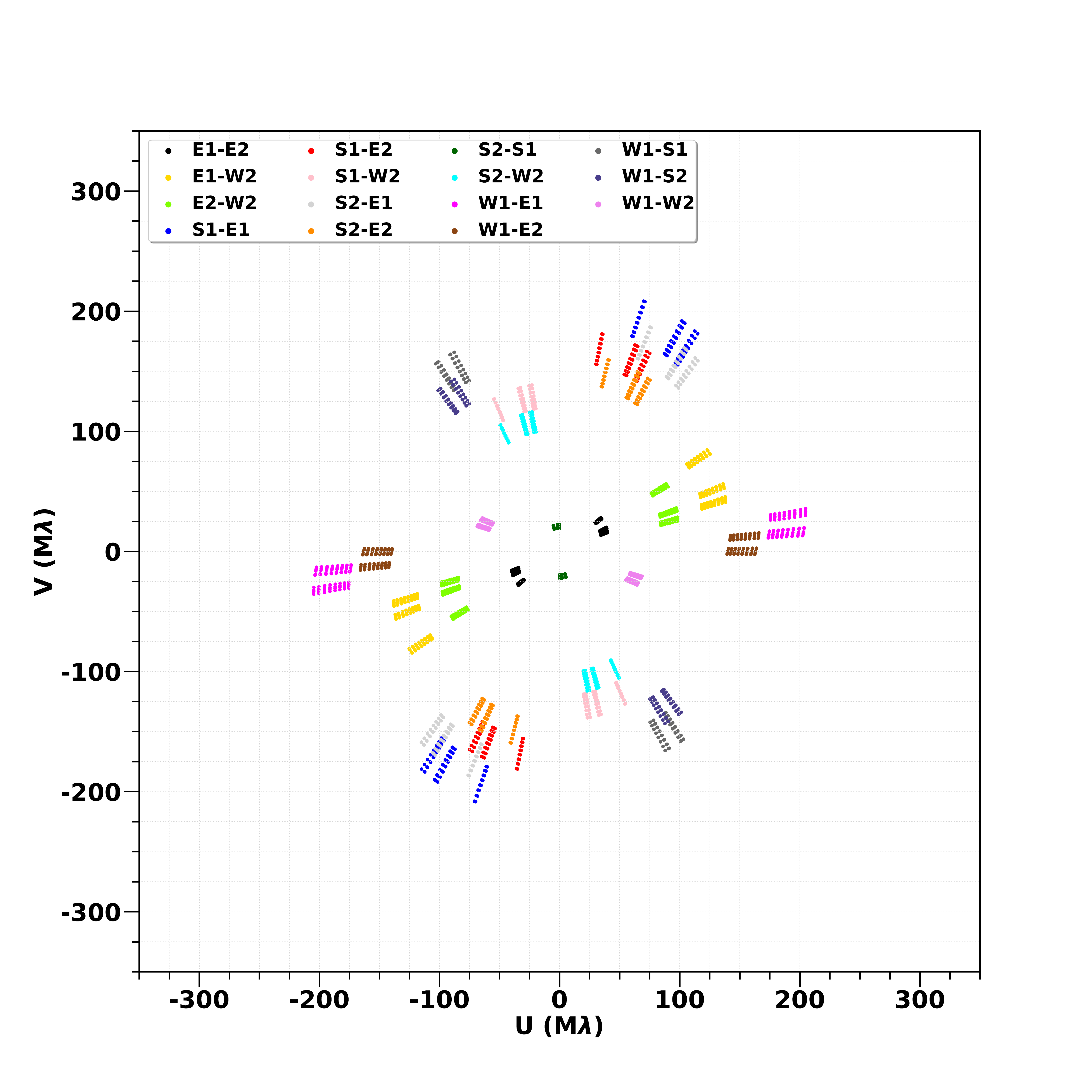}{0.46\textwidth}{2014}}
\gridline{\fig{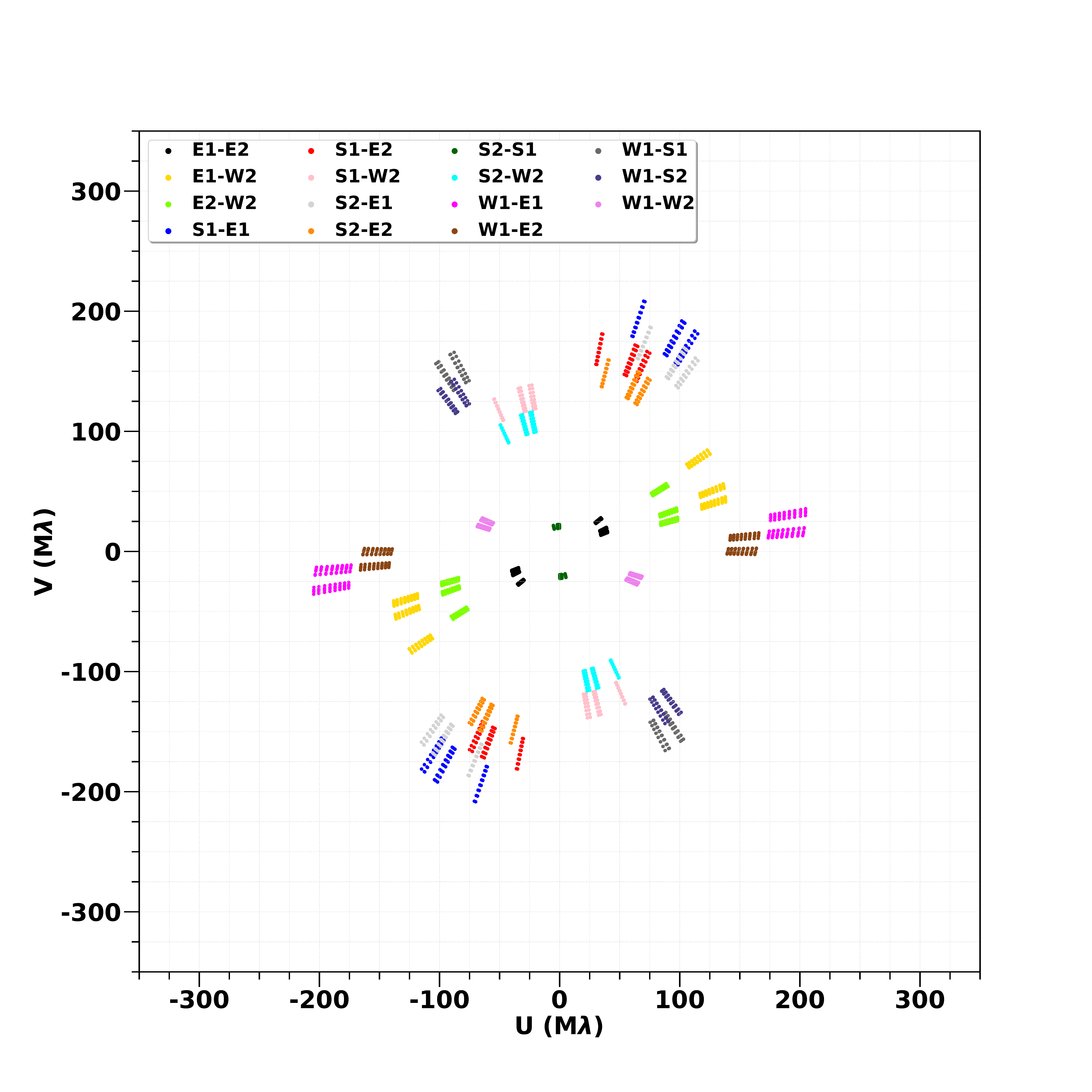}{0.46\textwidth}{2015} 
\fig{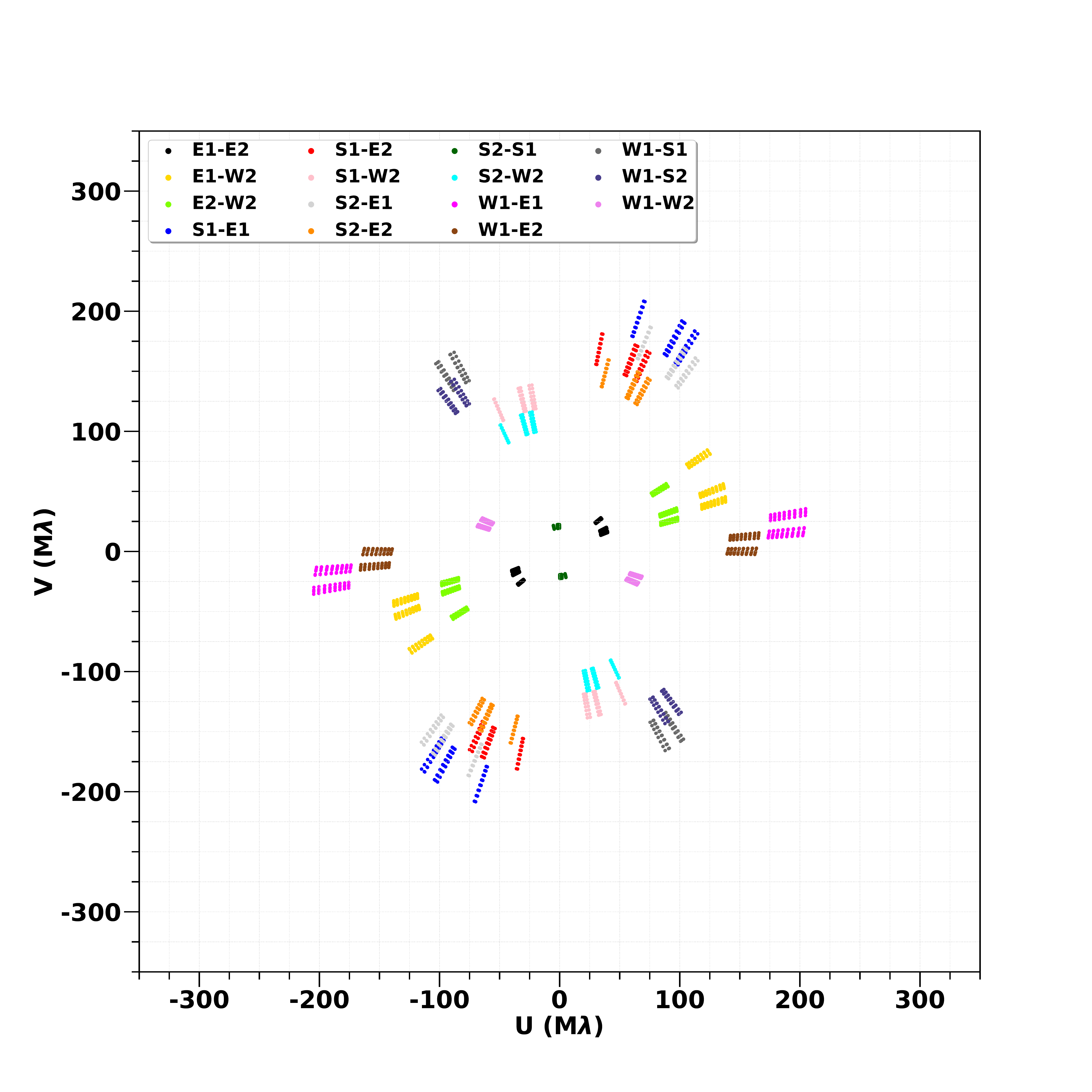}{0.46\textwidth}{2016}}

\caption{(\textit{u},\textit{v}) coverage of AZ~Cyg for four epochs over a five year time period. Different baseline combinations are identified by colors. \label{fig:uvall}}
\end{figure}

%% file: Figures/spectra.tex
\begin{figure*}[h]
\gridline{\fig{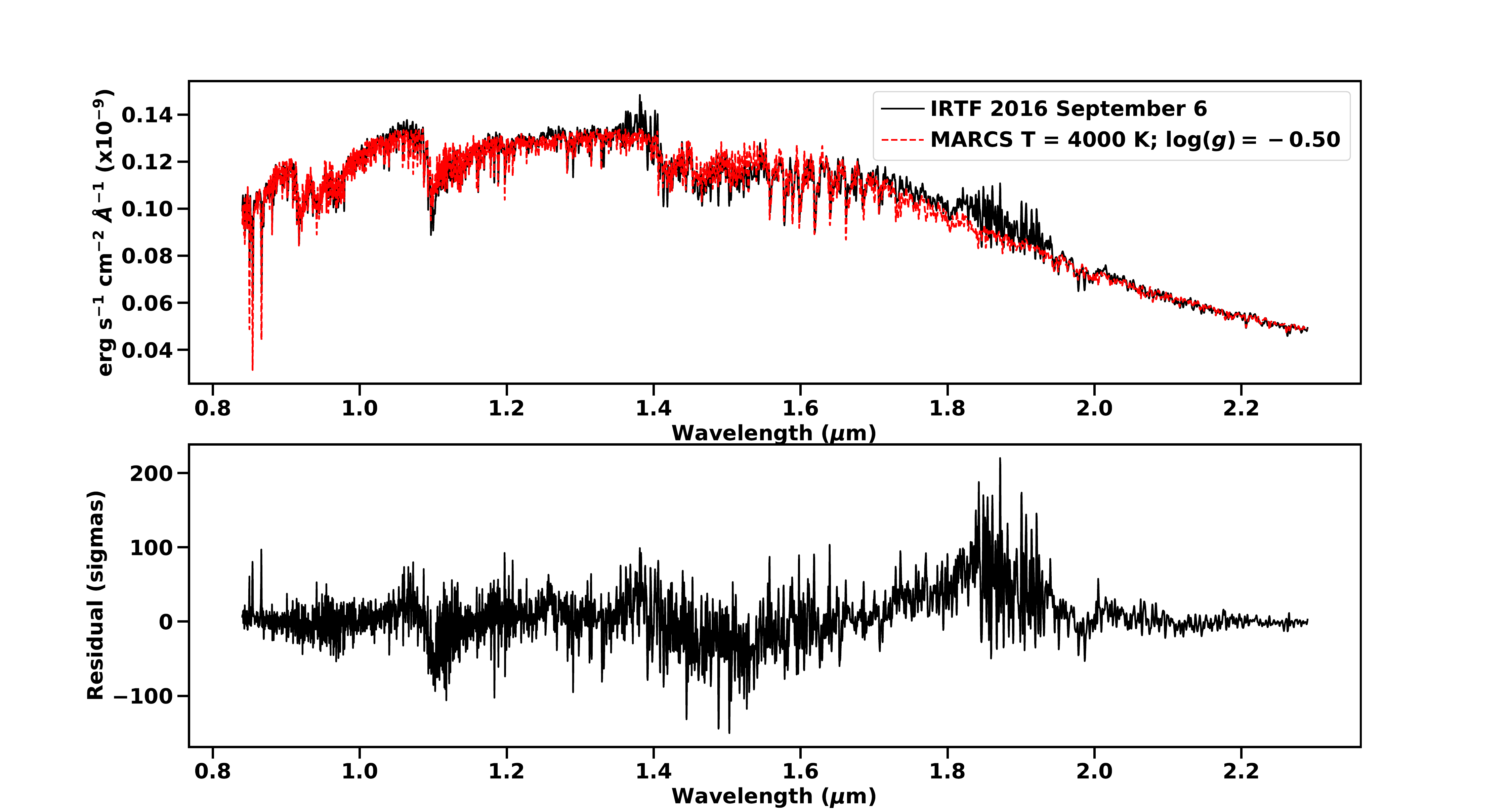}{0.7\textwidth}{MARCS fit}}
\gridline{\fig{Files/phoenix_final2update.pdf}{0.7\textwidth}{PHOENIX fit}}
\gridline{\fig{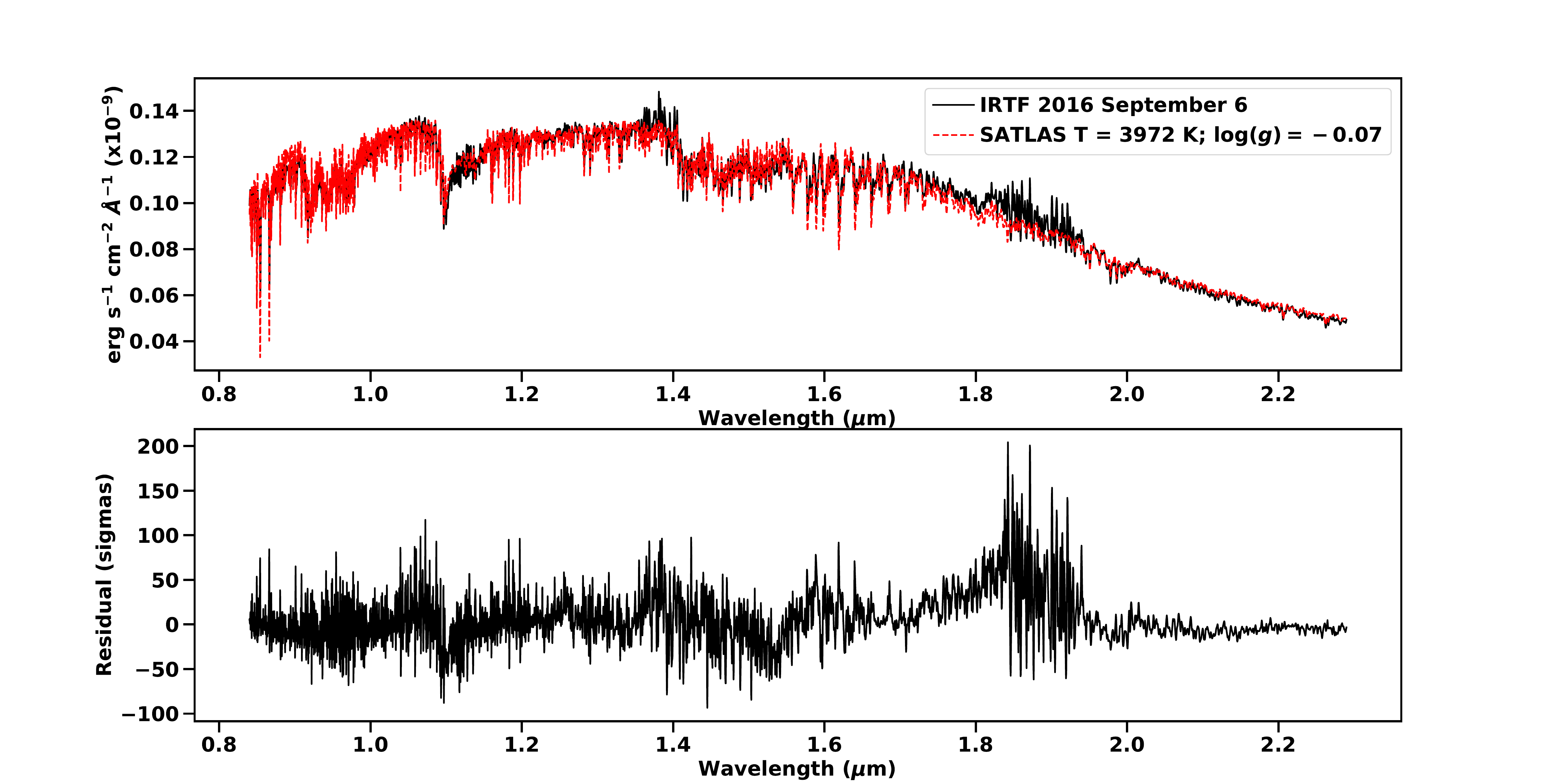}{0.7\textwidth}{SATLAS fit}}
\caption{Best fitting model spectrum (red) compared to observed spectrum.\label{fig:spectra}}
\end{figure*}

%% file: Figures/compare.tex
\begin{figure*}[!h]
\gridline{
\fig{Files/V2model2011_2.pdf}{0.5\textwidth}{AZ~Cyg 2011 Squared Visibilities}
\fig{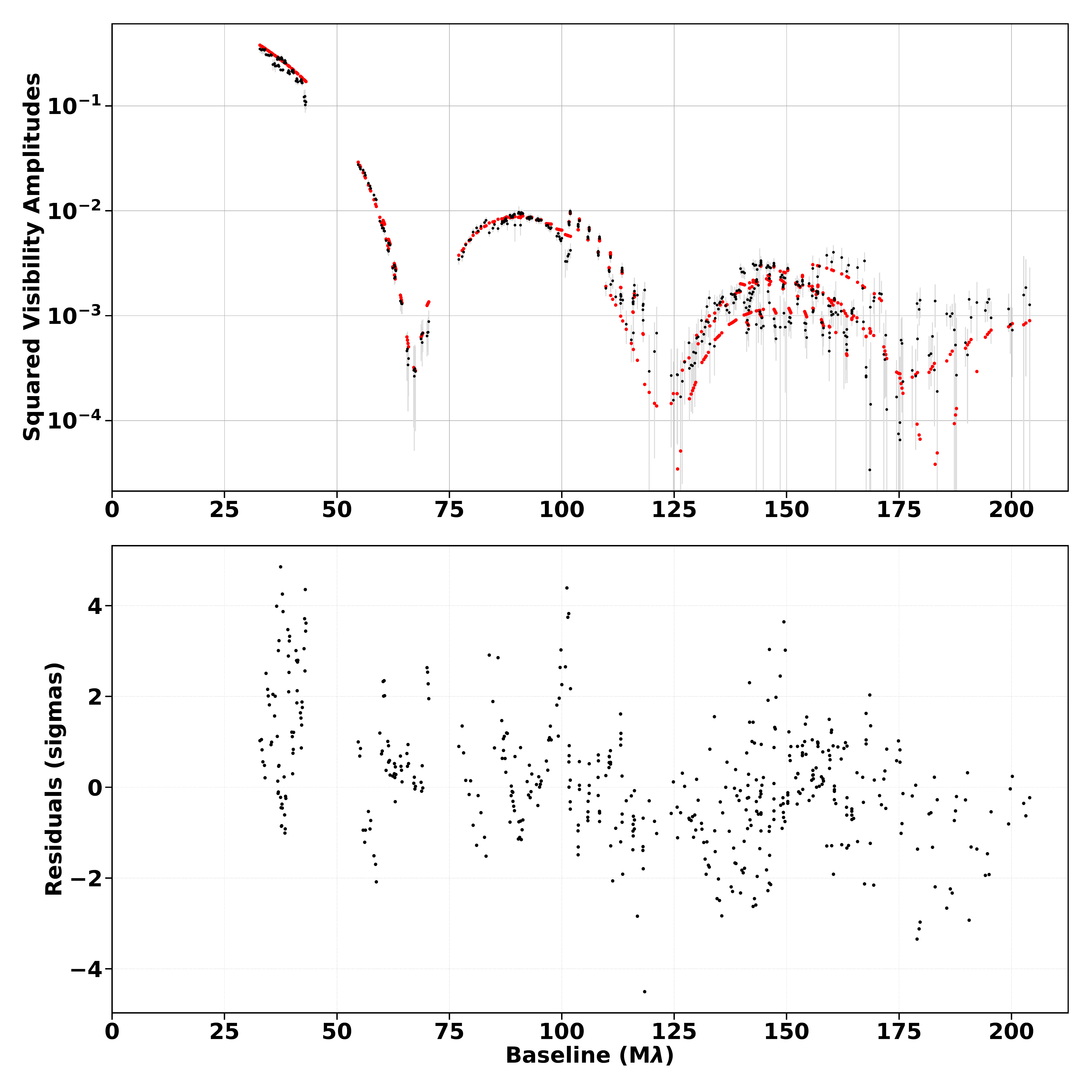}{0.5\textwidth}{AZ~Cyg 2014 Squared Visibilities}}
\gridline{ \fig{Files/t3model2011_2.pdf}{0.5\textwidth}{AZ~Cyg 2011 Closure Phases}
\fig{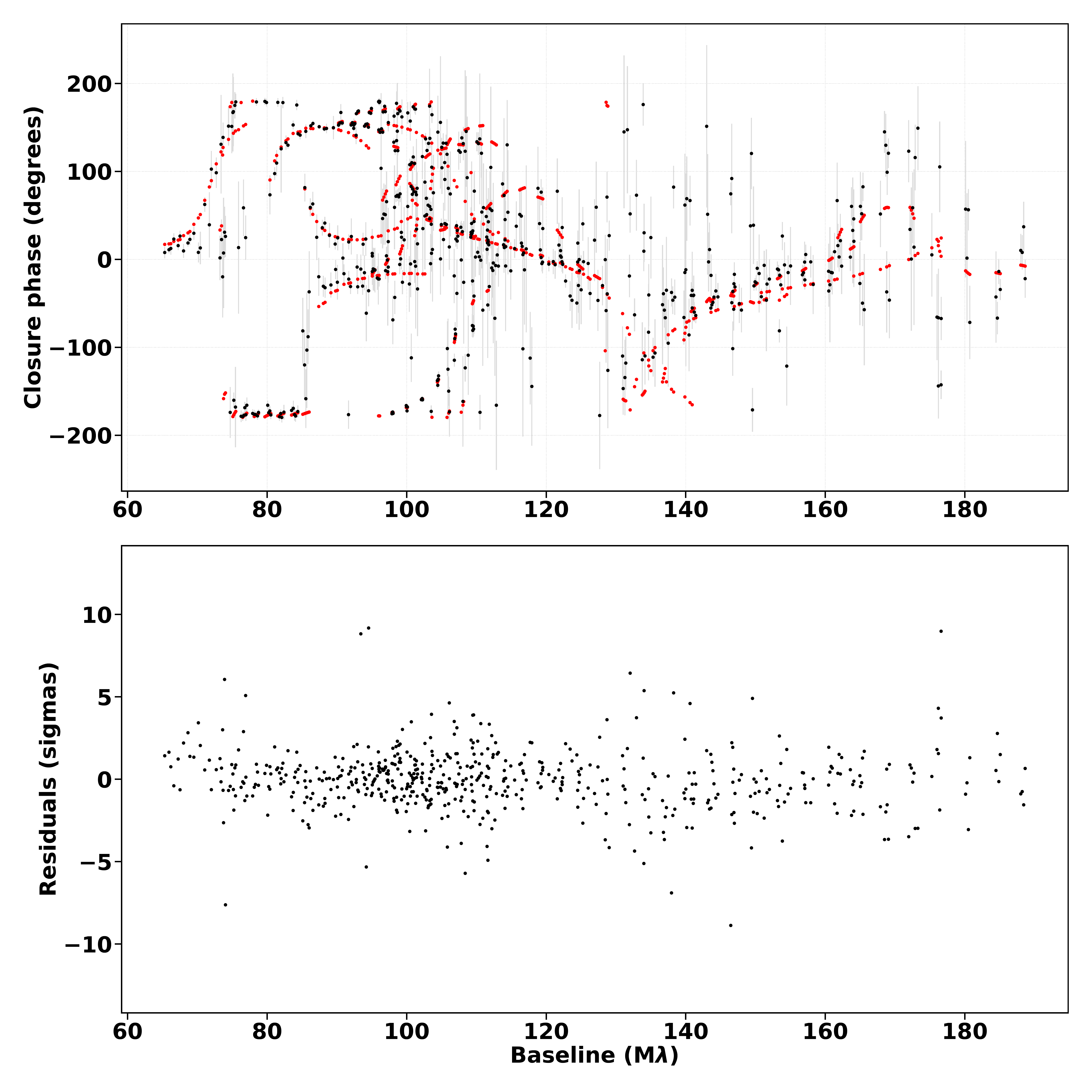}{0.5\textwidth}{AZ~Cyg 2014 Closure Phases}}

\end{figure*}
\begin{figure*}[!h]

\gridline{
\fig{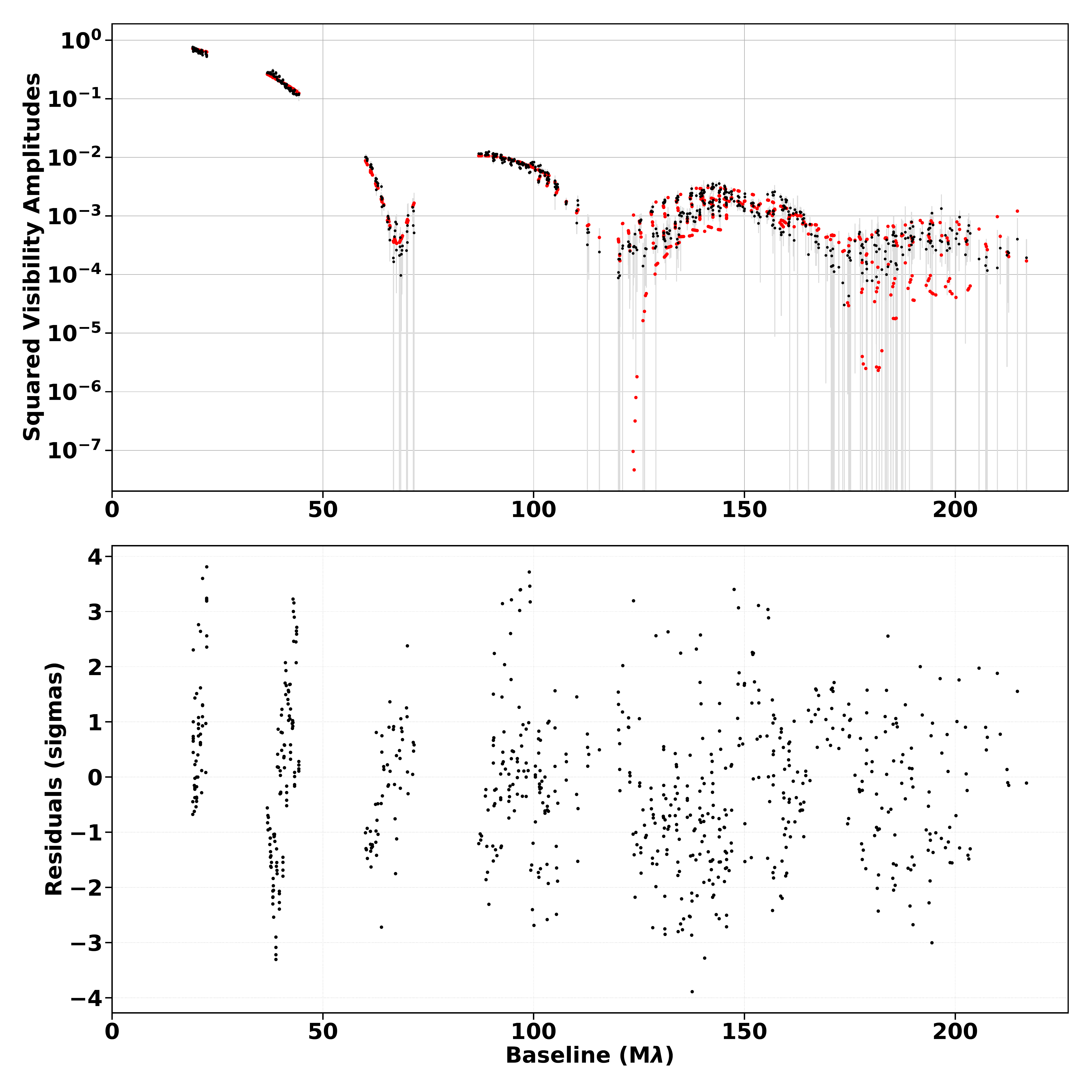}{0.5\textwidth}{AZ~Cyg 2015 Squared Visibilities}
\fig{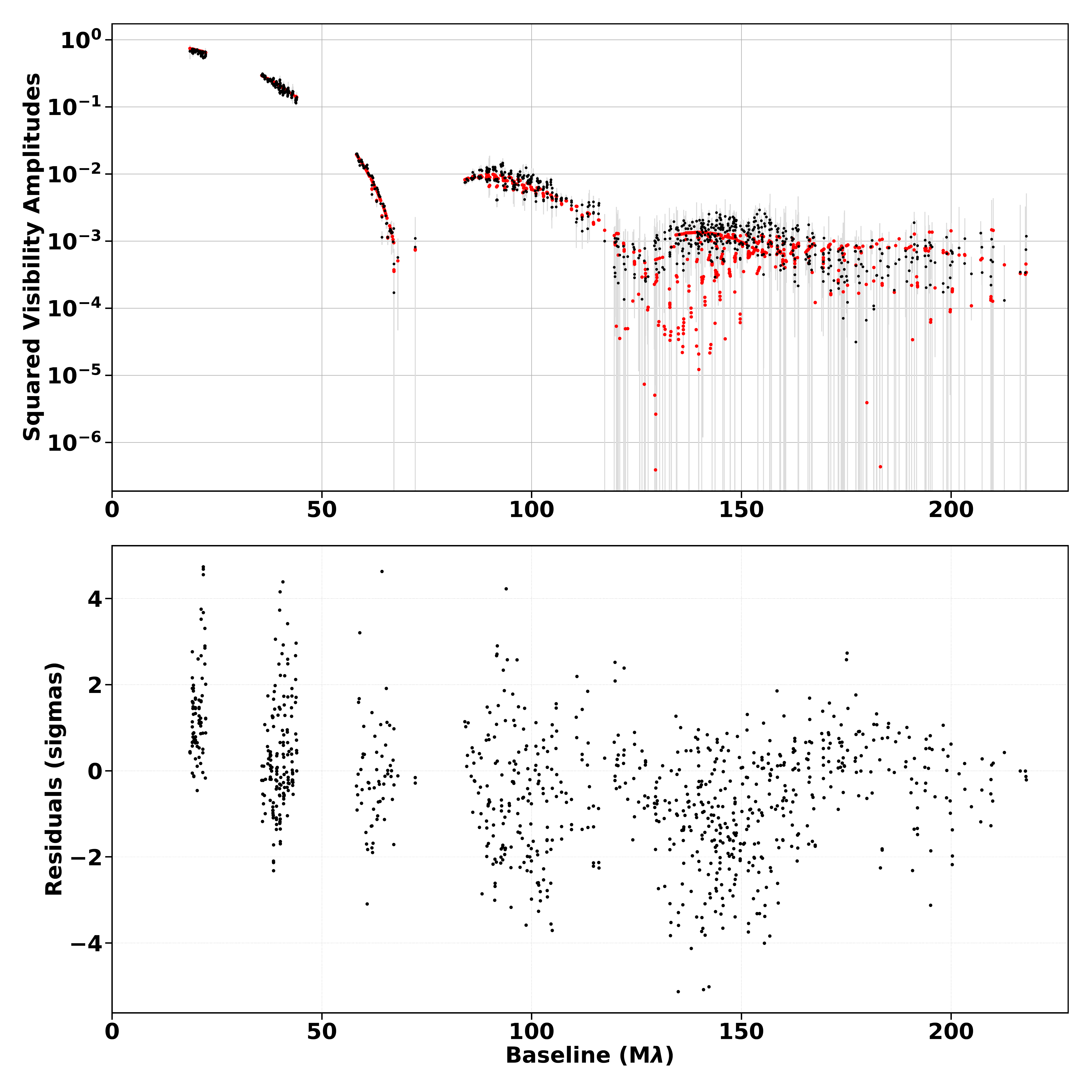}{0.5\textwidth}{AZ~Cyg 2016 Squared Visibilities}
}
\gridline{
\fig{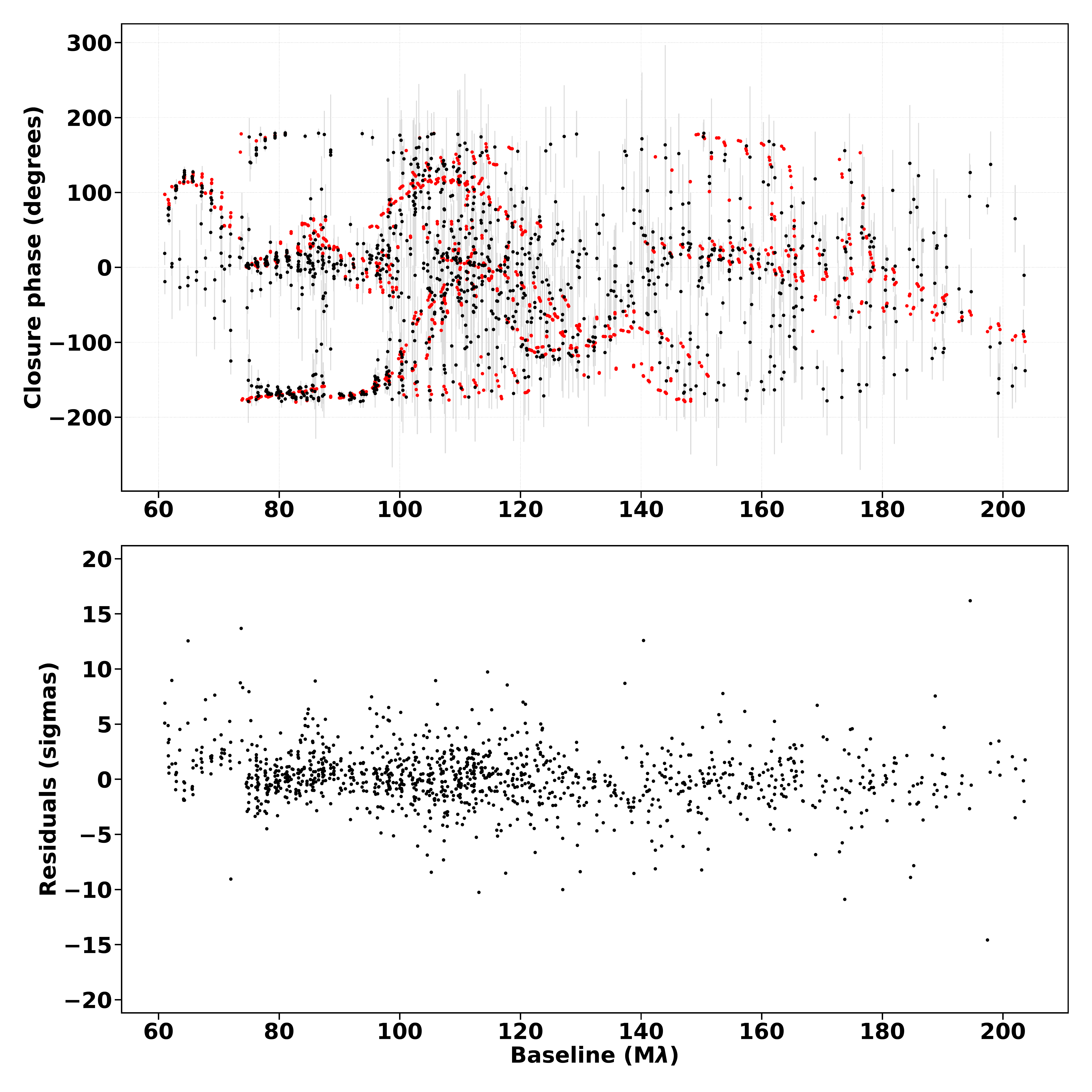}{0.5\textwidth}{AZ~Cyg 2015 Closure Phases}
\fig{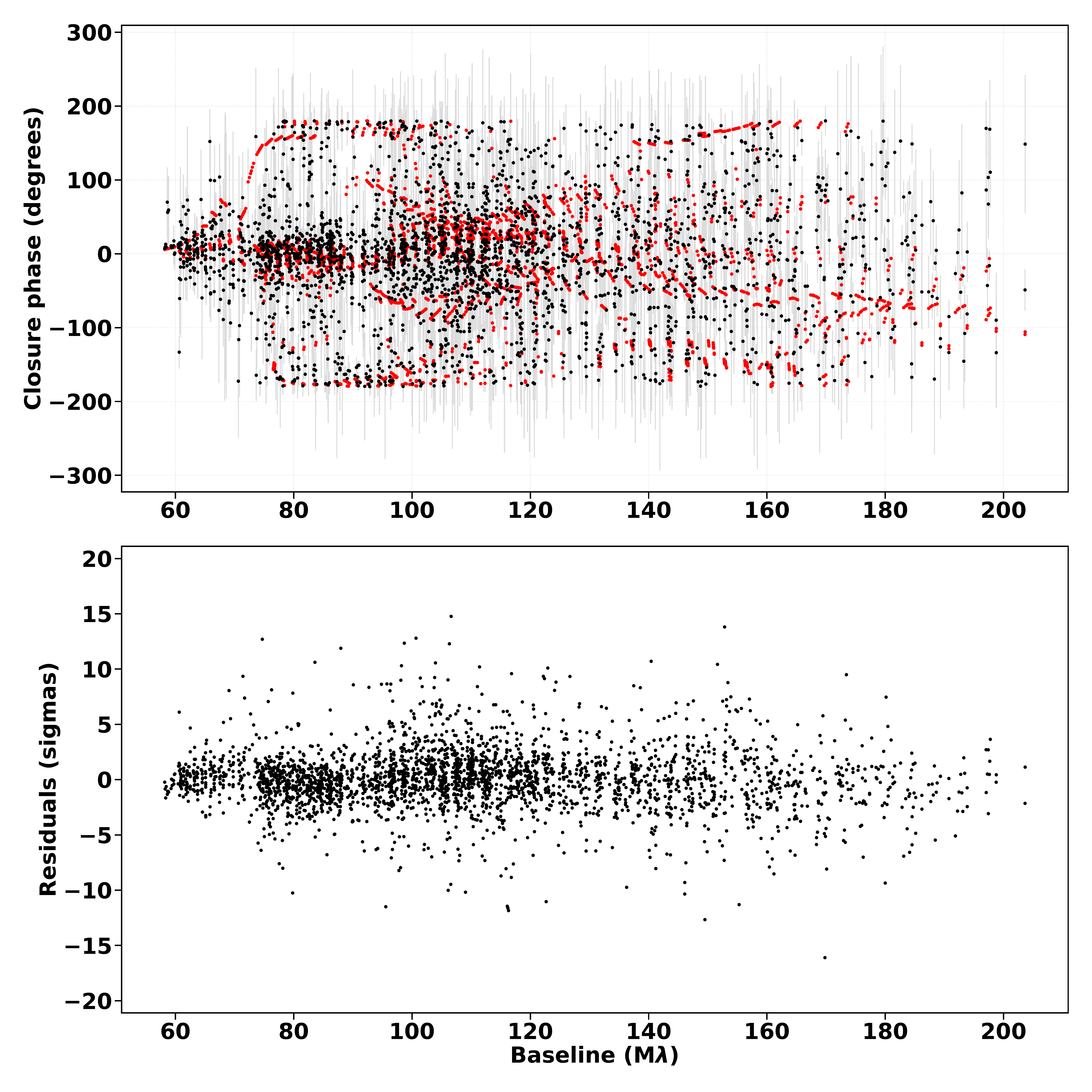}{0.5\textwidth}{AZ~Cyg 2016 Closure Phases}
          }
          \caption{Upper: Comparison of AZ~Cyg observations (black) to squared visibilities and closure phases calculated from the mean SQUEEZE images (red). Lower: Residuals (in sigmas) between the quantities in each upper graph. 
\label{fig:compare}}
\end{figure*}

%% file: Figures/wavelength_1.tex
\begin{figure*}[p]
\gridline{\fig{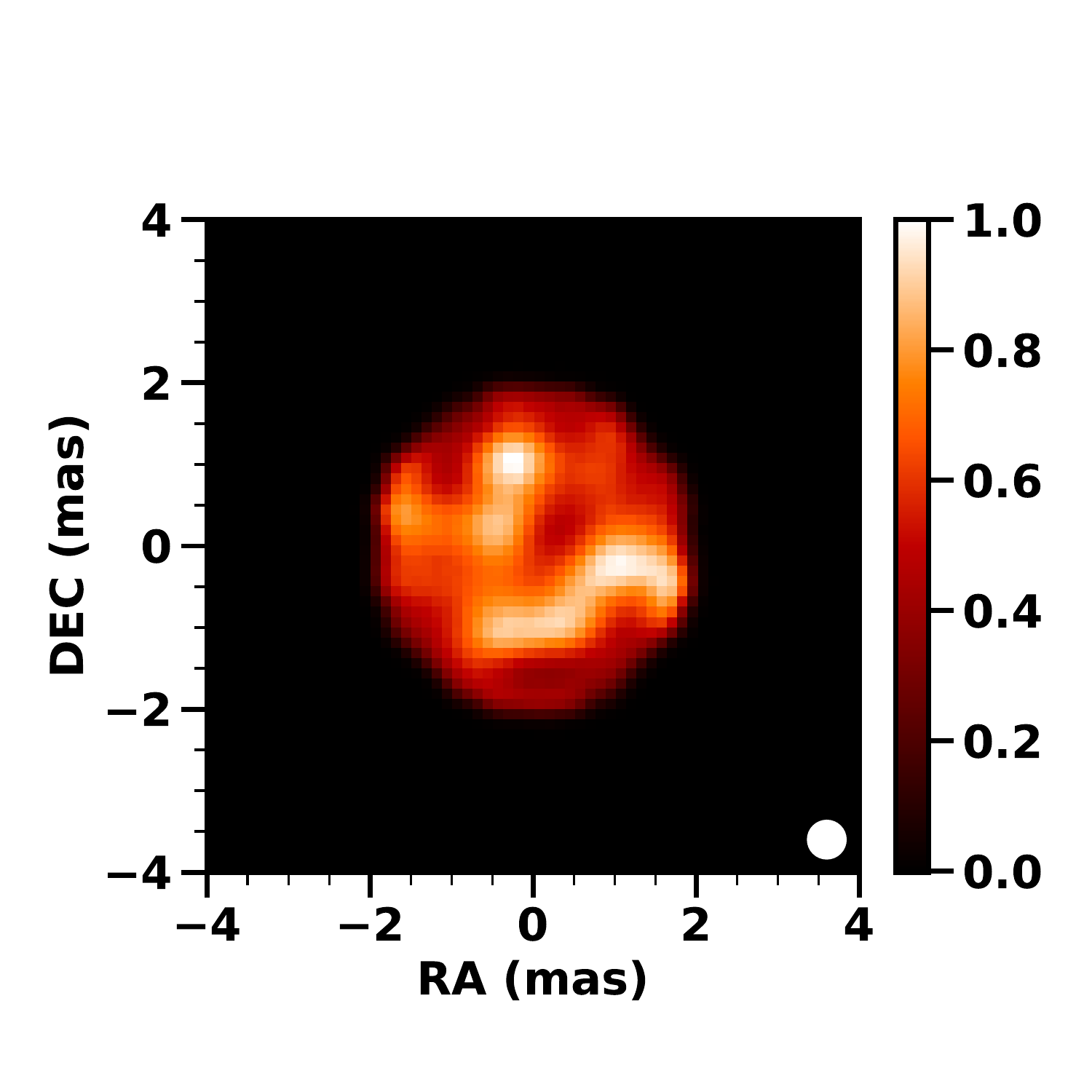}{0.2\textwidth}{AZ~Cyg 2011 1.47$u$m-1.52$u$m}
          \fig{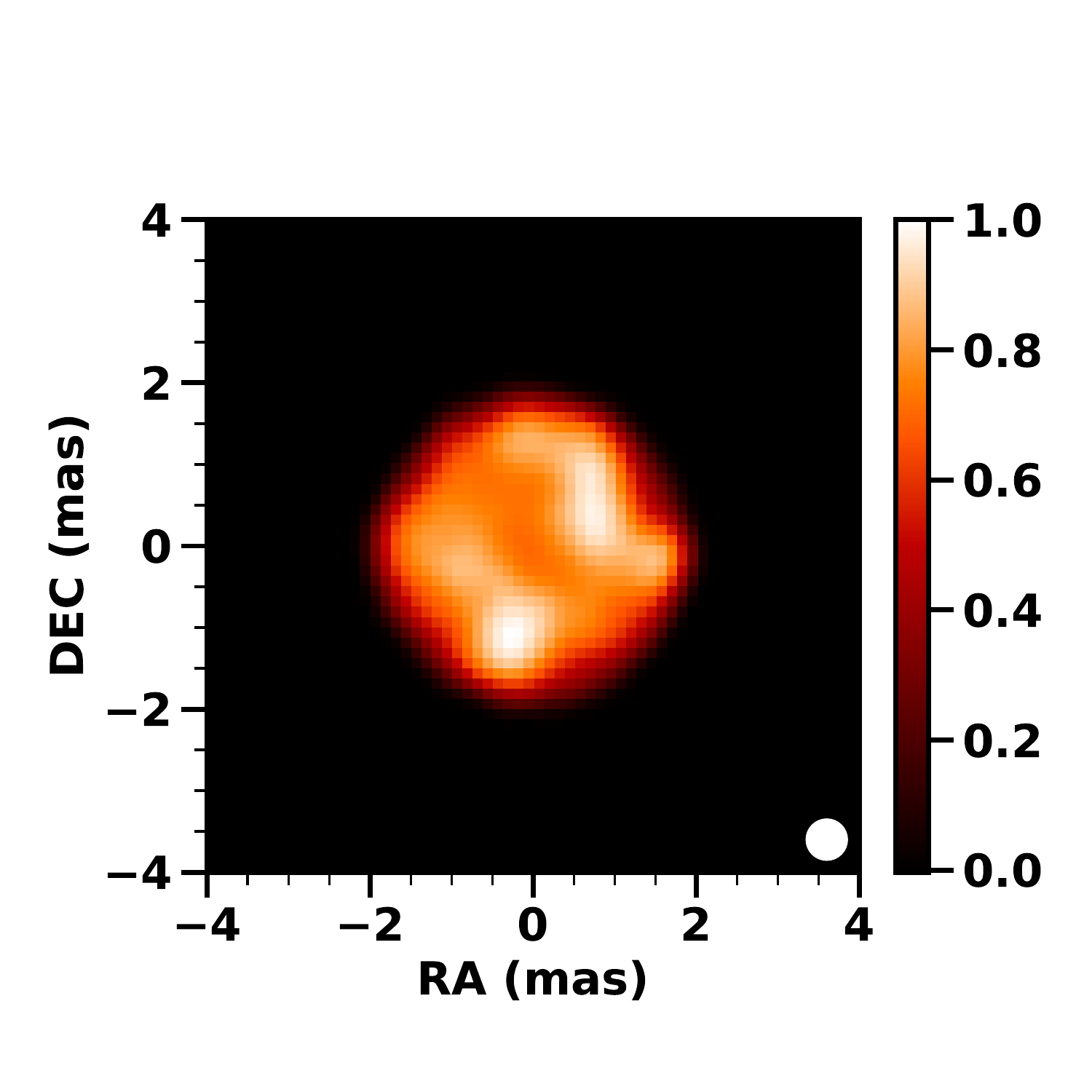}{0.2\textwidth}{AZ~Cyg 2014 1.47$u$m-1.52$u$m}
          \fig{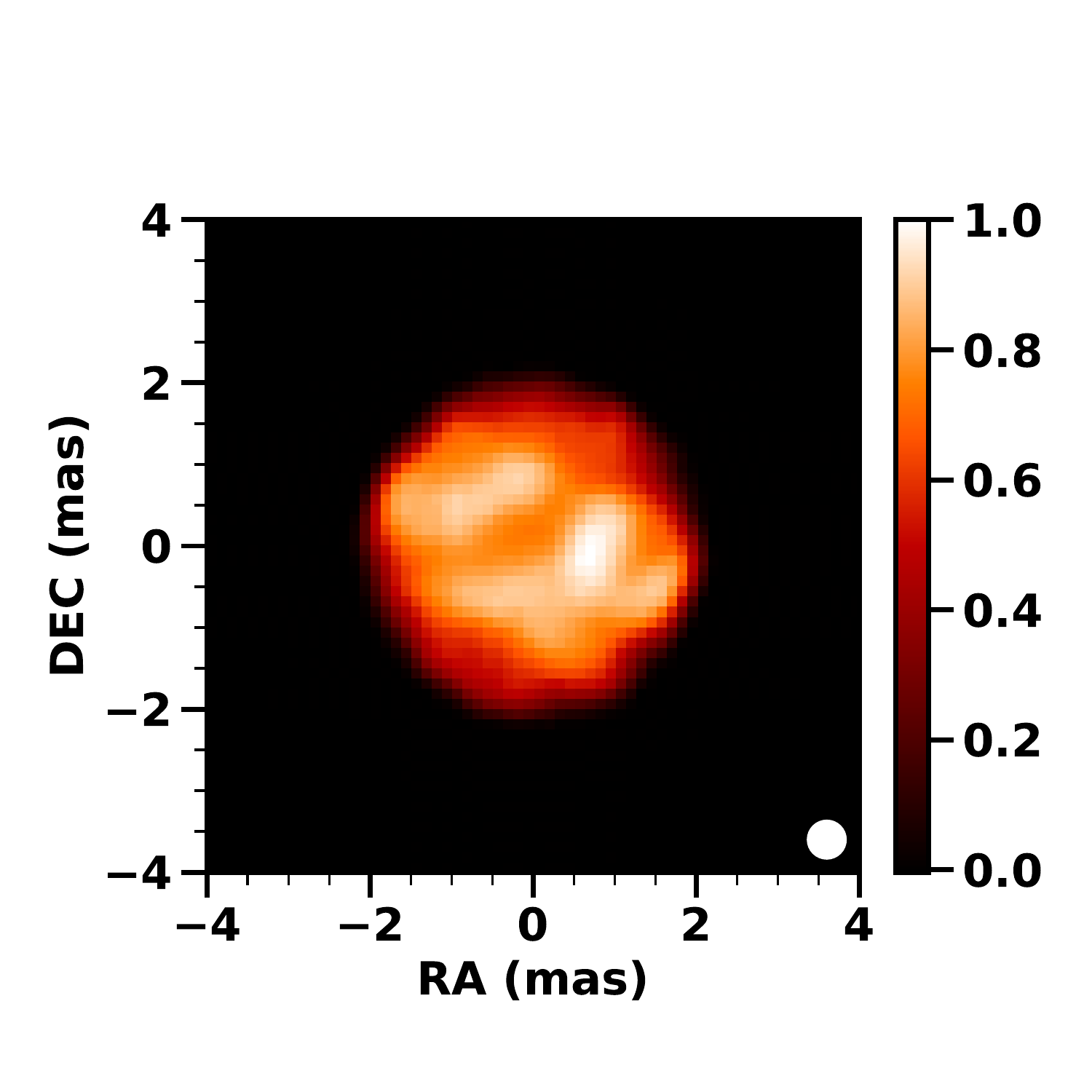}{0.2\textwidth}{AZ~Cyg 2015 1.47$u$m-1.52$u$m}
          \fig{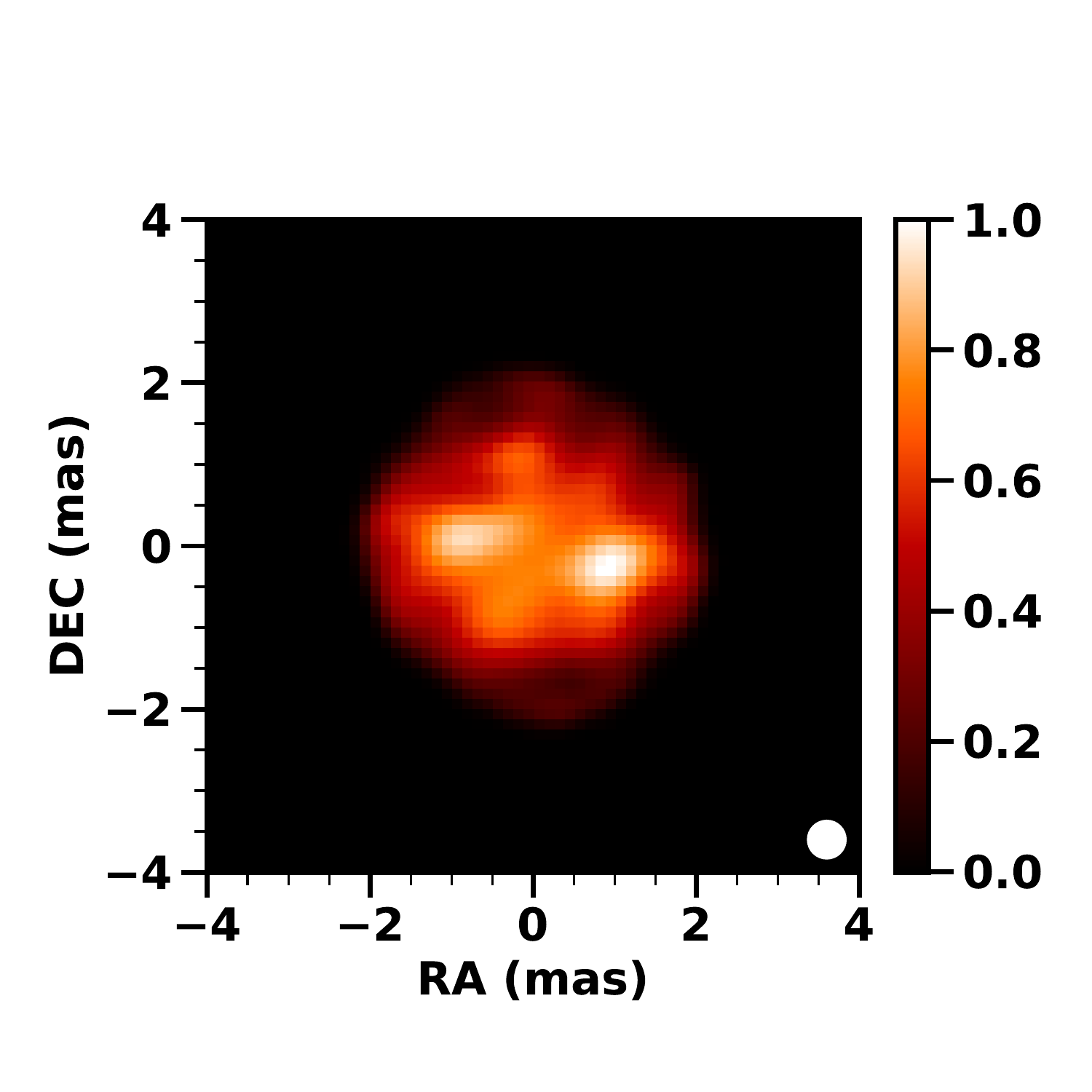}{0.2\textwidth}{AZ~Cyg 2016 1.47$u$m-1.52$u$m}
          }
        \gridline{\fig{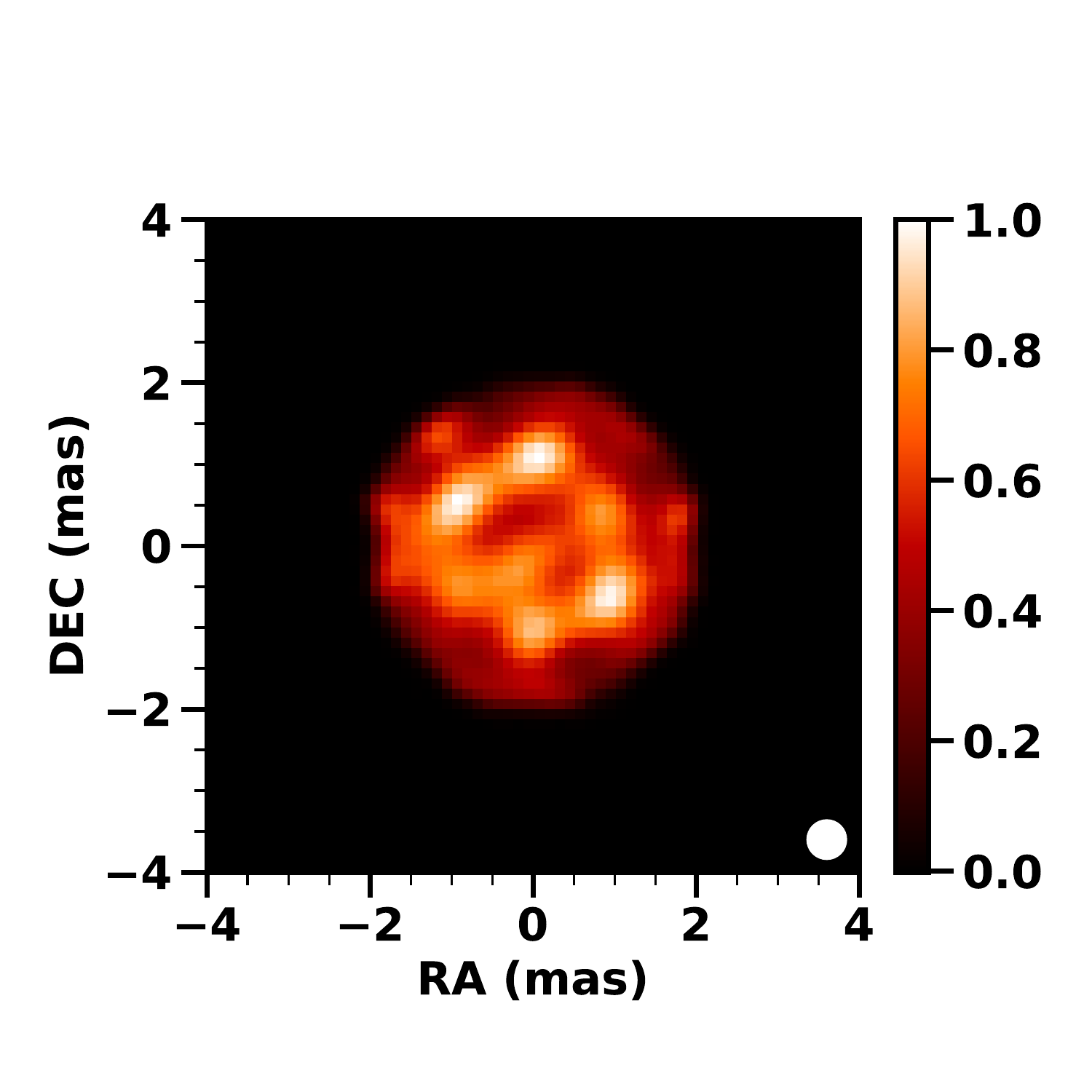}{0.2\textwidth}{AZ~Cyg 2011 1.52$u$m-1.56$u$m}
          \fig{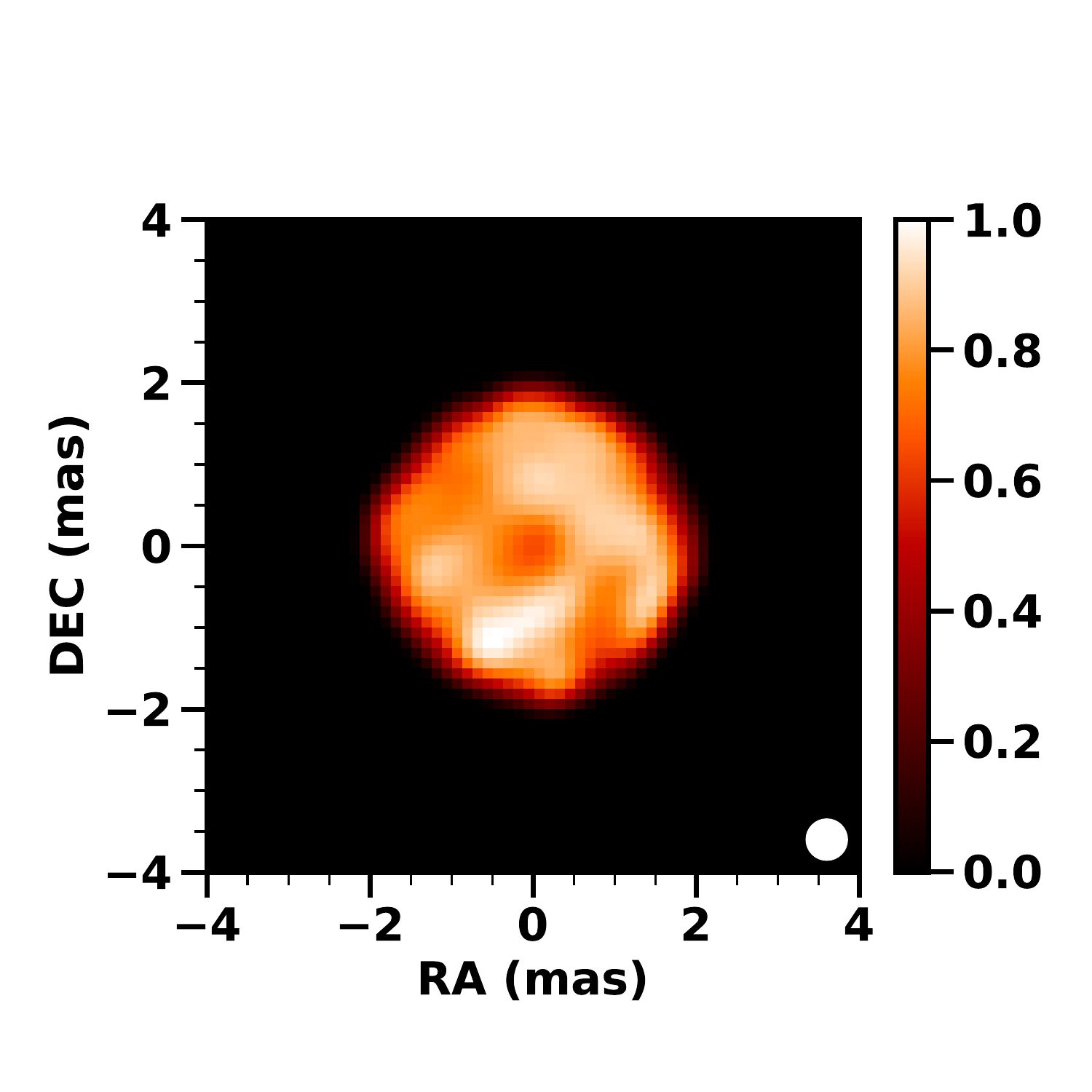}{0.2\textwidth}{AZ~Cyg 2014 1.52$u$m-1.56$u$m}
          \fig{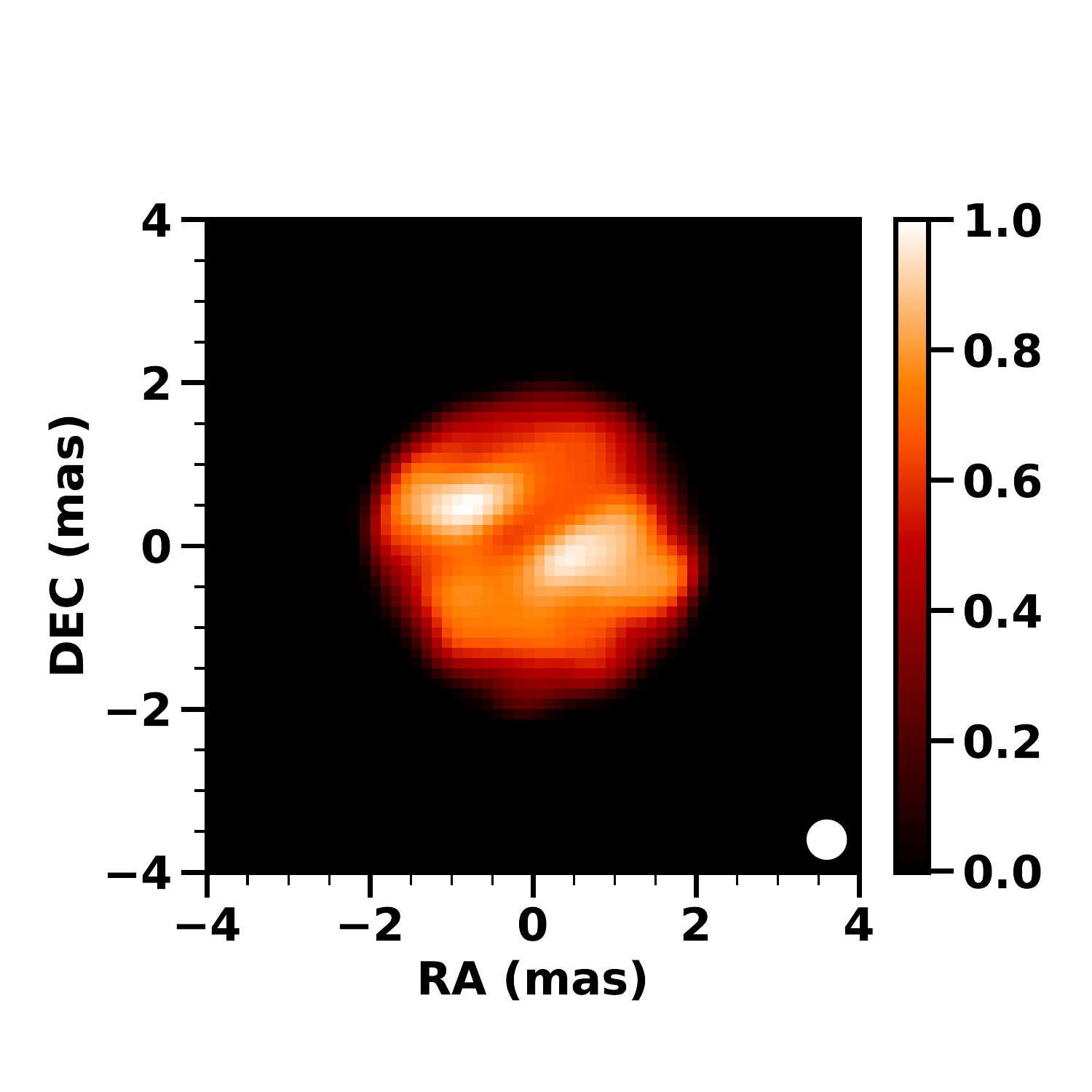}{0.2\textwidth}{AZ~Cyg 2015 1.52$u$m-1.56$u$m}
          \fig{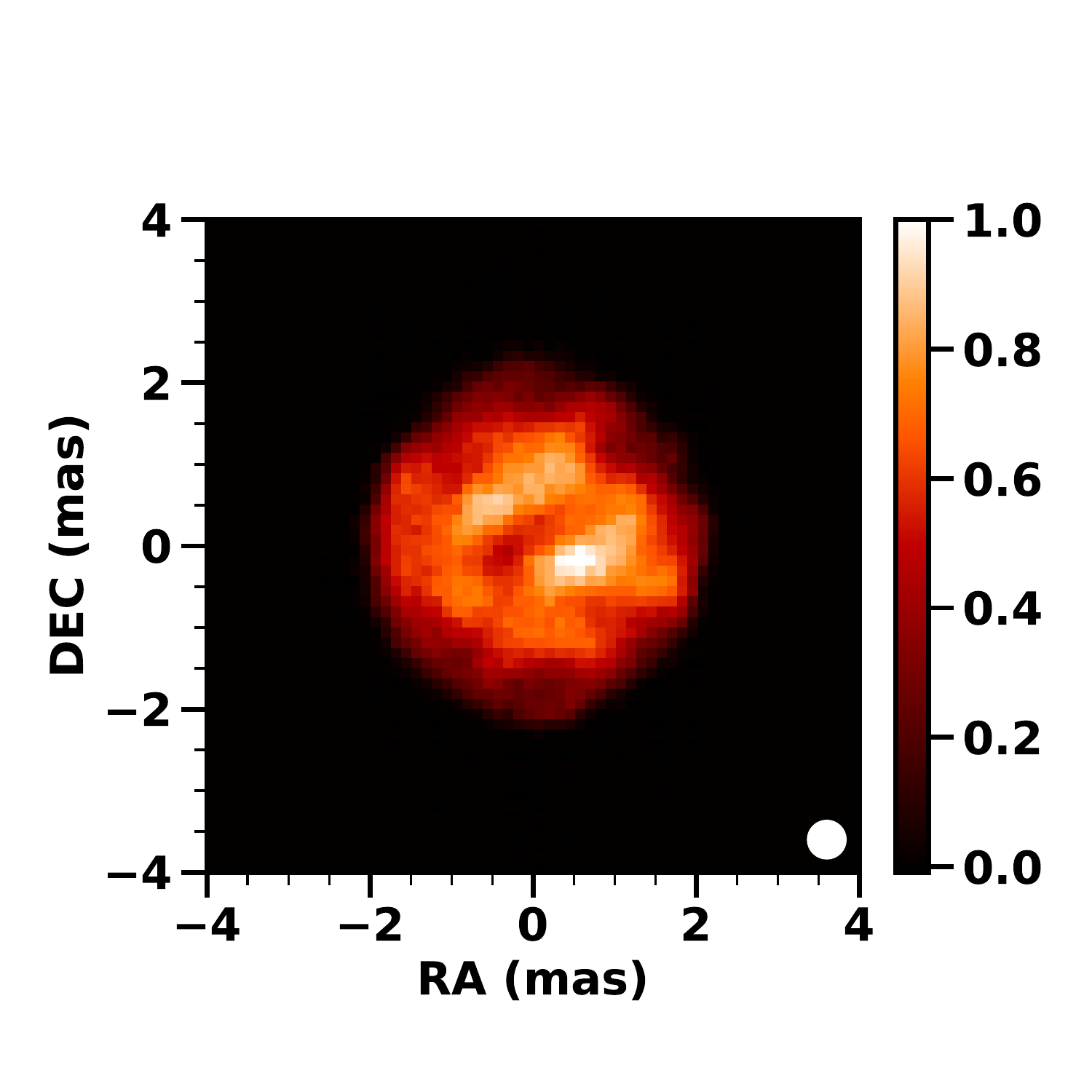}{0.2\textwidth}{AZ~Cyg 2016 1.52$u$m-1.56$u$m}
          }
            \gridline{\fig{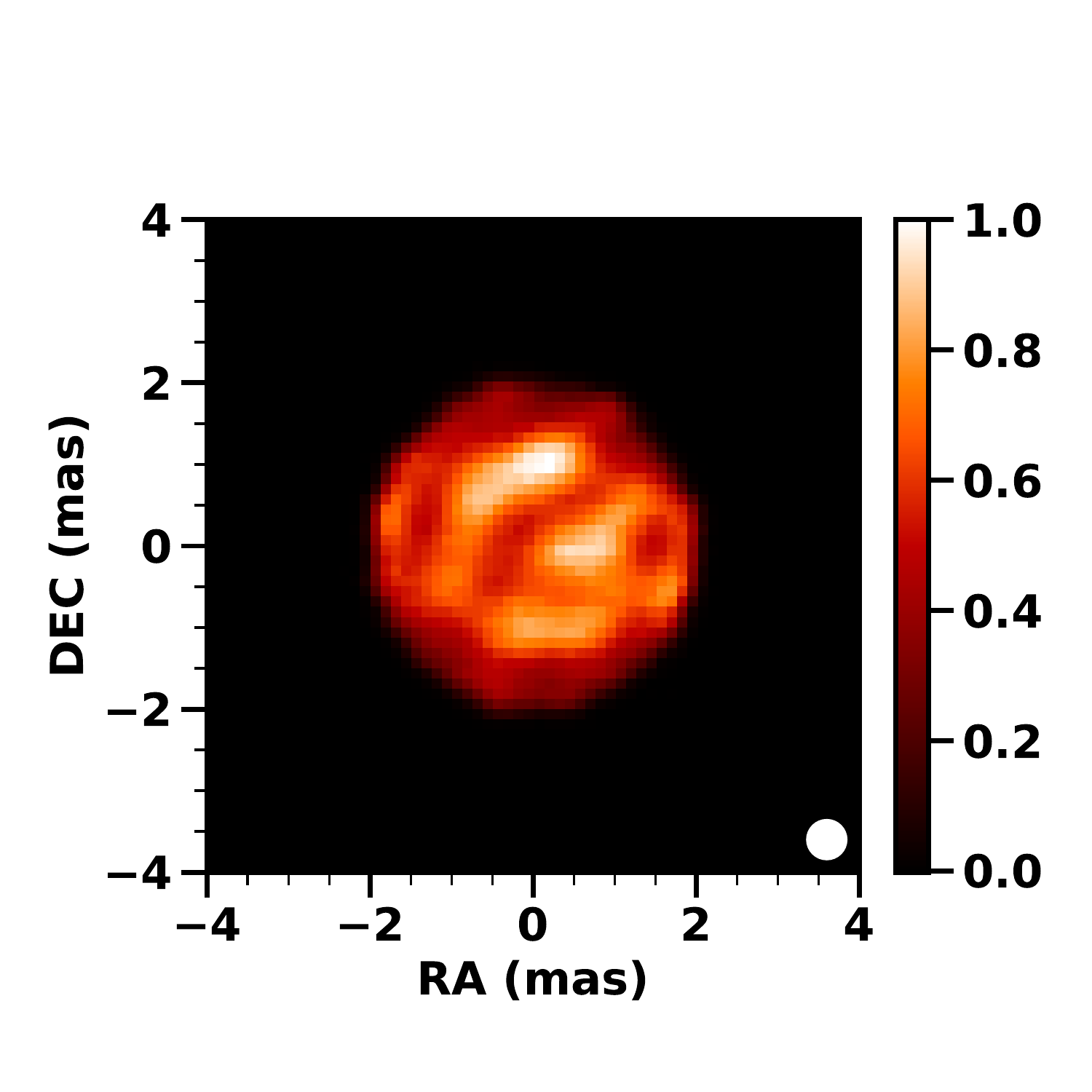}{0.2\textwidth}{AZ~Cyg 2011 1.56$u$m-1.60$u$m}
          \fig{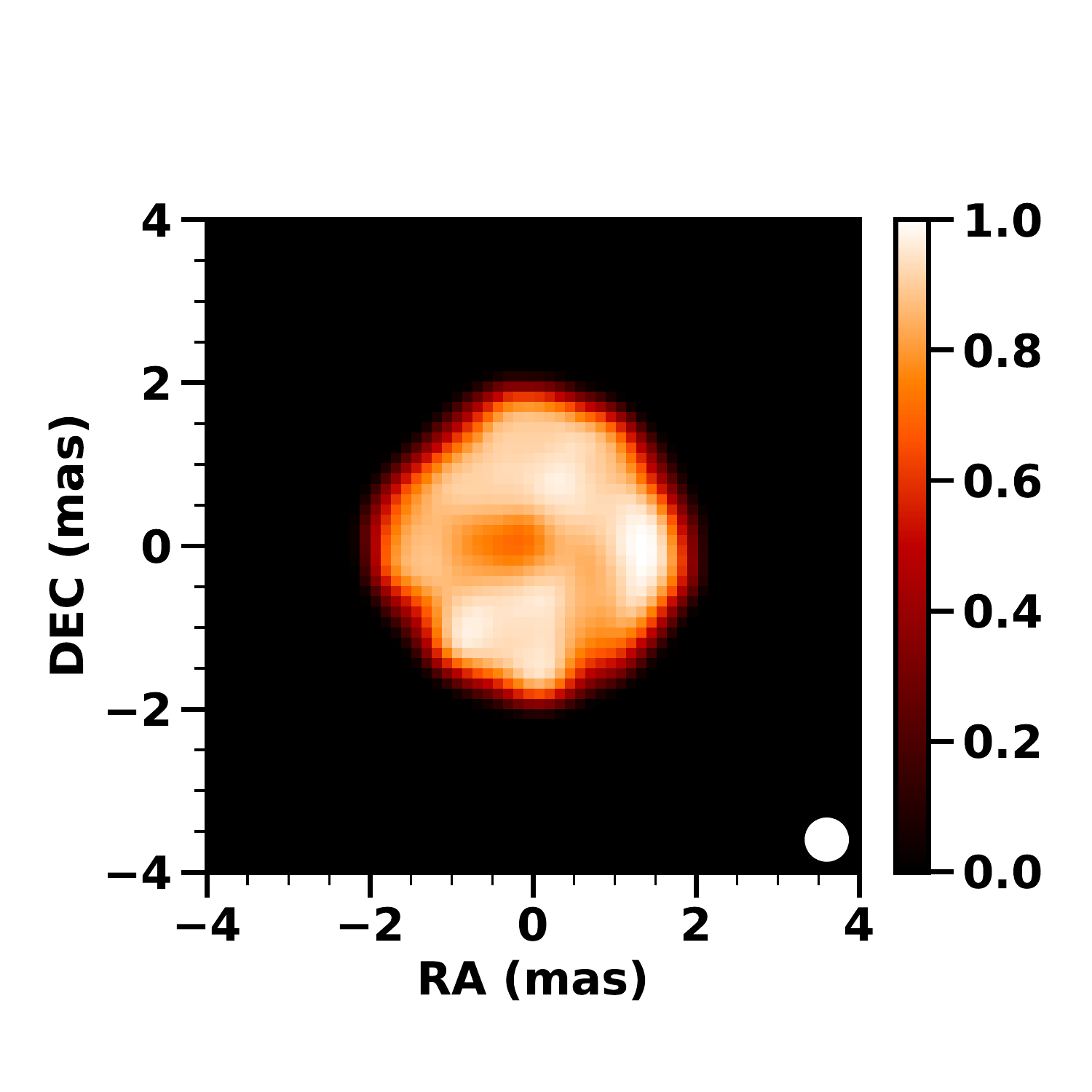}{0.2\textwidth}{AZ~Cyg 2014 1.56$u$m-1.60$u$m}
          \fig{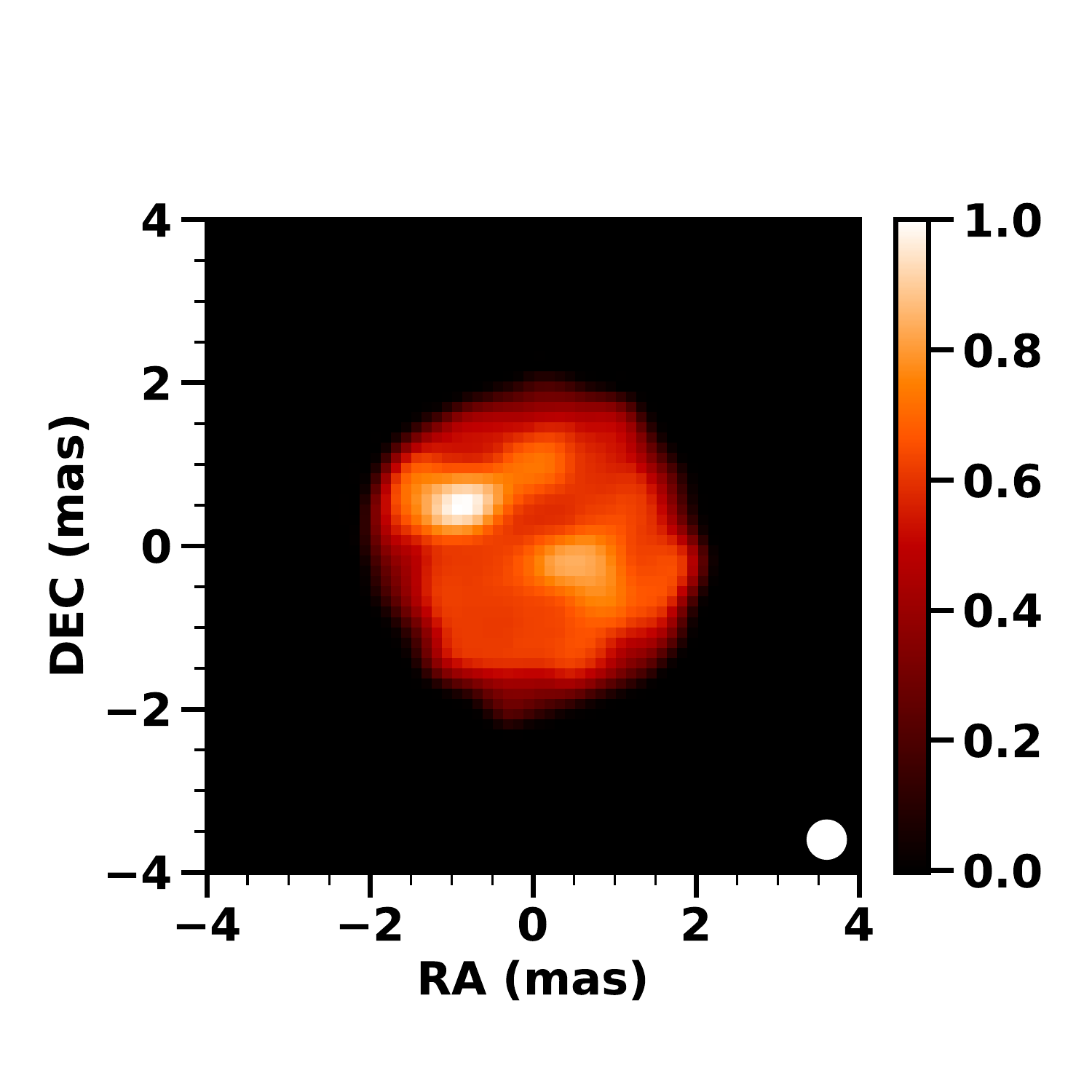}{0.2\textwidth}{AZ~Cyg 2015 1.56$u$m-1.60$u$m}
          \fig{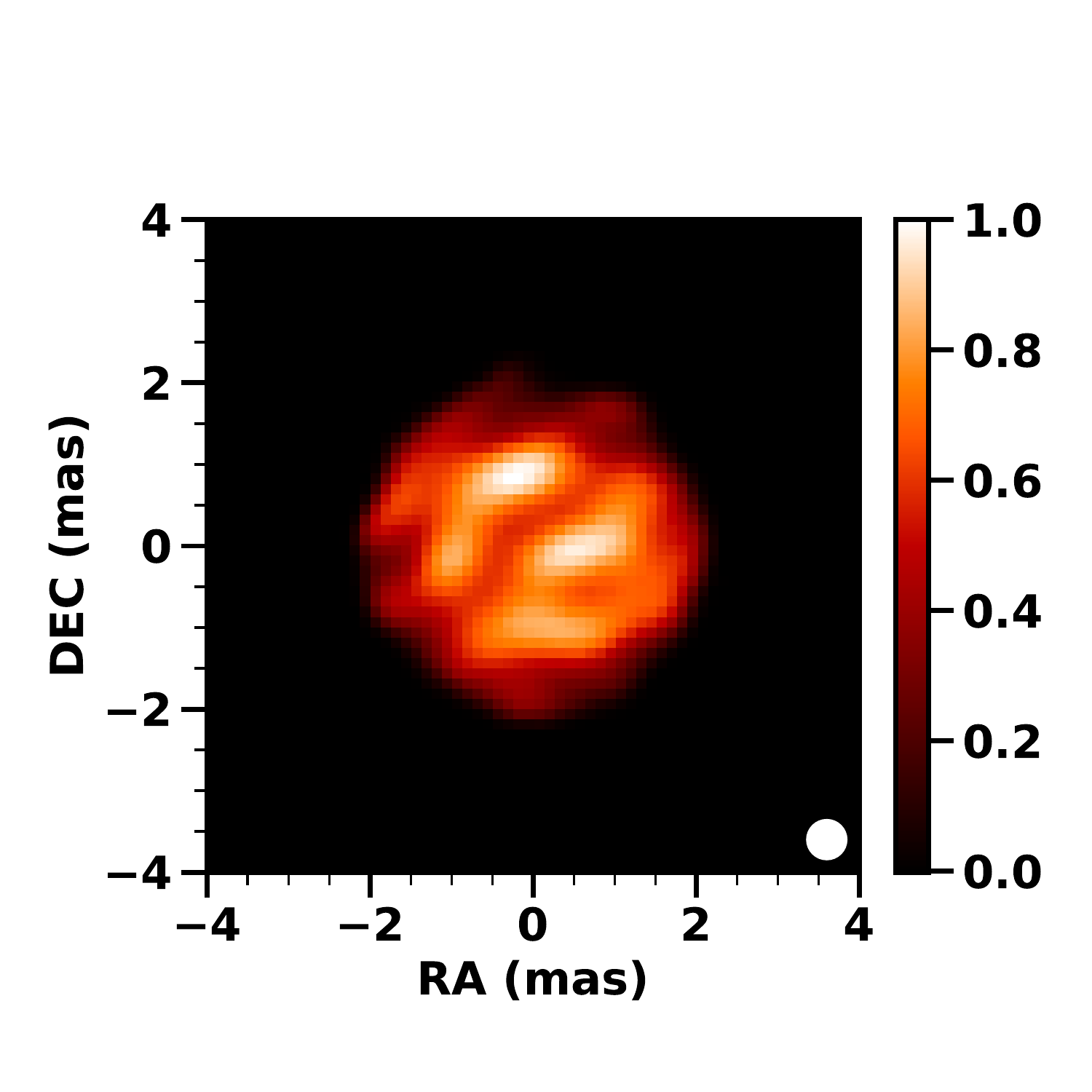}{0.2\textwidth}{AZ~Cyg 2016 1.56$u$m-1.60$u$m}
          }
           \gridline{\fig{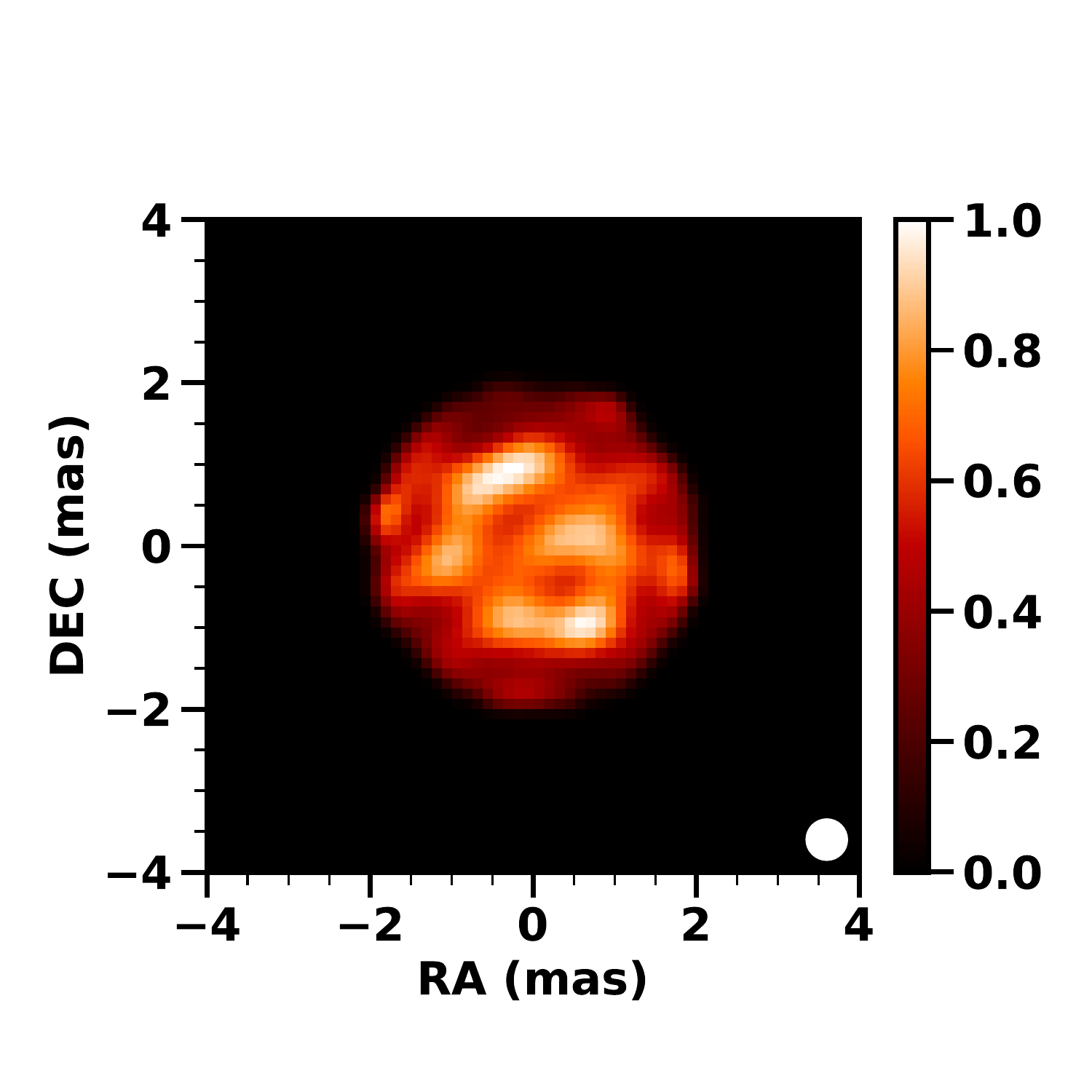}{0.2\textwidth}{AZ~Cyg 2011 1.60$u$m-1.63$u$m}
          \fig{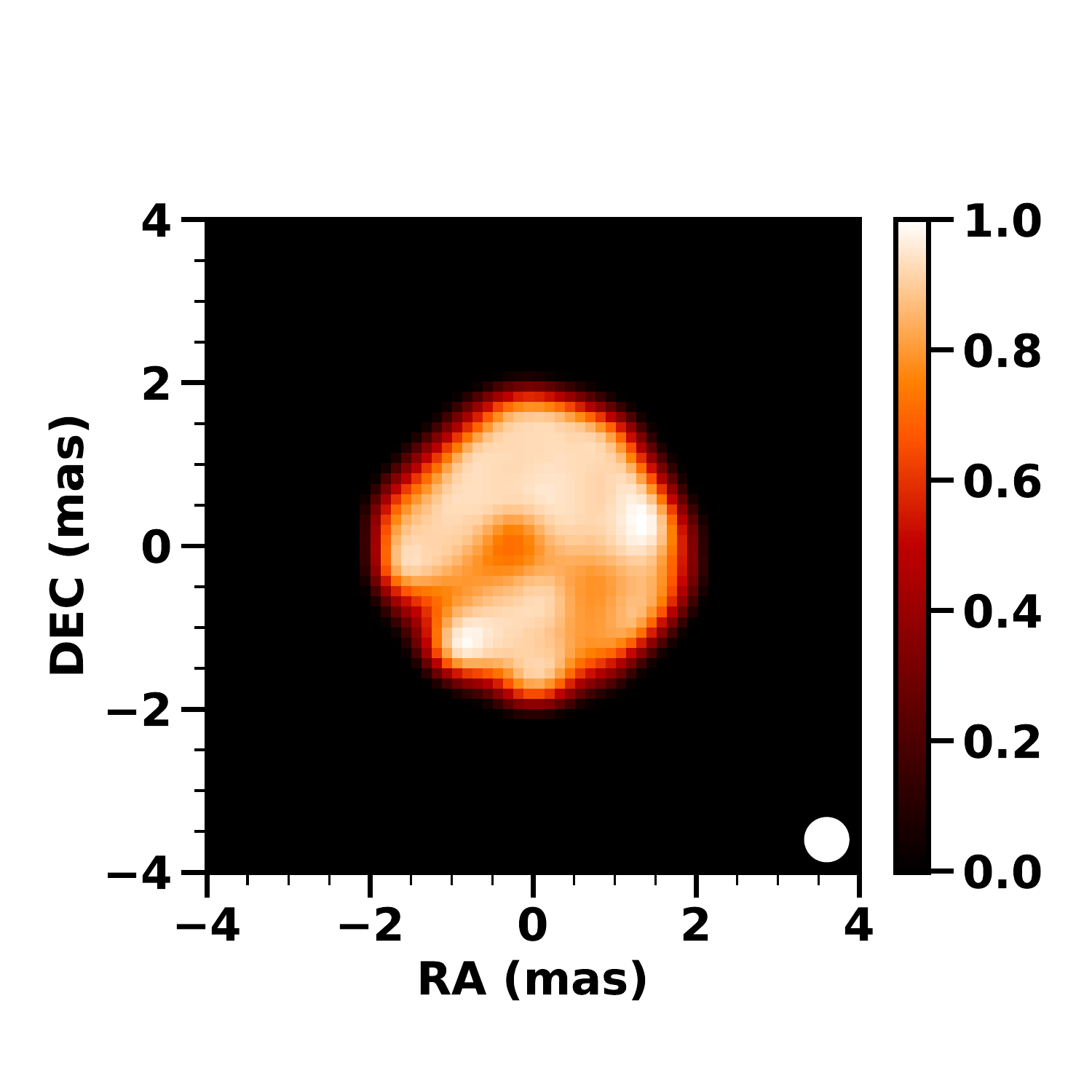}{0.2\textwidth}{AZ~Cyg 2014 1.60$u$m-1.63$u$m}
          \fig{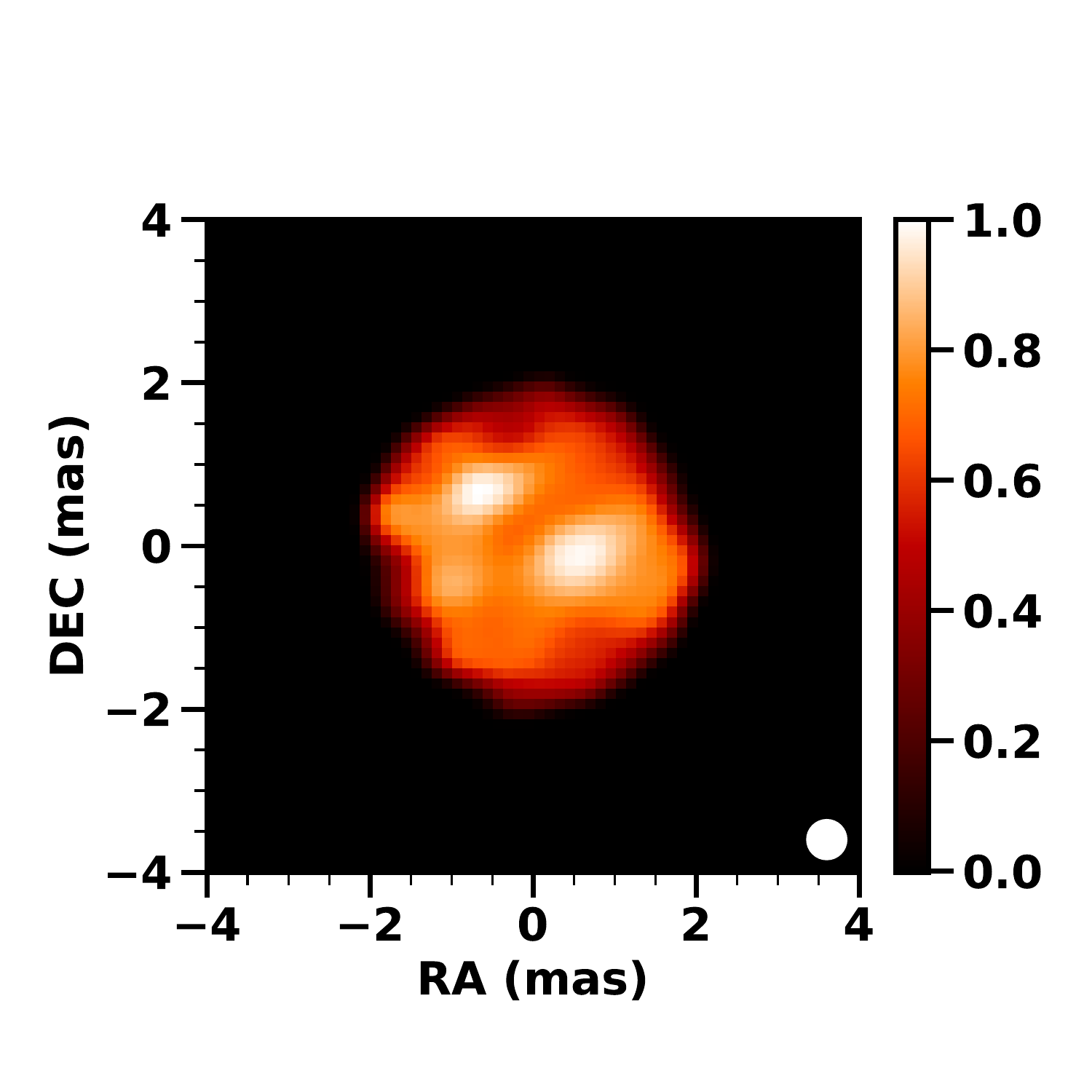}{0.2\textwidth}{AZ~Cyg 2015 1.60$u$m-1.63$u$m}
          \fig{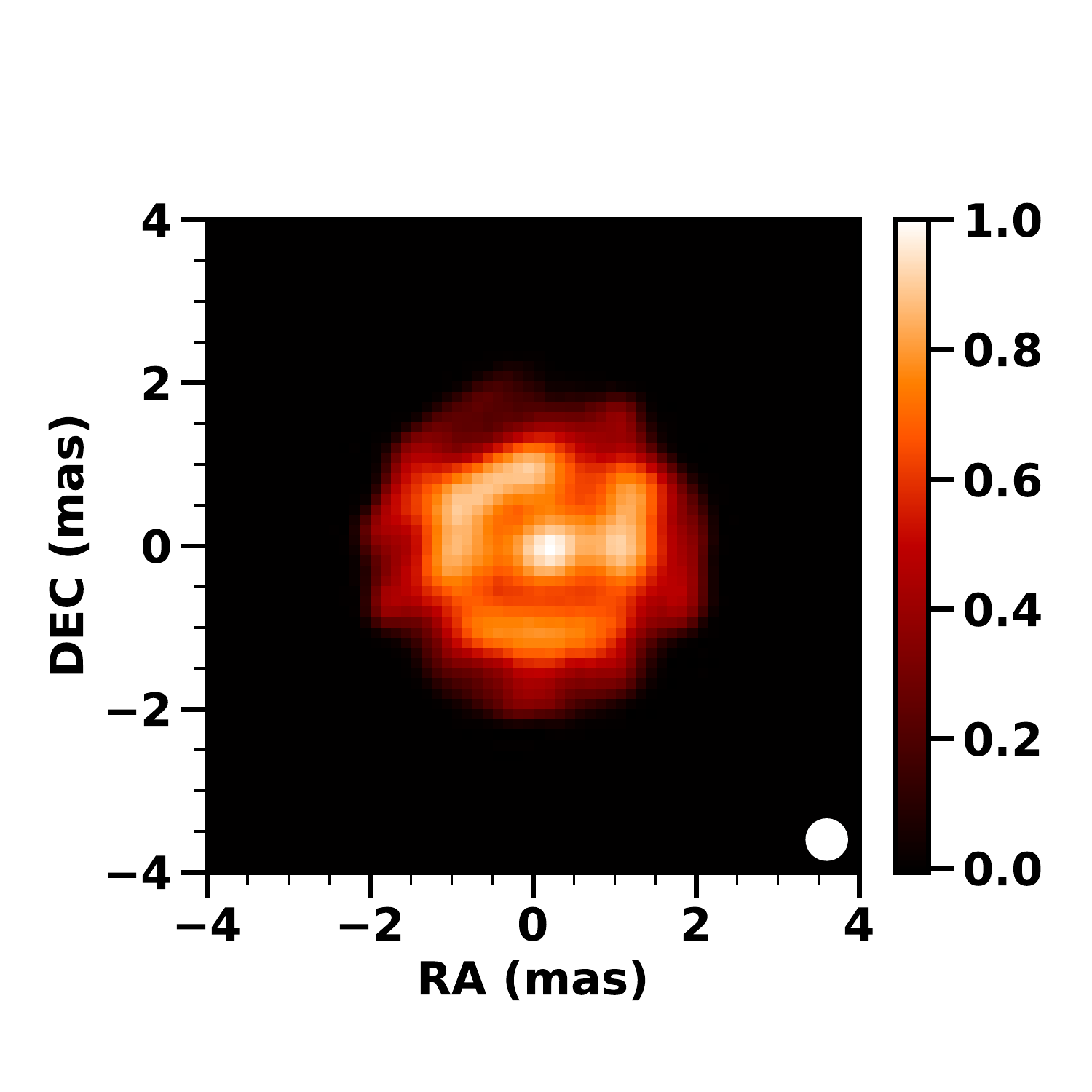}{0.2\textwidth}{AZ~Cyg 2016 1.60$u$m-1.63$u$m}
          }
          \end{figure*}
          
          \begin{figure*}[p]
            \gridline{\fig{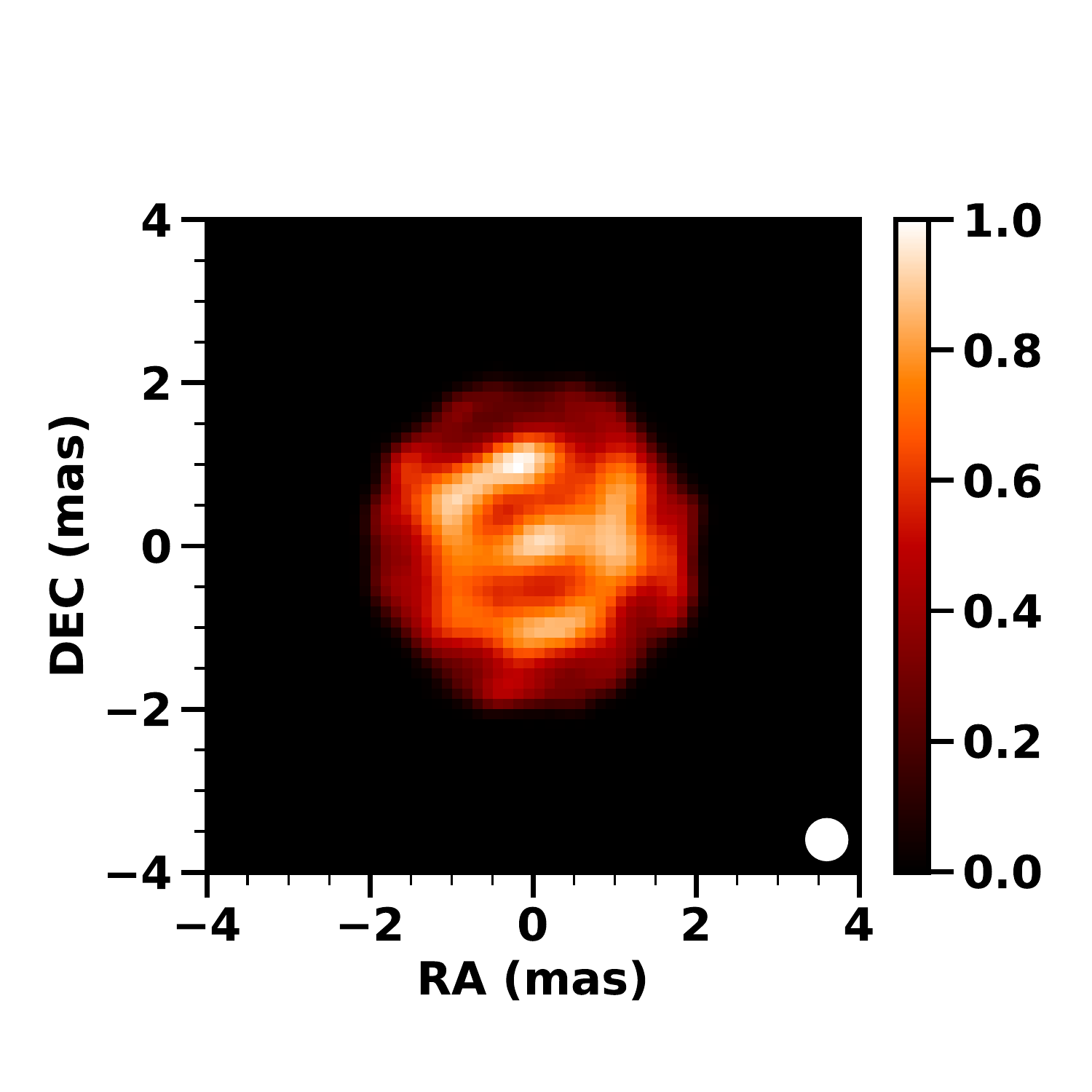}{0.2\textwidth}{AZ~Cyg 2011 1.63$u$m-1.67$u$m}
          \fig{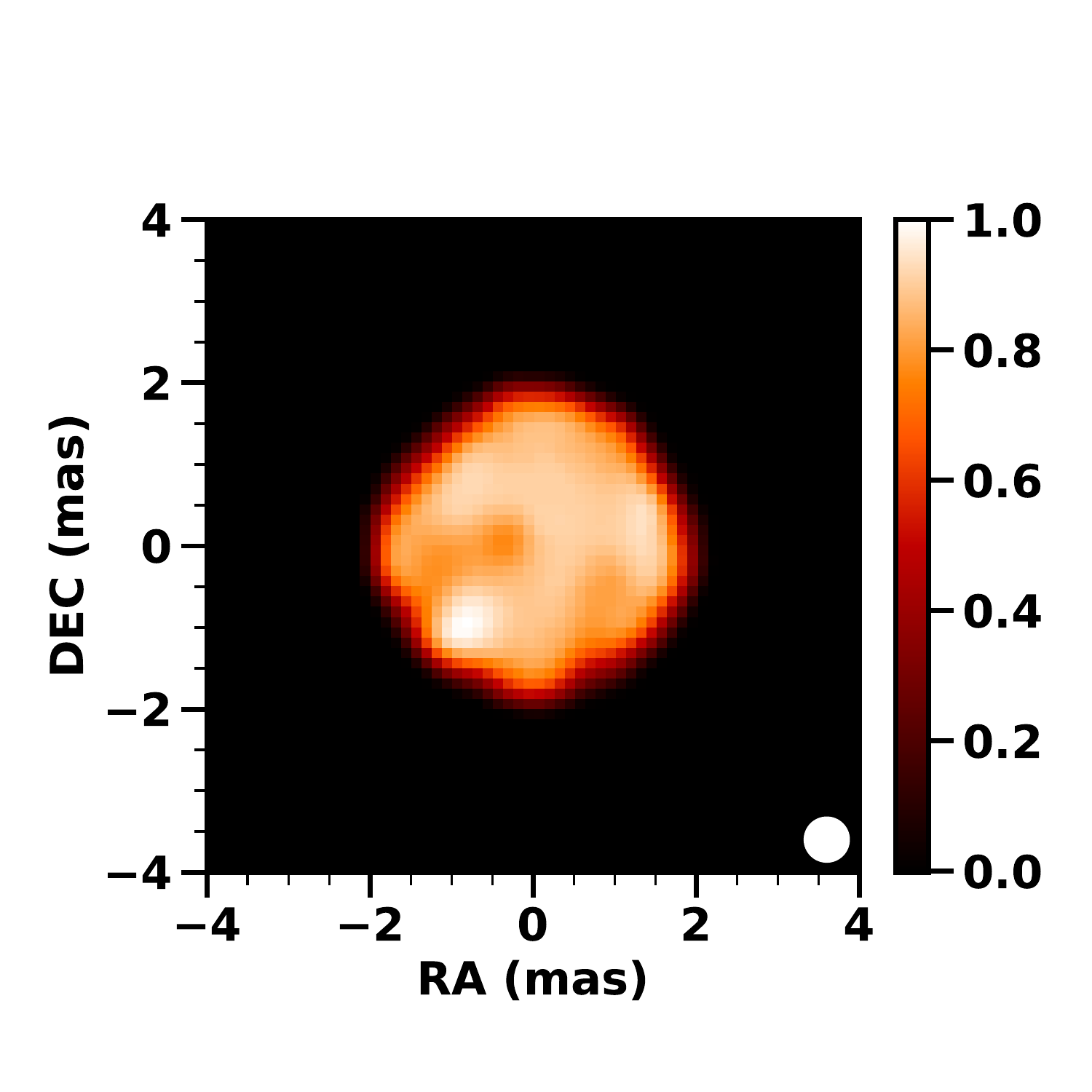}{0.2\textwidth}{AZ~Cyg 2014 1.63$u$m-1.67$u$m}
          \fig{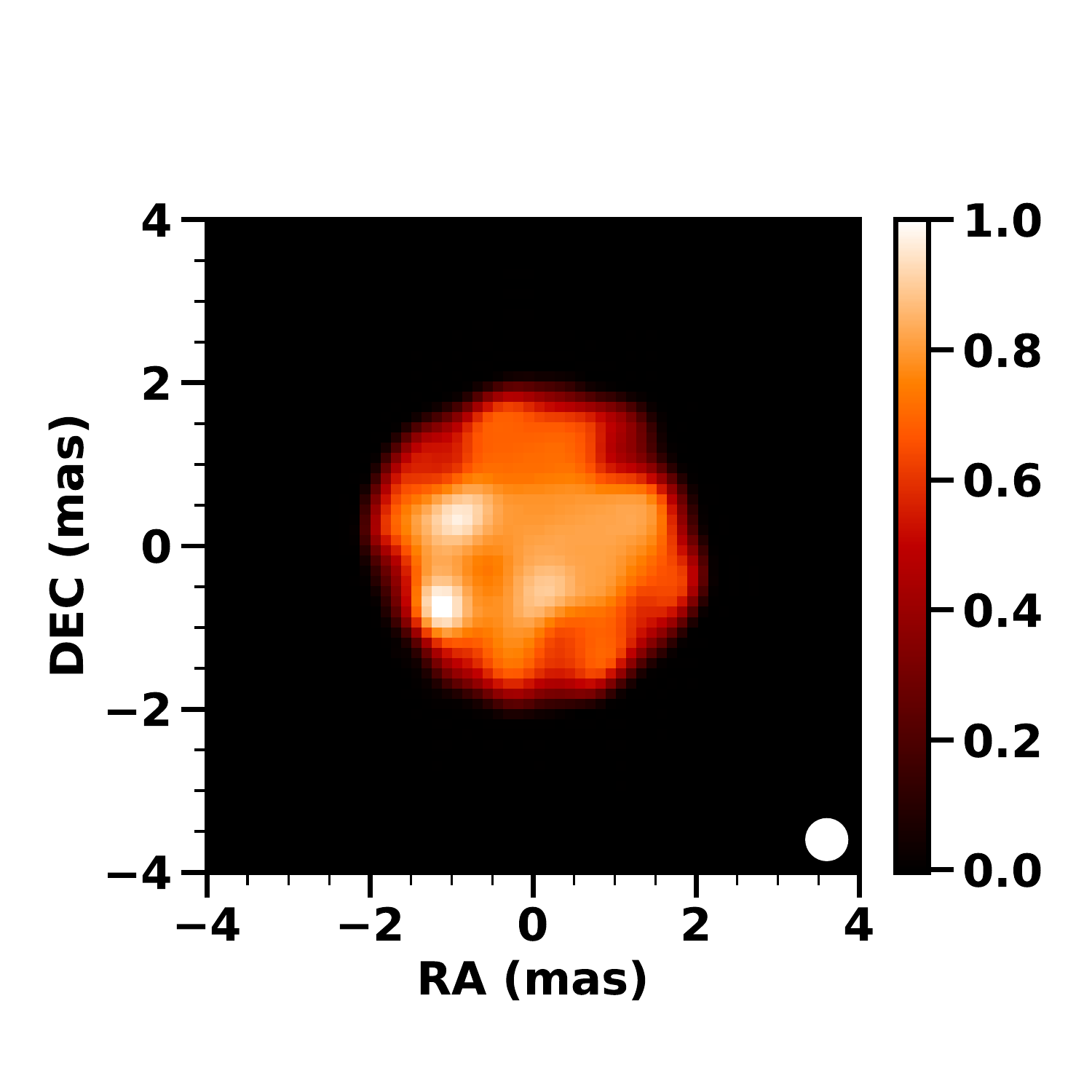}{0.2\textwidth}{AZ~Cyg 2015 1.63$u$m-1.67$u$m}
          \fig{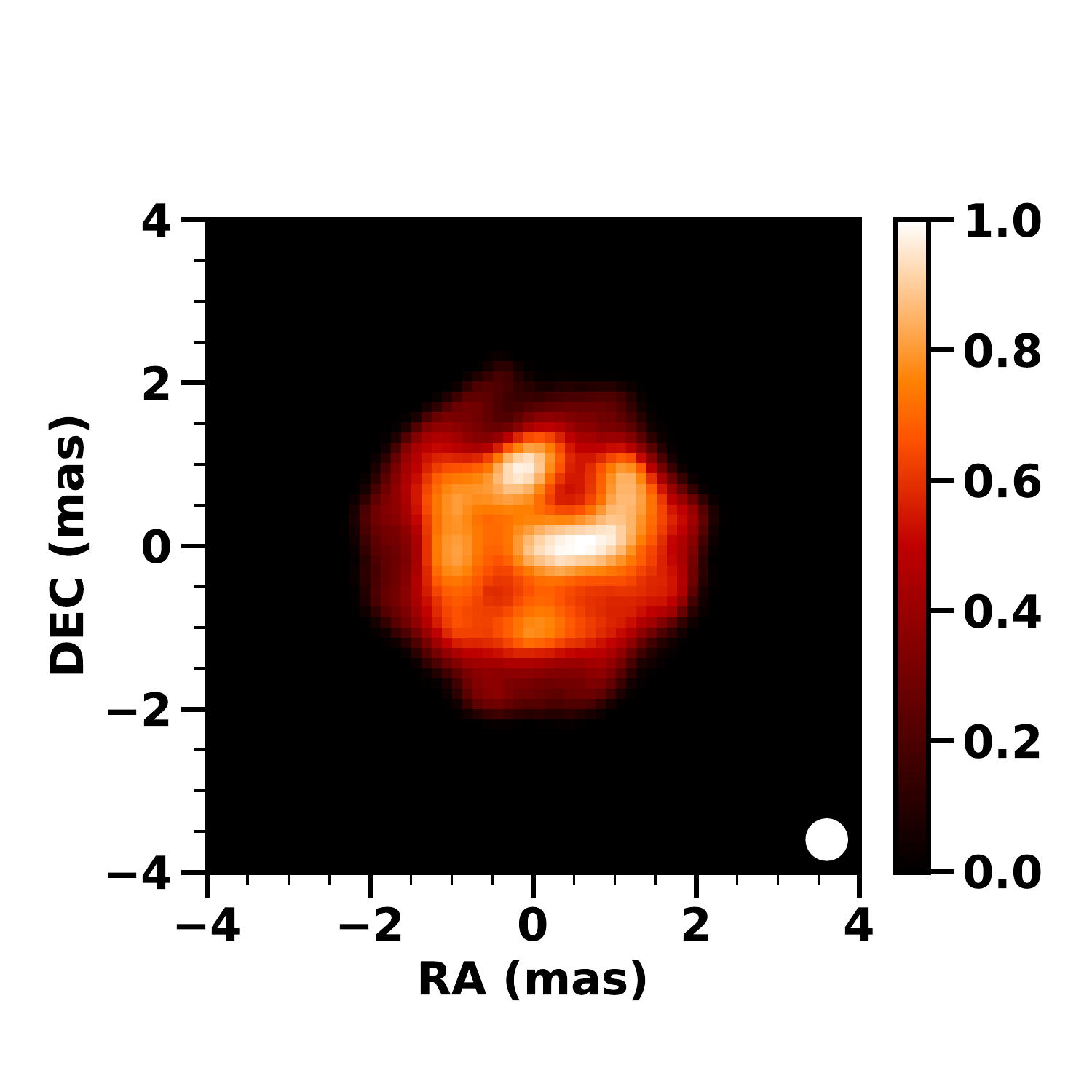}{0.2\textwidth}{AZ~Cyg 2016 1.63$u$m-1.67$u$m}
          }
           \gridline{\fig{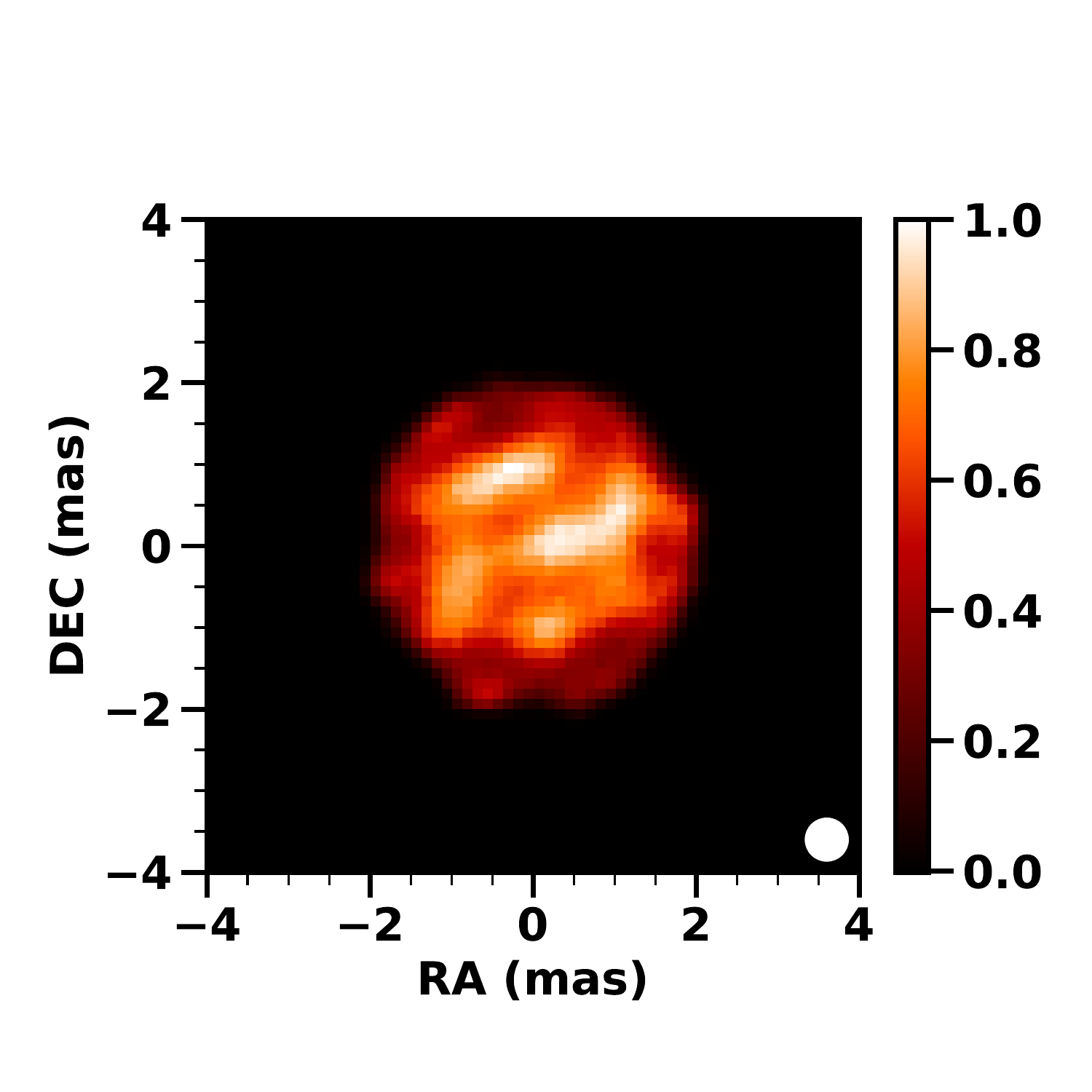}{0.2\textwidth}{AZ~Cyg 2011 1.67$u$m-1.70$u$m}
          \fig{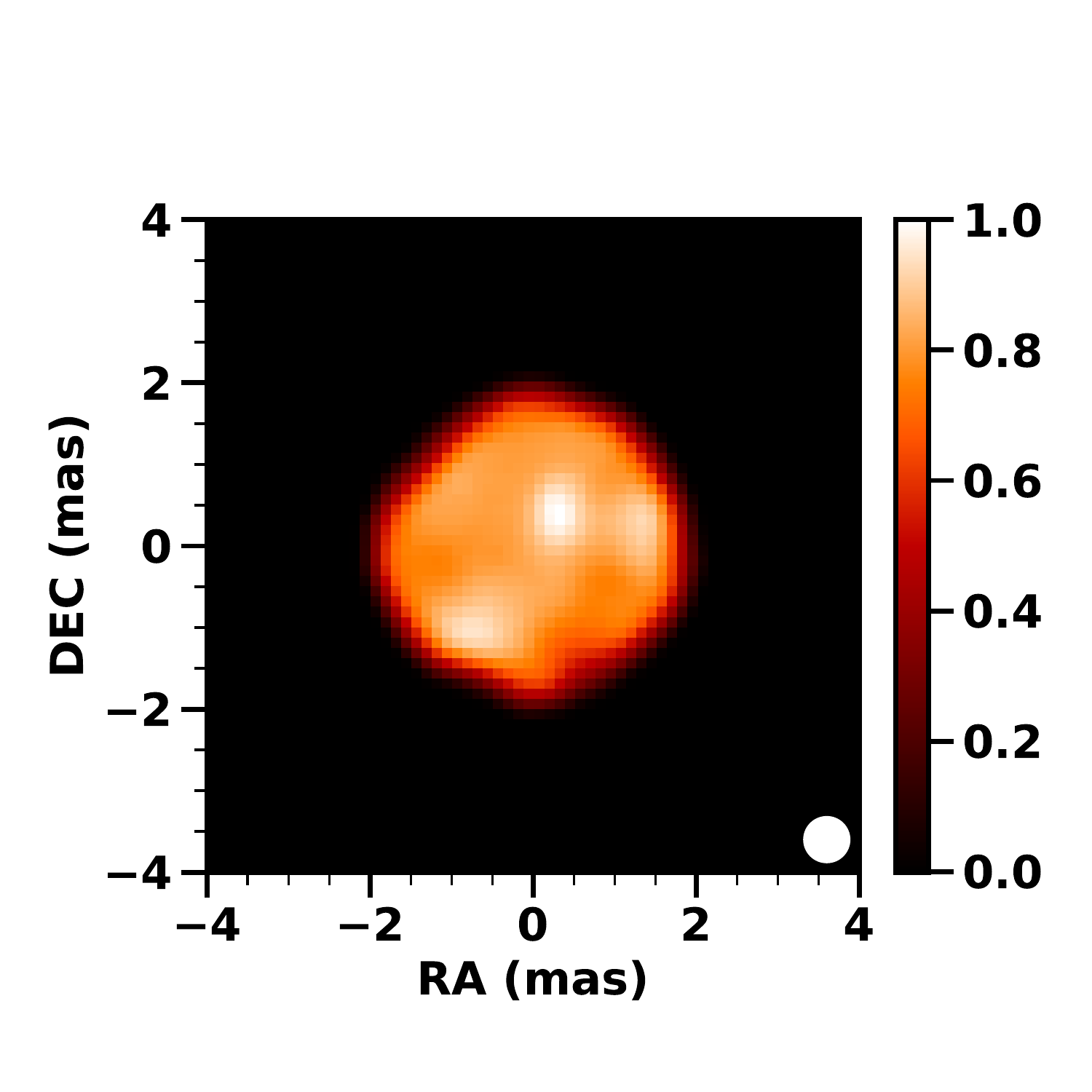}{0.2\textwidth}{AZ~Cyg 2014 1.67$u$m-1.70$u$m}
          \fig{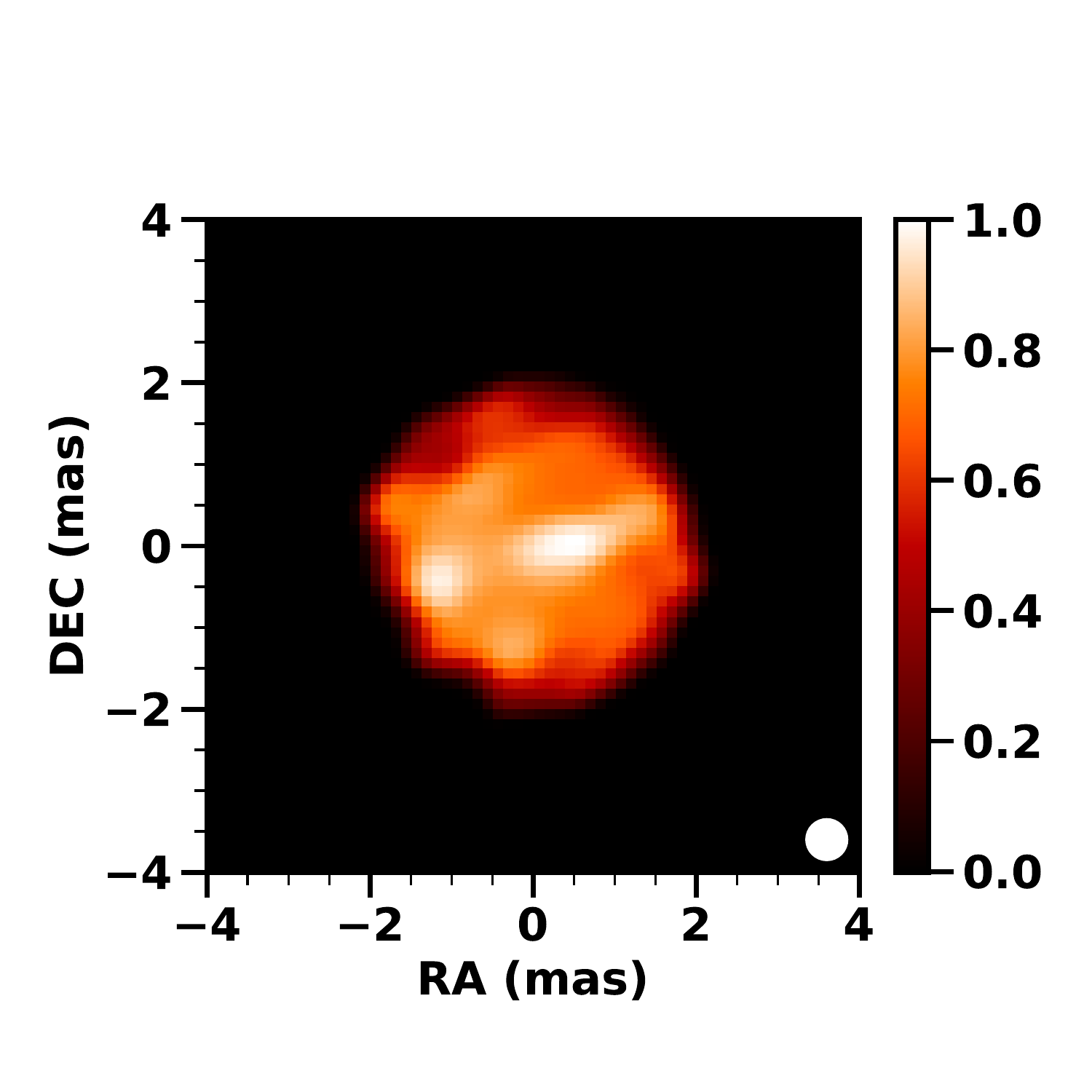}{0.2\textwidth}{AZ~Cyg 2015 1.67$u$m-1.70$u$m}
          \fig{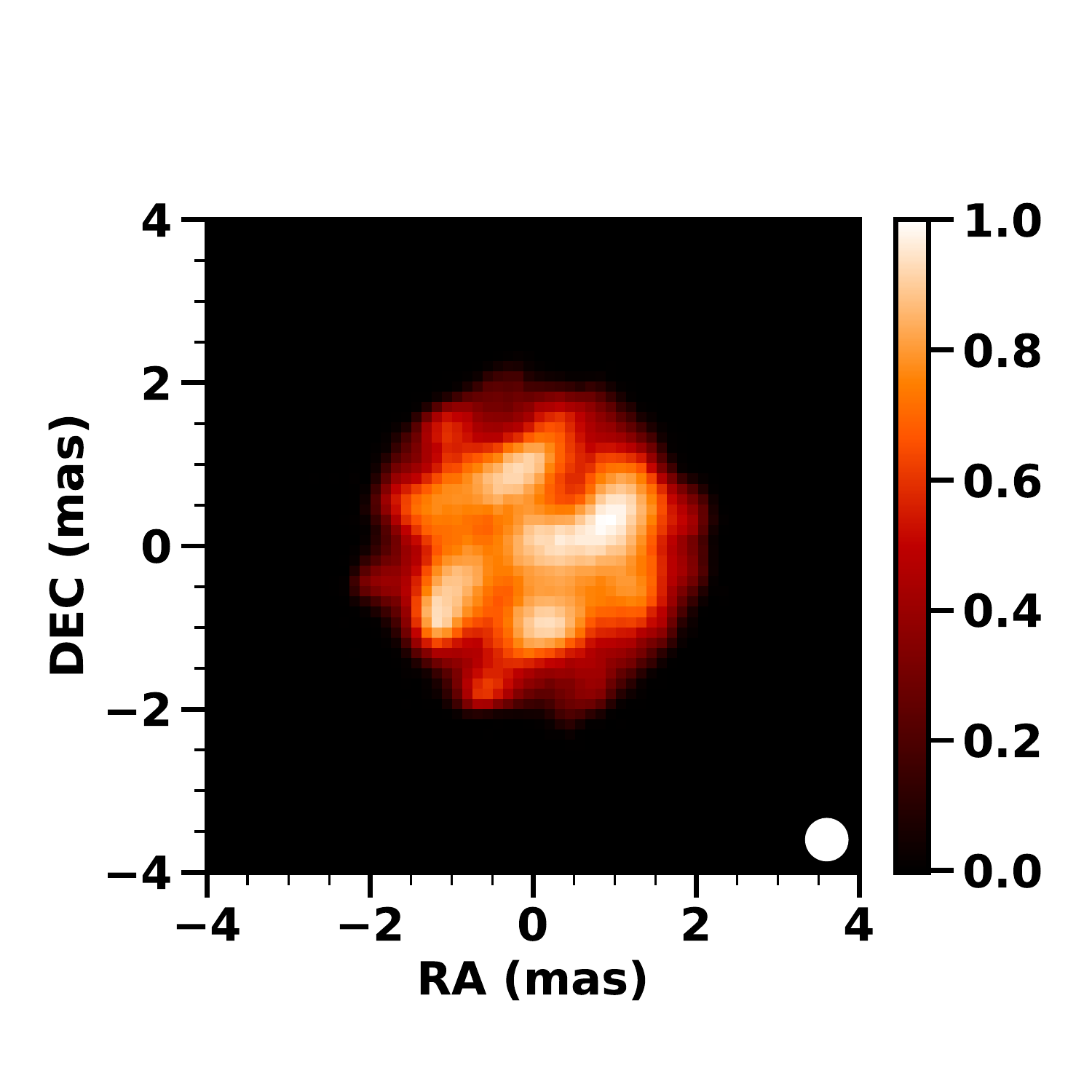}{0.2\textwidth}{AZ~Cyg 2016 1.67$u$m-1.70$u$m}
          }
           \gridline{\fig{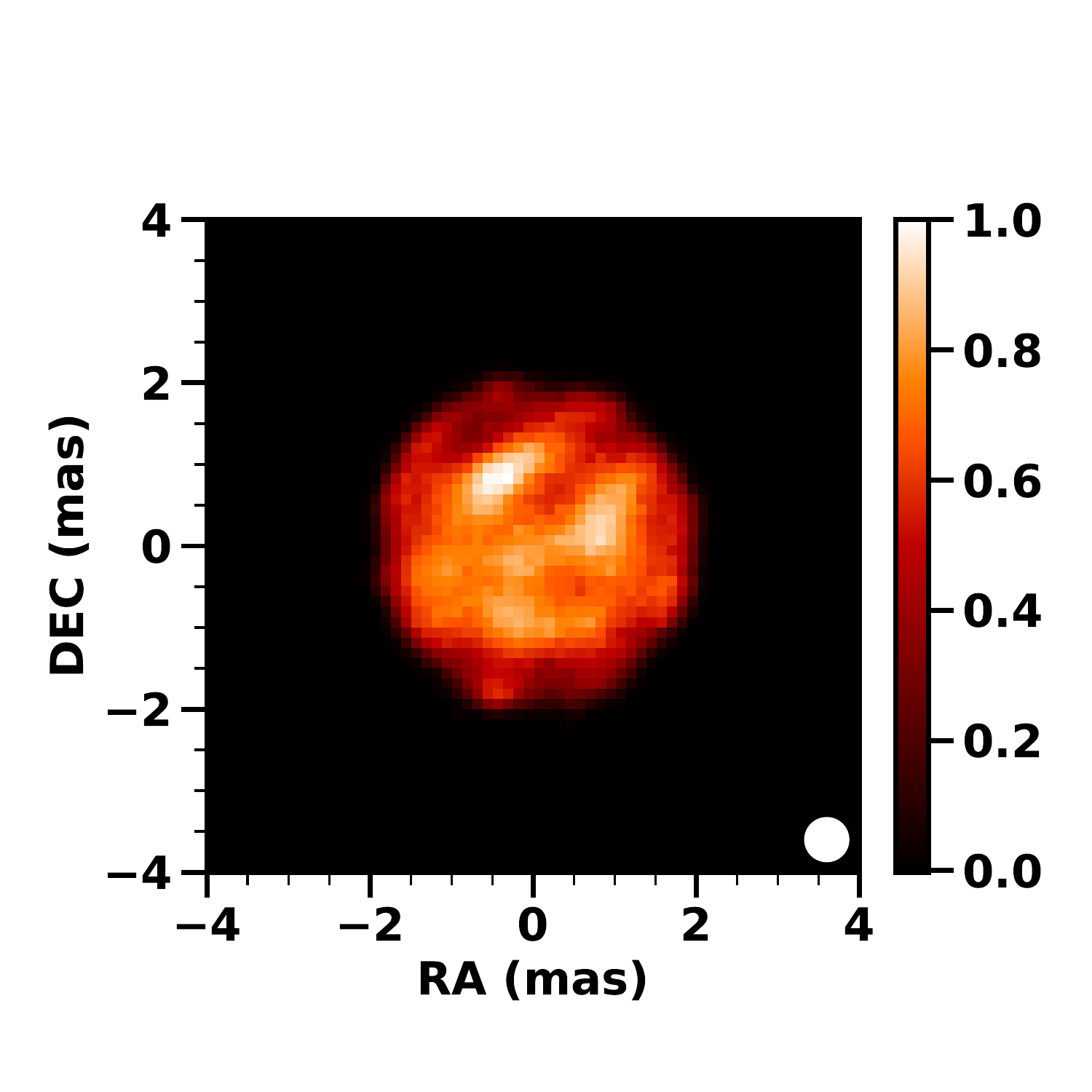}{0.2\textwidth}{AZ~Cyg 2011 1.70$u$m-1.73$u$m}
          \fig{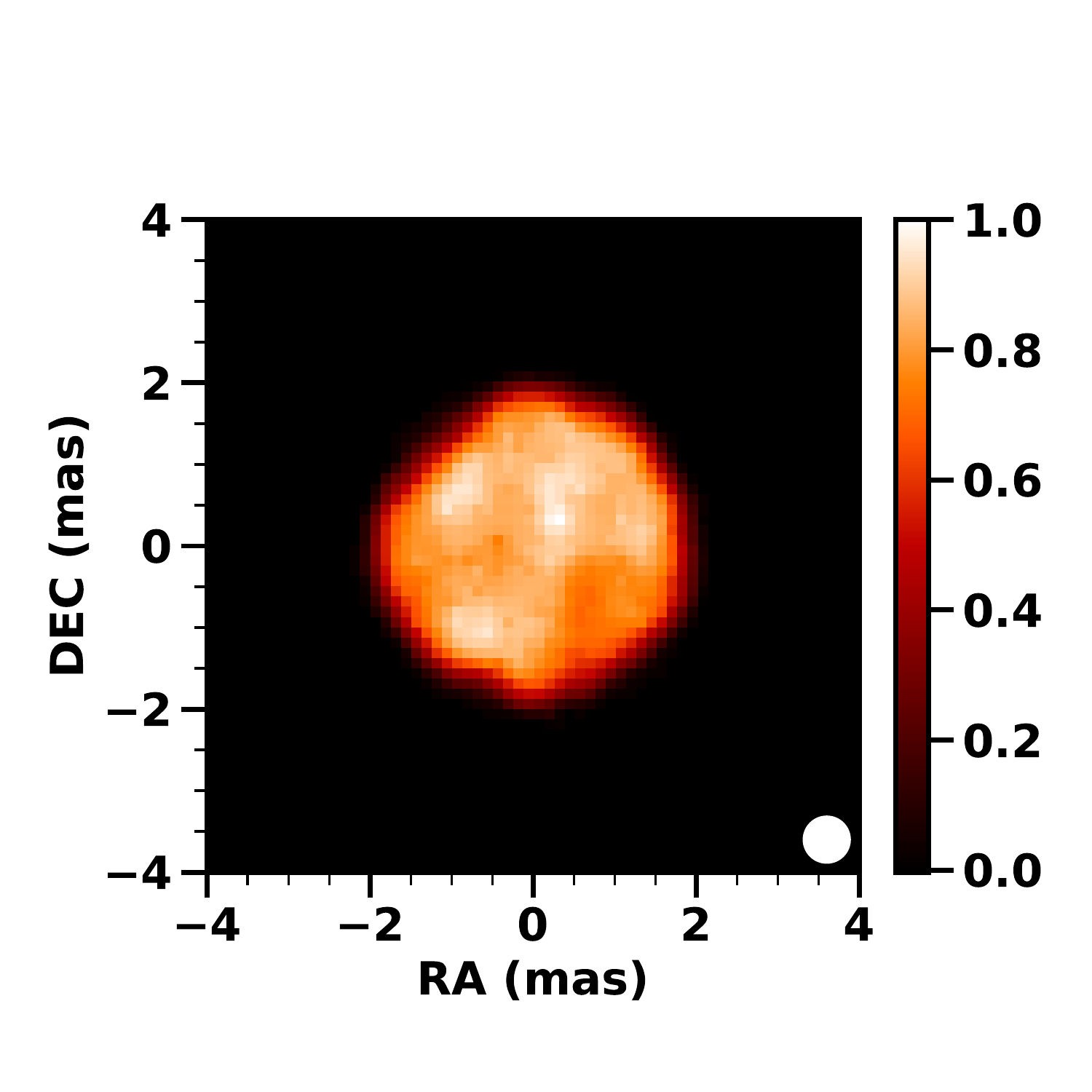}{0.2\textwidth}{AZ~Cyg 2014 1.70$u$m-1.73$u$m}
          \fig{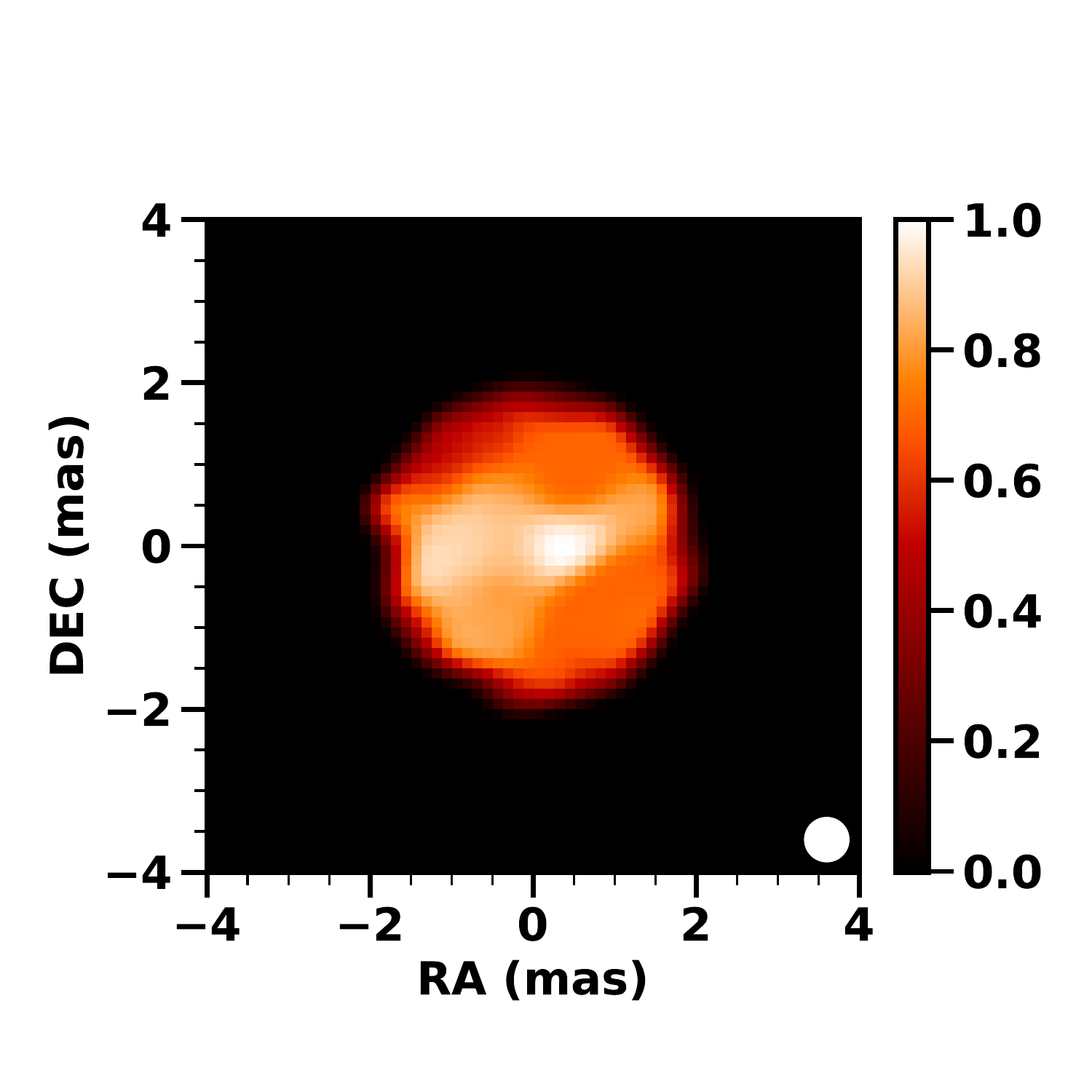}{0.2\textwidth}{AZ~Cyg 2015 1.70$u$m-1.73$u$m}
          \fig{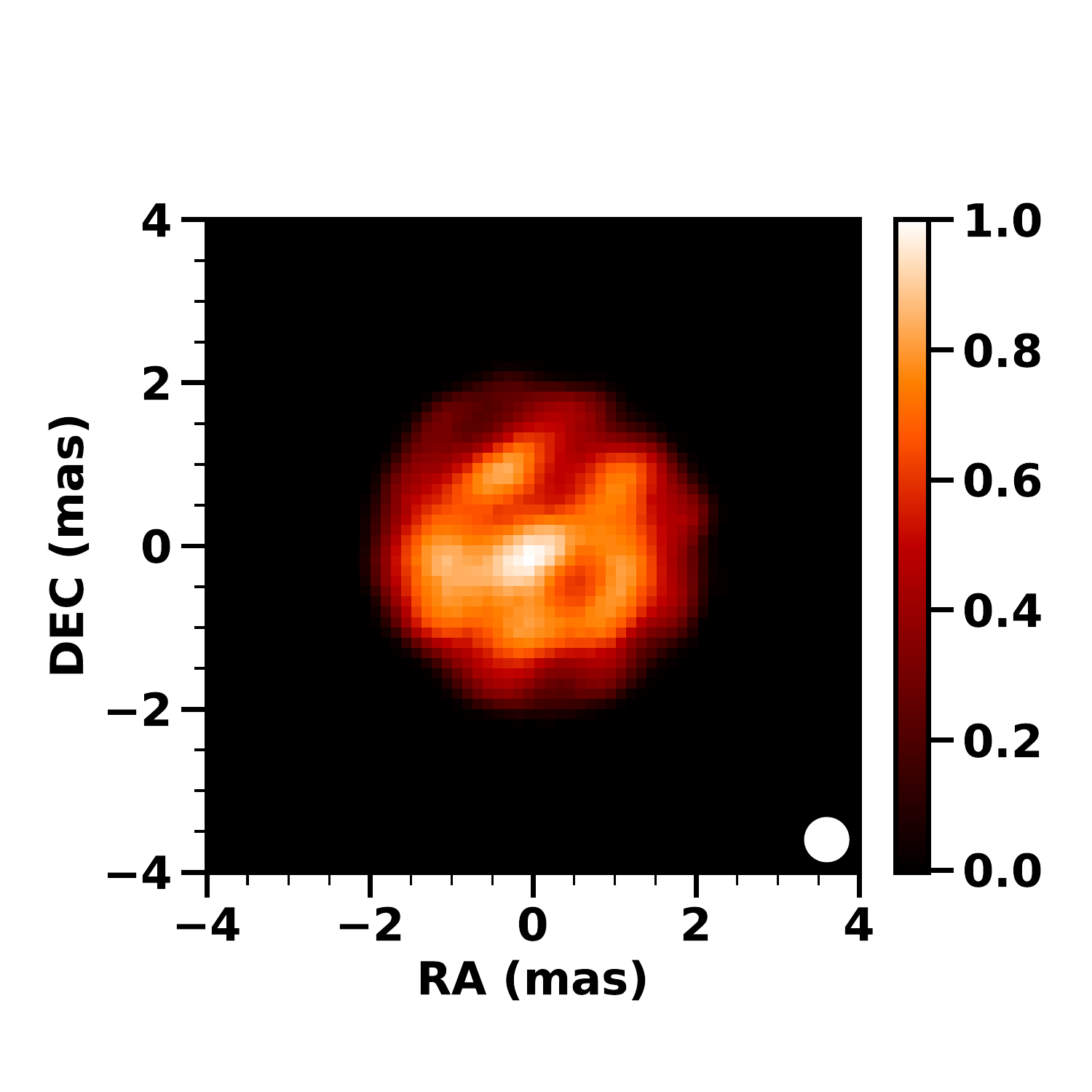}{0.2\textwidth}{AZ~Cyg 2016 1.70$u$m-1.73$u$m}
          }
           \gridline{\fig{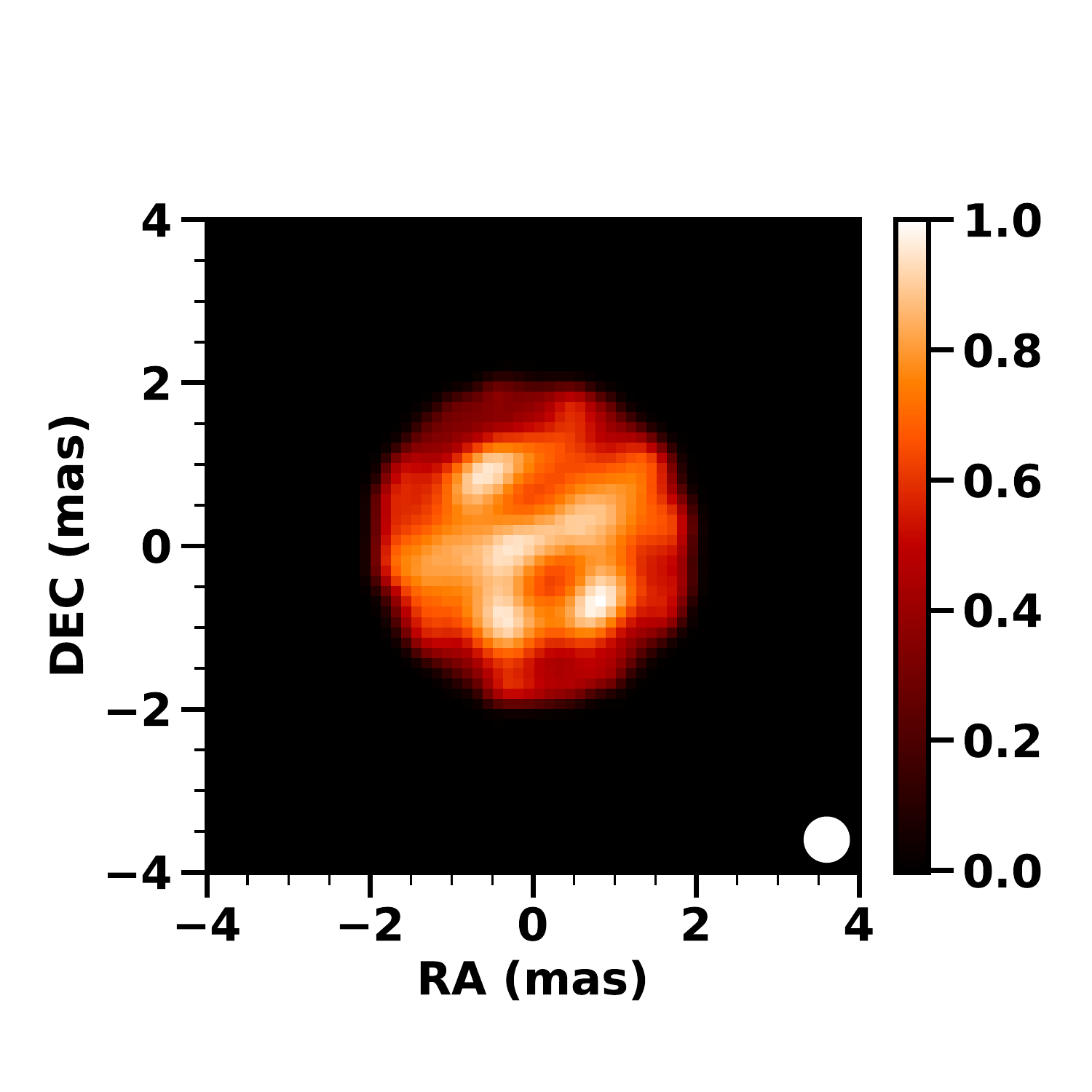}{0.2\textwidth}{AZ~Cyg 2011 1.73$u$m-1.76$u$m}
          \fig{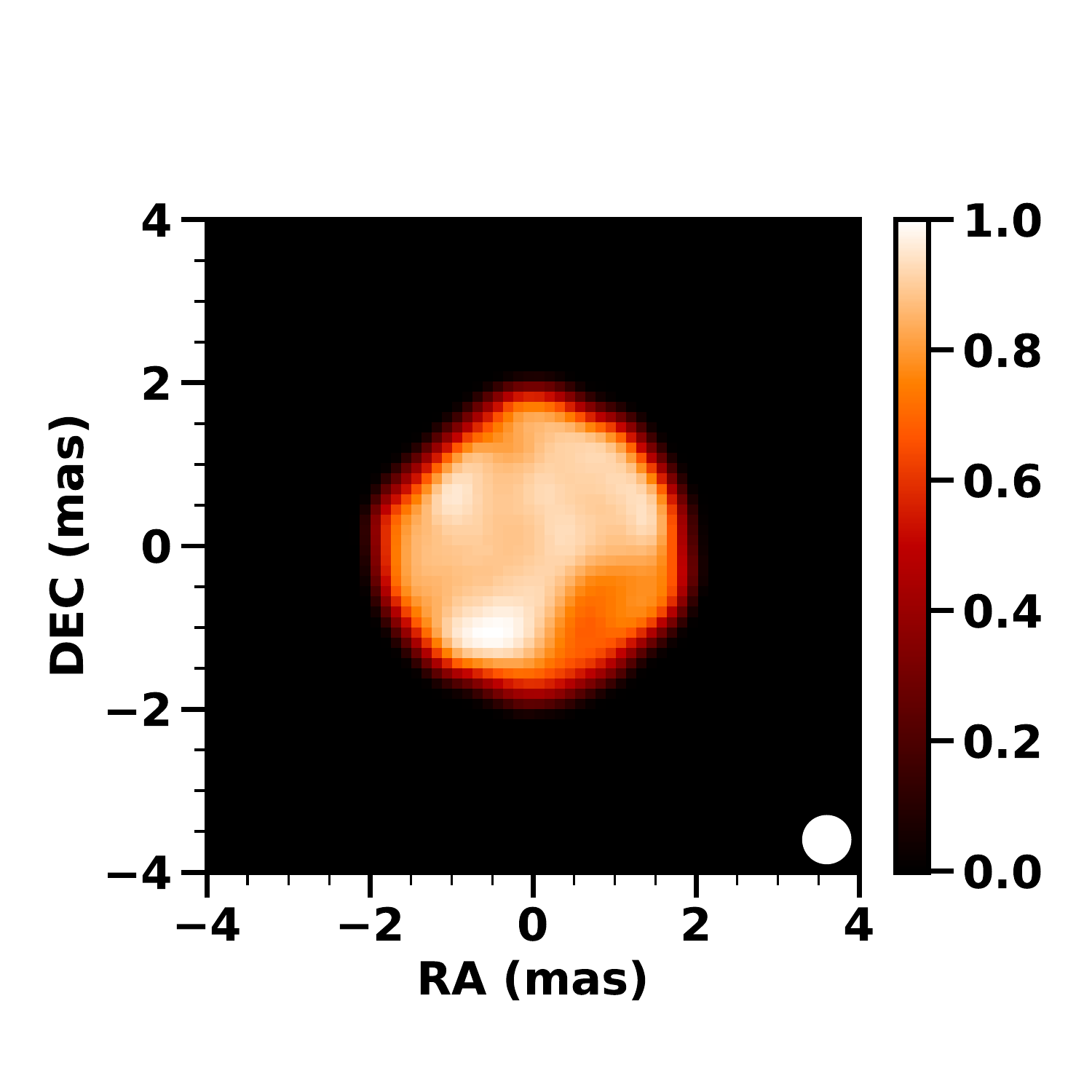}{0.2\textwidth}{AZ~Cyg 2014 1.73$u$m-1.76$u$m}
          \fig{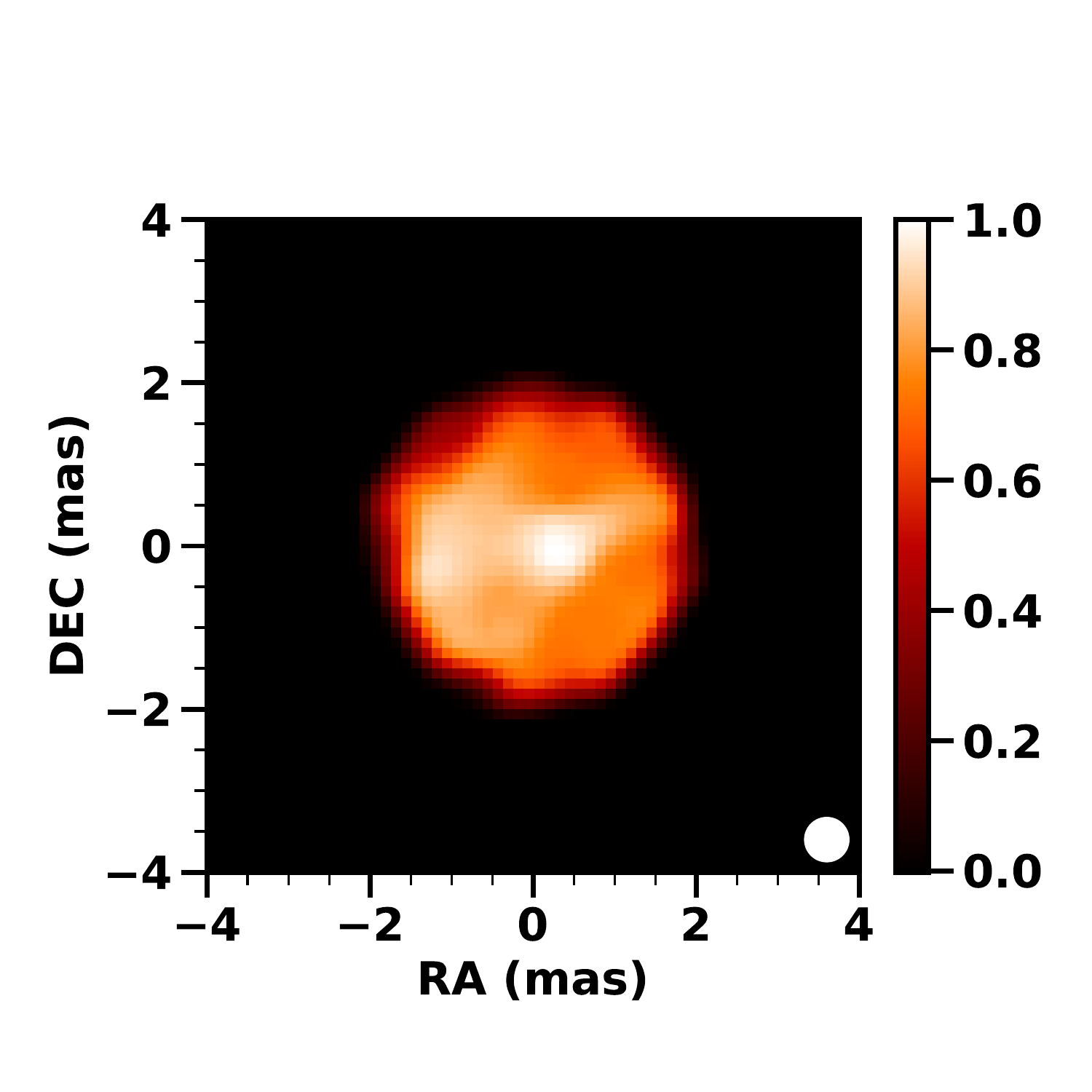}{0.2\textwidth}{AZ~Cyg 2015 1.73$u$m-1.76$u$m}
          \fig{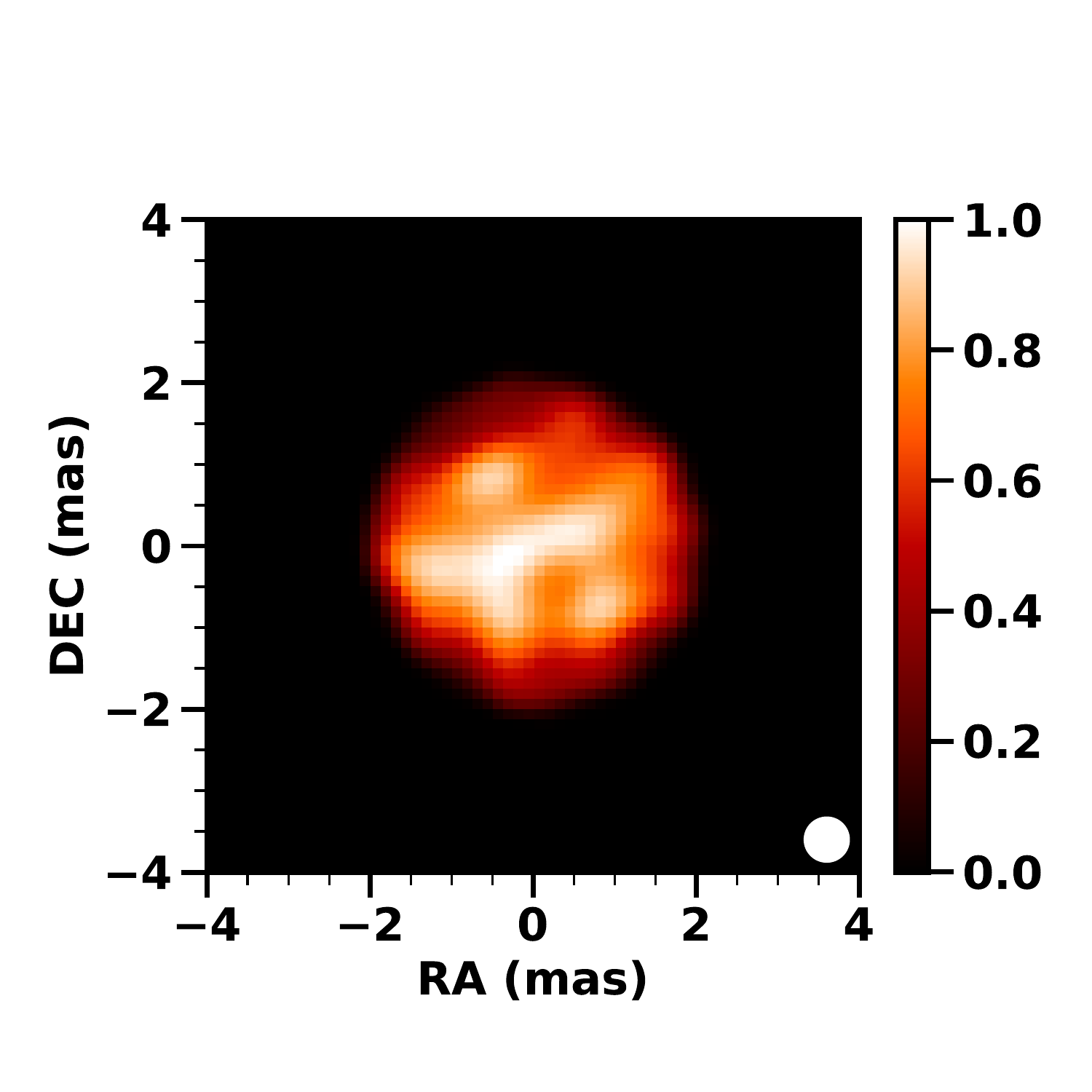}{0.2\textwidth}{AZ~Cyg 2016 1.73$u$m-1.76$u$m}
          }
          
\caption{Channel separated images of AZ~Cyg. The beamsize in the right hand corner represents the resolution given by the maximum projected baseline of each dataset. 
\label{fig:azcygallwave}}
\end{figure*}

%% file: azcyg.bbl
\begin{thebibliography}{}
\expandafter\ifx\csname natexlab\endcsname\relax\def\natexlab#1{#1}\fi
\providecommand{\url}[1]{\href{#1}{#1}}
\providecommand{\dodoi}[1]{doi:~\href{http://doi.org/#1}{\nolinkurl{#1}}}
\providecommand{\doeprint}[1]{\href{http://ascl.net/#1}{\nolinkurl{http://ascl.net/#1}}}
\providecommand{\doarXiv}[1]{\href{https://arxiv.org/abs/#1}{\nolinkurl{https://arxiv.org/abs/#1}}}

\bibitem[{{Alvarez} \& {Plez}(1998)}]{turbospectrum}
{Alvarez}, R., \& {Plez}, B. 1998, \aap, 330, 1109

\bibitem[{{Antia} {et~al.}(1984){Antia}, {Chitre}, \& {Narasimha}}]{antia1984}
{Antia}, H.~M., {Chitre}, S.~M., \& {Narasimha}, D. 1984, \apj, 282, 574,
  \dodoi{10.1086/162236}

\bibitem[{{Arroyo-Torres} {et~al.}(2015){Arroyo-Torres}, {Wittkowski},
  {Chiavassa}, {Scholz}, {Freytag}, {Marcaide}, {Hauschildt}, {Wood}, \&
  {Abellan}}]{torres_paper3}
{Arroyo-Torres}, B., {Wittkowski}, M., {Chiavassa}, A., {et~al.} 2015, \aap,
  575, A50, \dodoi{10.1051/0004-6361/201425212}

\bibitem[{{Bailer-Jones} {et~al.}(2021){Bailer-Jones}, {Rybizki}, {Fouesneau},
  {Demleitner}, \& {Andrae}}]{distgaia}
{Bailer-Jones}, C.~A.~L., {Rybizki}, J., {Fouesneau}, M., {Demleitner}, M., \&
  {Andrae}, R. 2021, \aj, 161, 147, \dodoi{10.3847/1538-3881/abd806}

\bibitem[{{Baron} {et~al.}(2010){Baron}, {Monnier}, \&
  {Kloppenborg}}]{baron2010}
{Baron}, F., {Monnier}, J.~D., \& {Kloppenborg}, B. 2010, in \procspie, Vol.
  7734, Optical and Infrared Interferometry II, 77342I,
  \dodoi{10.1117/12.857364}

\bibitem[{{Baron} {et~al.}(2012){Baron}, {Monnier}, {Pedretti}, {Zhao},
  {Schaefer}, {Parks}, {Che}, {Thureau}, {ten Brummelaar}, {McAlister},
  {Ridgway}, {Farrington}, {Sturmann}, {Sturmann}, \& {Turner}}]{baron2012}
{Baron}, F., {Monnier}, J.~D., {Pedretti}, E., {et~al.} 2012, \apj, 752, 20,
  \dodoi{10.1088/0004-637X/752/1/20}

\bibitem[{{Baron} {et~al.}(2014){Baron}, {Monnier}, {Kiss}, {Neilson}, {Zhao},
  {Anderson}, {Aarnio}, {Pedretti}, {Thureau}, {ten Brummelaar}, {Ridgway},
  {McAlister}, {Sturmann}, {Sturmann}, \& {Turner}}]{baronrsg}
{Baron}, F., {Monnier}, J.~D., {Kiss}, L.~L., {et~al.} 2014, \apj, 785, 46,
  \dodoi{10.1088/0004-637X/785/1/46}

\bibitem[{{Buscher} {et~al.}(1990){Buscher}, {Haniff}, {Baldwin}, \&
  {Warner}}]{BuscherBetelgeuse}
{Buscher}, D.~F., {Haniff}, C.~A., {Baldwin}, J.~E., \& {Warner}, P.~J. 1990,
  \mnras, 245, 7P

\bibitem[{{Cardelli} {et~al.}(1989){Cardelli}, {Clayton}, \&
  {Mathis}}]{cardelli}
{Cardelli}, J.~A., {Clayton}, G.~C., \& {Mathis}, J.~S. 1989, \apj, 345, 245,
  \dodoi{10.1086/167900}

\bibitem[{{Castelli} \& {Kurucz}(2003)}]{castelli_atlas09}
{Castelli}, F., \& {Kurucz}, R.~L. 2003, in Modelling of Stellar Atmospheres,
  ed. N.~{Piskunov}, W.~W. {Weiss}, \& D.~F. {Gray}, Vol. 210, A20.
\newblock \doarXiv{astro-ph/0405087}

\bibitem[{{Chatys} {et~al.}(2019){Chatys}, {Bedding}, {Murphy}, {Kiss},
  {Dobie}, \& {Grindlay}}]{rsggaia}
{Chatys}, F.~W., {Bedding}, T.~R., {Murphy}, S.~J., {et~al.} 2019, \mnras,
  1518, \dodoi{10.1093/mnras/stz1584}

\bibitem[{{Chiavassa} {et~al.}(2011{\natexlab{a}}){Chiavassa}, {Freytag},
  {Masseron}, \& {Plez}}]{chiavassa2011}
{Chiavassa}, A., {Freytag}, B., {Masseron}, T., \& {Plez}, B.
  2011{\natexlab{a}}, \aap, 535, A22, \dodoi{10.1051/0004-6361/201117463}

\bibitem[{{Chiavassa} {et~al.}(2010){Chiavassa}, {Haubois}, {Young}, {Plez},
  {Josselin}, {Perrin}, \& {Freytag}}]{chiavassaII}
{Chiavassa}, A., {Haubois}, X., {Young}, J.~S., {et~al.} 2010, \aap, 515, A12,
  \dodoi{10.1051/0004-6361/200913907}

\bibitem[{{Chiavassa} {et~al.}(2009){Chiavassa}, {Plez}, {Josselin}, \&
  {Freytag}}]{chiavassa1}
{Chiavassa}, A., {Plez}, B., {Josselin}, E., \& {Freytag}, B. 2009, \aap, 506,
  1351, \dodoi{10.1051/0004-6361/200911780}

\bibitem[{{Chiavassa} {et~al.}(2011{\natexlab{b}}){Chiavassa}, {Pasquato},
  {Jorissen}, {Sacuto}, {Babusiaux}, {Freytag}, {Ludwig}, {Cruzal{\`e}bes},
  {Rabbia}, {Spang}, \& {Chesneau}}]{chiavassa2010}
{Chiavassa}, A., {Pasquato}, E., {Jorissen}, A., {et~al.} 2011{\natexlab{b}},
  \aap, 528, A120, \dodoi{10.1051/0004-6361/201015768}

\bibitem[{{Climent} {et~al.}(2020){Climent}, {Wittkowski}, {Chiavassa},
  {Baron}, {Marcaide}, {Guirado}, {Freytag}, {H{\"o}fner}, {Haubois}, \&
  {Woillez}}]{climent2020}
{Climent}, J.~B., {Wittkowski}, M., {Chiavassa}, A., {et~al.} 2020, \aap, 635,
  A160, \dodoi{10.1051/0004-6361/201936734}

\bibitem[{{Cotton} {et~al.}(2020){Cotton}, {Bailey}, {Horta}, {Norris}, \&
  {Lomax}}]{polarimetry}
{Cotton}, D.~V., {Bailey}, J., {Horta}, A.~D., {Norris}, B. R.~M., \& {Lomax},
  J.~R. 2020, Research Notes of the American Astronomical Society, 4, 39,
  \dodoi{10.3847/2515-5172/ab7f2f}

\bibitem[{{Cushing} {et~al.}(2004){Cushing}, {Vacca}, \&
  {Rayner}}]{cushing_spextool}
{Cushing}, M.~C., {Vacca}, W.~D., \& {Rayner}, J.~T. 2004, \pasp, 116, 362,
  \dodoi{10.1086/382907}

\bibitem[{{Cutri} {et~al.}(2003){Cutri}, {Skrutskie}, {van Dyk}, {Beichman},
  {Carpenter}, {Chester}, {Cambresy}, {Evans}, {Fowler}, {Gizis}, {Howard},
  {Huchra}, {Jarrett}, {Kopan}, {Kirkpatrick}, {Light}, {Marsh}, {McCallon},
  {Schneider}, {Stiening}, {Sykes}, {Weinberg}, {Wheaton}, {Wheelock}, \&
  {Zacarias}}]{2masscutri}
{Cutri}, R.~M., {Skrutskie}, M.~F., {van Dyk}, S., {et~al.} 2003, VizieR Online
  Data Catalog, II/246

\bibitem[{{Davies} {et~al.}(2013){Davies}, {Kudritzki}, {Plez}, {Trager},
  {Lan{\c c}on}, {Gazak}, {Bergemann}, {Evans}, \&
  {Chiavassa}}]{davies_temperatures}
{Davies}, B., {Kudritzki}, R.-P., {Plez}, B., {et~al.} 2013, \apj, 767, 3,
  \dodoi{10.1088/0004-637X/767/1/3}

\bibitem[{{De Beck} {et~al.}(2010){De Beck}, {Decin}, {de Koter}, {Justtanont},
  {Verhoelst}, {Kemper}, \& {Menten}}]{debeck2010}
{De Beck}, E., {Decin}, L., {de Koter}, A., {et~al.} 2010, \aap, 523, A18,
  \dodoi{10.1051/0004-6361/200913771}

\bibitem[{{Dharmawardena} {et~al.}(2020){Dharmawardena}, {Mairs}, {Scicluna},
  {Bell}, {McDonald}, {Menten}, {Weiss}, \& {Zijlstra}}]{betelcoolspot}
{Dharmawardena}, T.~E., {Mairs}, S., {Scicluna}, P., {et~al.} 2020, \apjl, 897,
  L9, \dodoi{10.3847/2041-8213/ab9ca6}

\bibitem[{{Freytag} {et~al.}(1997){Freytag}, {Holweger}, {Steffen}, \&
  {Ludwig}}]{freytag1}
{Freytag}, B., {Holweger}, H., {Steffen}, M., \& {Ludwig}, H.-G. 1997, in
  Science with the VLT Interferometer, ed. F.~{Paresce}, 316

\bibitem[{{Gaia Collaboration}(2018)}]{gaia2cat}
{Gaia Collaboration}. 2018, VizieR Online Data Catalog, I/345

\bibitem[{{Gaia Collaboration} {et~al.}(2018){Gaia Collaboration}, {Brown},
  {Vallenari}, {Prusti}, {de Bruijne}, {Babusiaux}, {Bailer-Jones}, {Biermann},
  {Evans}, \& {Eyer}}]{gaia2}
{Gaia Collaboration}, {Brown}, A.~G.~A., {Vallenari}, A., {et~al.} 2018, \aap,
  616, A1, \dodoi{10.1051/0004-6361/201833051}

\bibitem[{{Gomes} {et~al.}(2017){Gomes}, {Garcia}, \&
  {Thi{\'e}baut}}]{gomes2017}
{Gomes}, N., {Garcia}, P. J.~V., \& {Thi{\'e}baut}, {\'E}. 2017, \mnras, 465,
  3823, \dodoi{10.1093/mnras/stw2896}

\bibitem[{{Guinan} \& {Wasatonic}(2020)}]{dimming}
{Guinan}, E.~F., \& {Wasatonic}, R.~J. 2020, The Astronomer's Telegram, 13410,
  1

\bibitem[{{Gustafsson} {et~al.}(2008){Gustafsson}, {Edvardsson}, {Eriksson},
  {J{\o}rgensen}, {Nordlund}, \& {Plez}}]{gustafsson_grid_2008-1}
{Gustafsson}, B., {Edvardsson}, B., {Eriksson}, K., {et~al.} 2008, \aap, 486,
  951, \dodoi{10.1051/0004-6361:200809724}

\bibitem[{{Harvey}(1985)}]{Harvey}
{Harvey}, J. 1985, in ESA Special Publication, Vol. 235, Future Missions in
  Solar, Heliospheric \& Space Plasma Physics, ed. E.~{Rolfe} \& B.~{Battrick},
  199

\bibitem[{{Haubois} {et~al.}(2009){Haubois}, {Perrin}, {Lacour}, {Verhoelst},
  {Meimon}, {Mugnier}, {Thi{\'e}baut}, {Berger}, {Ridgway}, {Monnier},
  {Millan-Gabet}, \& {Traub}}]{haubois}
{Haubois}, X., {Perrin}, G., {Lacour}, S., {et~al.} 2009, \aap, 508, 923,
  \dodoi{10.1051/0004-6361/200912927}

\bibitem[{{Hestroffer}(1997)}]{hestlaw}
{Hestroffer}, D. 1997, \aap, 327, 199

\bibitem[{{Josselin} \& {Plez}(2007)}]{josselin_plez}
{Josselin}, E., \& {Plez}, B. 2007, \aap, 469, 671,
  \dodoi{10.1051/0004-6361:20066353}

\bibitem[{{Joyce} {et~al.}(2020){Joyce}, {Leung}, {Moln{\'a}r}, {Ireland},
  {Kobayashi}, \& {Nomoto}}]{joyce}
{Joyce}, M., {Leung}, S.-C., {Moln{\'a}r}, L., {et~al.} 2020, \apj, 902, 63,
  \dodoi{10.3847/1538-4357/abb8db}

\bibitem[{{Kiss} {et~al.}(2010){Kiss}, {Monnier}, {Bedding}, {Tuthill}, {Zhao},
  {Ireland}, \& {ten Brummelaar}}]{kissrsg}
{Kiss}, L.~L., {Monnier}, J.~D., {Bedding}, T.~R., {et~al.} 2010, in
  Astronomical Society of the Pacific Conference Series, Vol. 425, Hot and
  Cool: Bridging Gaps in Massive Star Evolution, ed. C.~{Leitherer}, P.~D.
  {Bennett}, P.~W. {Morris}, \& J.~T. {Van Loon}, 140.
\newblock \doarXiv{0902.2602}

\bibitem[{{Kiss} {et~al.}(2006){Kiss}, {Szab{\'o}}, \&
  {Bedding}}]{kiss_variability}
{Kiss}, L.~L., {Szab{\'o}}, G.~M., \& {Bedding}, T.~R. 2006, \mnras, 372, 1721,
  \dodoi{10.1111/j.1365-2966.2006.10973.x}

\bibitem[{{Kravchenko} {et~al.}(2019){Kravchenko}, {Chiavassa}, {Van Eck},
  {Jorissen}, {Merle}, {Freytag}, \& {Plez}}]{kravchenko2019}
{Kravchenko}, K., {Chiavassa}, A., {Van Eck}, S., {et~al.} 2019, arXiv
  e-prints, arXiv:1910.04657.
\newblock \doarXiv{1910.04657}

\bibitem[{{Kurucz}(1993)}]{syntheorg}
{Kurucz}, R.~L. 1993, {SYNTHE spectrum synthesis programs and line data}

\bibitem[{{Lacour} {et~al.}(2008){Lacour}, {Meimon}, {Thi{\'e}baut}, {Perrin},
  {Verhoelst}, {Pedretti}, {Schuller}, {Mugnier}, {Monnier}, {Berger},
  {Haubois}, {Poncelet}, {Le Besnerais}, {Eriksson}, {Millan-Gabet}, {Ragland},
  {Lacasse}, \& {Traub}}]{arcturusiota}
{Lacour}, S., {Meimon}, S., {Thi{\'e}baut}, E., {et~al.} 2008, \aap, 485, 561,
  \dodoi{10.1051/0004-6361:200809611}

\bibitem[{{Lan{\c{c}}on} {et~al.}(2007){Lan{\c{c}}on}, {Hauschildt}, {Ladjal},
  \& {Mouhcine}}]{phoenixrsg}
{Lan{\c{c}}on}, A., {Hauschildt}, P.~H., {Ladjal}, D., \& {Mouhcine}, M. 2007,
  \aap, 468, 205, \dodoi{10.1051/0004-6361:20065824}

\bibitem[{{Lancon} \& {Rocca-Volmerange}(1992)}]{oldazcyg}
{Lancon}, A., \& {Rocca-Volmerange}, B. 1992, \aaps, 96, 593

\bibitem[{{Lester} \& {Neilson}(2008)}]{SAtlas}
{Lester}, J.~B., \& {Neilson}, H.~R. 2008, \aap, 491, 633,
  \dodoi{10.1051/0004-6361:200810578}

\bibitem[{{Levesque} \& {Massey}(2020)}]{levesequebetel}
{Levesque}, E.~M., \& {Massey}, P. 2020, \apjl, 891, L37,
  \dodoi{10.3847/2041-8213/ab7935}

\bibitem[{{Levesque} {et~al.}(2005){Levesque}, {Massey}, {Olsen}, {Plez},
  {Josselin}, {Maeder}, \& {Meynet}}]{levesque_2005}
{Levesque}, E.~M., {Massey}, P., {Olsen}, K.~A.~G., {et~al.} 2005, \apj, 628,
  973, \dodoi{10.1086/430901}

\bibitem[{{L{\'o}pez Ariste} {et~al.}(2018){L{\'o}pez Ariste}, {Mathias},
  {Tessore}, {L{\`e}bre}, {Auri{\`e}re}, {Petit}, {Ikhenache}, {Josselin},
  {Morin}, \& {Montarg{\`e}s}}]{spectropolar}
{L{\'o}pez Ariste}, A., {Mathias}, P., {Tessore}, B., {et~al.} 2018, \aap, 620,
  A199, \dodoi{10.1051/0004-6361/201834178}

\bibitem[{{Massey} {et~al.}(2005){Massey}, {Plez}, {Levesque}, {Olsen},
  {Clayton}, \& {Josselin}}]{reddening}
{Massey}, P., {Plez}, B., {Levesque}, E.~M., {et~al.} 2005, \apj, 634, 1286,
  \dodoi{10.1086/497065}

\bibitem[{{Mauron} \& {Josselin}(2011)}]{mauron_josselin}
{Mauron}, N., \& {Josselin}, E. 2011, \aap, 526, A156,
  \dodoi{10.1051/0004-6361/201013993}

\bibitem[{{Meynet} {et~al.}(2015){Meynet}, {Chomienne}, {Ekstr{\"o}m},
  {Georgy}, {Granada}, {Groh}, {Maeder}, {Eggenberger}, {Levesque}, \&
  {Massey}}]{meynet_evolution}
{Meynet}, G., {Chomienne}, V., {Ekstr{\"o}m}, S., {et~al.} 2015, \aap, 575,
  A60, \dodoi{10.1051/0004-6361/201424671}

\bibitem[{{Monnier} {et~al.}(2004){Monnier}, {Berger}, {Millan-Gabet}, \& {ten
  Brummelaar}}]{monnier_michigan_2004}
{Monnier}, J.~D., {Berger}, J.-P., {Millan-Gabet}, R., \& {ten Brummelaar},
  T.~A. 2004, in \procspie, Vol. 5491, New Frontiers in Stellar Interferometry,
  ed. W.~A. {Traub}, 1370, \dodoi{10.1117/12.550804}

\bibitem[{{Monnier} {et~al.}(2007){Monnier}, {Zhao}, {Pedretti}, {Thureau},
  {Ireland}, {Muirhead}, {Berger}, {Millan-Gabet}, {Van Belle}, {ten
  Brummelaar}, {McAlister}, {Ridgway}, {Turner}, {Sturmann}, {Sturmann}, \&
  {Berger}}]{monnier_imaging_2007}
{Monnier}, J.~D., {Zhao}, M., {Pedretti}, E., {et~al.} 2007, Science, 317, 342,
  \dodoi{10.1126/science.1143205}

\bibitem[{{Monnier} {et~al.}(2012){Monnier}, {Che}, {Zhao}, {Ekstr{\"o}m},
  {Maestro}, {Aufdenberg}, {Baron}, {Georgy}, {Kraus}, {McAlister}, {Pedretti},
  {Ridgway}, {Sturmann}, {Sturmann}, {ten Brummelaar}, {Thureau}, {Turner}, \&
  {Tuthill}}]{monnier_resolving_2012}
{Monnier}, J.~D., {Che}, X., {Zhao}, M., {et~al.} 2012, \apjl, 761, L3,
  \dodoi{10.1088/2041-8205/761/1/L3}

\bibitem[{{Montarg{\`e}s} {et~al.}(2014){Montarg{\`e}s}, {Kervella}, {Perrin},
  {Ohnaka}, {Chiavassa}, {Ridgway}, \& {Lacour}}]{montarges2014}
{Montarg{\`e}s}, M., {Kervella}, P., {Perrin}, G., {et~al.} 2014, \aap, 572,
  A17, \dodoi{10.1051/0004-6361/201423538}

\bibitem[{{Montarg{\`e}s} {et~al.}(2018){Montarg{\`e}s}, {Norris}, {Chiavassa},
  {Tessore}, {L{\`e}bre}, \& {Baron}}]{montargescetau}
{Montarg{\`e}s}, M., {Norris}, R., {Chiavassa}, A., {et~al.} 2018, \aap, 614,
  A12, \dodoi{10.1051/0004-6361/201731471}

\bibitem[{{Montarg{\`e}s} {et~al.}(2021){Montarg{\`e}s}, {Cannon}, {Lagadec},
  {de Koter}, {Kervella}, {Sanchez-Bermudez}, {Paladini}, {Cantalloube},
  {Decin}, {Scicluna}, {Kravchenko}, {Dupree}, {Ridgway}, {Wittkowski},
  {Anugu}, {Norris}, {Rau}, {Perrin}, {Chiavassa}, {Kraus}, {Monnier},
  {Millour}, {Le Bouquin}, {Haubois}, {Lopez}, {Stee}, \&
  {Danchi}}]{montarges2021}
{Montarg{\`e}s}, M., {Cannon}, E., {Lagadec}, E., {et~al.} 2021, \nat, 594,
  365, \dodoi{10.1038/s41586-021-03546-8}

\bibitem[{{Mourard} {et~al.}(2015){Mourard}, {Monnier}, {Meilland}, {Gies},
  {Millour}, {Benisty}, {Che}, {Grundstrom}, {Ligi}, \&
  {Schaefer}}]{2015mourard}
{Mourard}, D., {Monnier}, J.~D., {Meilland}, A., {et~al.} 2015, \aap, 577, A51,
  \dodoi{10.1051/0004-6361/201425141}

\bibitem[{{Ochsenbein} {et~al.}(2000){Ochsenbein}, {Bauer}, \&
  {Marcout}}]{vizier}
{Ochsenbein}, F., {Bauer}, P., \& {Marcout}, J. 2000, \aaps, 143, 23,
  \dodoi{10.1051/aas:2000169}

\bibitem[{{Paladini} {et~al.}(2018){Paladini}, {Baron}, {Jorissen}, {Le
  Bouquin}, {Freytag}, {van Eck}, {Wittkowski}, {Hron}, {Chiavassa}, \&
  {Berger}}]{paladiniagb}
{Paladini}, C., {Baron}, F., {Jorissen}, A., {et~al.} 2018, \nat, 553, 310,
  \dodoi{10.1038/nature25001}

\bibitem[{{Pauls} {et~al.}(2005){Pauls}, {Young}, {Cotton}, \&
  {Monnier}}]{pauls2005}
{Pauls}, T.~A., {Young}, J.~S., {Cotton}, W.~D., \& {Monnier}, J.~D. 2005,
  \pasp, 117, 1255, \dodoi{10.1086/444523}

\bibitem[{{Percy} \& {Sato}(2009)}]{percy1}
{Percy}, J.~R., \& {Sato}, H. 2009, \jrasc, 103, 11

\bibitem[{{Pickles}(1998)}]{pickles}
{Pickles}, A.~J. 1998, \pasp, 110, 863, \dodoi{10.1086/316197}

\bibitem[{{Plez}(2012)}]{turbspectrumcode}
{Plez}, B. 2012, {Turbospectrum: Code for spectral synthesis}, Astrophysics
  Source Code Library,  Astrophysics Source Code Library.
\newblock \doeprint{1205.004}

\bibitem[{{Rayner} {et~al.}(2009){Rayner}, {Cushing}, \& {Vacca}}]{irtflibrary}
{Rayner}, J.~T., {Cushing}, M.~C., \& {Vacca}, W.~D. 2009, \apjs, 185, 289,
  \dodoi{10.1088/0067-0049/185/2/289}

\bibitem[{{Rayner} {et~al.}(2003){Rayner}, {Toomey}, {Onaka}, {Denault},
  {Stahlberger}, {Vacca}, {Cushing}, \& {Wang}}]{rayner_spex}
{Rayner}, J.~T., {Toomey}, D.~W., {Onaka}, P.~M., {et~al.} 2003, \pasp, 115,
  362, \dodoi{10.1086/367745}

\bibitem[{{Ren} \& {Jiang}(2020)}]{RSGgranulation2020}
{Ren}, Y., \& {Jiang}, B.-W. 2020, \apj, 898, 24,
  \dodoi{10.3847/1538-4357/ab9c17}

\bibitem[{{Renard} {et~al.}(2011){Renard}, {Thi{\'e}baut}, \&
  {Malbet}}]{renard2011}
{Renard}, S., {Thi{\'e}baut}, E., \& {Malbet}, F. 2011, \aap, 533, A64,
  \dodoi{10.1051/0004-6361/201016263}

\bibitem[{{Sbordone}(2005)}]{synthelinux}
{Sbordone}, L. 2005, Memorie della Societa Astronomica Italiana Supplementi, 8,
  61.
\newblock \doarXiv{astro-ph/0509337}

\bibitem[{{Schwarzschild}(1975)}]{schwarzschild1975}
{Schwarzschild}, M. 1975, \apj, 195, 137, \dodoi{10.1086/153313}

\bibitem[{{Stothers}(1972)}]{stothers1972}
{Stothers}, R. 1972, \aap, 18, 325

\bibitem[{{ten Brummelaar} {et~al.}(2005){ten Brummelaar}, {McAlister},
  {Ridgway}, {Bagnuolo}, {Turner}, {Sturmann}, {Sturmann}, {Berger}, {Ogden},
  \& {Cadman}}]{ten_brummelaar_first_2005}
{ten Brummelaar}, T.~A., {McAlister}, H.~A., {Ridgway}, S.~T., {et~al.} 2005,
  \apj, 628, 453, \dodoi{10.1086/430729}

\bibitem[{{Thiebaut}(2002)}]{thiebaut2002}
{Thiebaut}, E. 2002, in \procspie, Vol. 4847, Astronomical Data Analysis II,
  ed. J.-L. {Starck} \& F.~D. {Murtagh}, 174--183, \dodoi{10.1117/12.461151}

\bibitem[{{Thi{\'e}baut} \& {Young}(2017)}]{thiebaut2017}
{Thi{\'e}baut}, {\'E}., \& {Young}, J. 2017, Journal of the Optical Society of
  America A, 34, 904, \dodoi{10.1364/JOSAA.34.000904}

\bibitem[{{Trampedach} {et~al.}(2013){Trampedach}, {Asplund}, {Collet},
  {Nordlund}, \& {Stein}}]{trempedach2013}
{Trampedach}, R., {Asplund}, M., {Collet}, R., {Nordlund}, {\AA}., \& {Stein},
  R.~F. 2013, \apj, 769, 18, \dodoi{10.1088/0004-637X/769/1/18}

\bibitem[{{Tremblay} {et~al.}(2013){Tremblay}, {Ludwig}, {Freytag}, {Steffen},
  \& {Caffau}}]{tremblay2013}
{Tremblay}, P.~E., {Ludwig}, H.~G., {Freytag}, B., {Steffen}, M., \& {Caffau},
  E. 2013, \aap, 557, A7, \dodoi{10.1051/0004-6361/201321878}

\bibitem[{{Tuthill} {et~al.}(1997){Tuthill}, {Haniff}, \&
  {Baldwin}}]{tuthill1997}
{Tuthill}, P.~G., {Haniff}, C.~A., \& {Baldwin}, J.~E. 1997, \mnras, 285, 529,
  \dodoi{10.1093/mnras/285.3.529}

\bibitem[{{Vacca} {et~al.}(2003){Vacca}, {Cushing}, \& {Rayner}}]{vacca2003}
{Vacca}, W.~D., {Cushing}, M.~C., \& {Rayner}, J.~T. 2003, \pasp, 115, 389,
  \dodoi{10.1086/346193}

\bibitem[{{van Belle} {et~al.}(2009){van Belle}, {Creech-Eakman}, \&
  {Hart}}]{vanbelle}
{van Belle}, G.~T., {Creech-Eakman}, M.~J., \& {Hart}, A. 2009, \mnras, 394,
  1925, \dodoi{10.1111/j.1365-2966.2008.14146.x}

\bibitem[{{van Loon} {et~al.}(2005){van Loon}, {Cioni}, {Zijlstra}, \&
  {Loup}}]{vanloon}
{van Loon}, J.~T., {Cioni}, M.-R.~L., {Zijlstra}, A.~A., \& {Loup}, C. 2005,
  \aap, 438, 273, \dodoi{10.1051/0004-6361:20042555}

\bibitem[{{Villaume} {et~al.}(2017){Villaume}, {Conroy}, {Johnson}, {Rayner},
  {Mann}, \& {van Dokkum}}]{fluxcal}
{Villaume}, A., {Conroy}, C., {Johnson}, B., {et~al.} 2017, \apjs, 230, 23,
  \dodoi{10.3847/1538-4365/aa72ed}

\bibitem[{{Wittkowski} {et~al.}(2017{\natexlab{a}}){Wittkowski}, {Abell{\'a}n},
  {Arroyo-Torres}, {Chiavassa}, {Guirado}, {Marcaide}, {Alberdi}, {de Wit},
  {Hofmann}, {Meilland }, {Millour}, {Mohamed}, \&
  {Sanchez-Bermudez}}]{wittkowski2017}
{Wittkowski}, M., {Abell{\'a}n}, F.~J., {Arroyo-Torres}, B., {et~al.}
  2017{\natexlab{a}}, \aap, 606, L1, \dodoi{10.1051/0004-6361/201731569}

\bibitem[{{Wittkowski} {et~al.}(2017{\natexlab{b}}){Wittkowski}, {Abell{\'a}n},
  {Arroyo-Torres}, {Chiavassa}, {Guirado}, {Marcaide}, {Alberdi}, {de Wit},
  {Hofmann}, \& {Meilland }}]{v776cen}
---. 2017{\natexlab{b}}, \aap, 606, L1, \dodoi{10.1051/0004-6361/201731569}

\bibitem[{{Young} {et~al.}(2000){Young}, {Baldwin}, {Boysen}, {Haniff},
  {Lawson}, {Mackay}, {Pearson}, {Rogers}, {St.-Jacques}, {Warner}, {Wilson},
  \& {Wilson}}]{youngbetelgeuse}
{Young}, J.~S., {Baldwin}, J.~E., {Boysen}, R.~C., {et~al.} 2000, \mnras, 315,
  635, \dodoi{10.1046/j.1365-8711.2000.03438.x}

\bibitem[{{Zhao} {et~al.}(2008){Zhao}, {Gies}, {Monnier}, {Thureau},
  {Pedretti}, {Baron}, {Merand}, {ten Brummelaar}, {McAlister}, {Ridgway},
  {Turner}, {Sturmann}, {Sturmann}, {Farrington}, \& {Goldfinger}}]{zhao2008}
{Zhao}, M., {Gies}, D., {Monnier}, J.~D., {et~al.} 2008, \apjl, 684, L95,
  \dodoi{10.1086/592146}

\bibitem[{{Zhao} {et~al.}(2011){Zhao}, {Monnier}, {Che}, {Pedretti}, {Thureau},
  {Schaefer}, {ten Brummelaar}, {M{\'e}rand}, {Ridgway}, {McAlister}, {Turner},
  {Sturmann}, {Sturmann}, {Goldfinger}, \& {Farrington}}]{zhao_toward_2011}
{Zhao}, M., {Monnier}, J.~D., {Che}, X., {et~al.} 2011, \pasp, 123, 964,
  \dodoi{10.1086/661762}

\end{thebibliography}
